\documentclass{aastex6}

\usepackage{color}
\usepackage{amsmath}
\usepackage{lineno}
\usepackage{longtable}
\newcounter{reaction}
 \usepackage{ulem}  

\def\simge{\mathrel{%
    \rlap{\raise 0.511ex \hbox{$>$}}{\lower 0.511ex \hbox{$\sim$}}}}
\def\simle{\mathrel{
    \rlap{\raise 0.511ex \hbox{$<$}}{\lower 0.511ex \hbox{$\sim$}}}}

\slugcomment{Accepted 19 April 2016 by The Astrophysical Journal}
\shorttitle{Photolytic Hazes in 51 Eri b}
\shortauthors{Zahnle et al.}

\begin{document}
\title{Photolytic Hazes in the Atmosphere of 51 Eri b}

\author{Kevin J. Zahnle\altaffilmark{1} and Mark S. Marley}
\affil{NASA Ames Research Center, Moffett Field, CA 94035}

\author{Caroline V.\ Morley}
\affil{Department of Astronomy and Astrophysics, University of California, Santa Cruz, CA 95064}

\and

\author{Julianne I.\ Moses}
\affil{Space Science Institute, 4750 Walnut Street, Suite 205, Boulder, CO 80301}

\altaffiltext{1}{Kevin.J.Zahnle@NASA.gov}


\begin{abstract}

We use a 1D model
to address photochemistry and possible haze formation in the irradiated warm Jupiter, 51 Eridani b.
The intended focus was to be carbon, but sulfur photochemistry turns out to be important. 
The case for organic photochemical hazes is intriguing but falls short of being compelling.
If organic hazes form, they are likeliest to do so if vertical mixing in 51 Eri b is weaker than in Jupiter,
and they would be found below the altitudes where methane and water are photolyzed.  
The more novel result is that photochemistry turns H$_2$S into elemental sulfur, here treated as S$_8$. 
In the cooler models, S$_8$ is predicted to condense in optically thick clouds of solid sulfur particles,
whilst in the warmer models S$_8$ remains a vapor along with several other sulfur allotropes 
that are both visually striking and potentially observable. 
 For 51 Eri b, the division between models with and without condensed sulfur is at 
 an effective temperature of 700 K, which is within error its actual effective temperature;
  the local temperature where sulfur condenses is between 280 and 320 K.
 The sulfur photochemistry we have discussed is quite general and ought to be found
in a wide variety of worlds over a broad temperature range, both colder and hotter
than the 650-750 K range studied here, 
and we show that products of sulfur photochemistry will be nearly as abundant  
on planets where the UV irradiation is orders of magnitude
 weaker than it is on 51 Eri b.  

\end{abstract}


\section{Introduction}

The star 51 Eridani is a pre-main-sequence F dwarf that is only 20 million years old.  
Direct-imaging observations with GPI (Gemini Planet Imager)
 reveal that the star is orbited by a self-radiant young Jupiter, designated 51 Eri b, that emits with
an effective temperature on the order of $T_{\rm eff} = 700 \pm 50$ K \citep{Macintosh2015}.
Thermal evolution models predict that a 20 Myr old jovian planet with 51 Eri b's luminosity will
have mass $\sim 2 M_{\rm Jup}$ and radius $\sim 1 R_{\rm Jup}$ \citep{Macintosh2015}. 

Comparison by \citet{Macintosh2015} of the available spectral and photometric data to spectral models reveal
that while the planet shows methane in absorption, methane is depleted compared to 
thermochemical equilibrium. Carbon monoxide is therefore expected to be abundant
but available data do not yet constrain it.
Spectral matching with radiative transfer models also strongly suggest that clouds, possibly patchy, are present in the atmosphere \citep{Macintosh2015}.
However, the planet is cool enough that silicate clouds 
if present would be confined to levels deep beneath the photosphere and thus unlikely to affect what can be seen. 
Clouds of salts like Na$_2$S and NaCl are possible, but even these would be expected
to be confined to levels beneath the photosphere 
by the low temperature of the planet \citep{Morley2012}.

In this study we use a 1D chemical kinetics model to ask whether, and under what conditions,
photochemical hazes are likely to form in the atmosphere of 51 Eri b
and perhaps be the agent responsible for the observed particulate opacity.
We consider two candidates, one familiar, the other more novel.
The familiar candidate is an organic photochemical haze loosely analogous to the hazes seen over Titan,
Pluto, or Beijing.
Such hazes have been proposed by many workers, but to date
the case for them has been inconclusive \citep{Moses2014}.
We will find here that a reasonable case for a photochemical organic haze in 51 Eri b can be made, but we do not follow the chain
of polymerization reactions to molecules big enough and refractory enough that we can prove that condensates actually form.   
The novel candidate is sulfur. 
With sulfur we can follow a much shorter chain of polymerization to the point where sulfur condenses.
We will show here that a good case
can be made for the presence of photochemical sulfur clouds in the atmosphere of 51 Eri b. 

This paper begins with a brief review of some related previous work.
We next reprise our own model.
In section \ref{Results} we present results for models that span
 the parameter space in which 51 Eri b probably resides.
 We will find that for some of these parameters organic hazes might form, and for some parameters sulfur clouds will form,
 for some parameters both might form, and for some parameters neither kind of haze is likely to form.
 The important role of sulfur raises the issue that much of the sulfur chemistry is very poorly known.
 In section \ref{section:sensitivity} we perform a series of sensitivity tests to examine how the model responds to alternative
 assumptions about sulfur's photochemistry.

\section{Previous Models}

The possibility that photochemical organic hazes might be important in irradiated brown dwarfs was first raised by \citet{Griffith1998}.
It remains an open question.

The first exoplanet photochemical models showed that small hydrocarbons would not condense 
in the solar composition atmospheres of hot Jupiters \citep{Liang2003,Liang2004}.
\citet{Line2010,Line2011} confirmed this result for hot Jupiters. 
They predicted the flowering of a rich disequilibrium non-methane hydrocarbon (NMHC) photochemistry
in the cooler ($\sim 800$ K) and presumptively metal-rich warm Neptune GJ 436b,
but stopped short of concluding that the chemistry would necessarily lead to smogs. 
Moses and coworkers \citep{Moses2011,Visscher2011,Moses2013a,Moses2013b,Moses2014} 
extended this model to bigger molecules, concluding   
that  ``complex hydrocarbons and nitriles might produce high-altitude photochemical hazes'' \citep{Moses2014}.
 On the other hand, as \citet{Moses2014} also points out, methane has not yet been seen in GJ 436b.

There are several other models of exoplanet thermochemistry and photochemistry that have been 
used to address a variety of hydrogen-rich exoplanets, from Jupiters to Neptunes to super-Earths,
but none of them go as far as predicting the photochemical production of organic hazes.
\citet{Venot2012} examined C-N-O photochemistry on HD 189733b and HD 209458b; 
\citet{Kopparapu2012} explored the effect of the C/O ratio on the hot Jupiter WASP-12b;
\citet{Venot2013} used high-temperature UV cross sections to study the effect of CO$_2$ photolysis on the warm Neptune GJ 436b;
\cite{Hu2014} addressed temperature and elemental abundances in super-Earths and mini-Neptunes, with application to GJ 1214b, HD 97658b, and 55 Cnc e; 
\citet{Agundez2014b} added tidal heating and metallicity variations to GJ 436b;
\citet{Venot2014} looked at temperature, metallicity, UV flux, tidal heating, and atmospheric mixing in warm Neptunes, with application to GJ 3470b and GJ 436b;
\cite{Miguel2014} took into account stellar type and orbital distance; 
\cite{Miguel2015} focused on Lyman $\alpha$ irradiation of GJ 436b and other warm Neptunes;
\citet{Koskinen2013} and \citet{Lavvas2014} addressed ion chemistry;
\citet{Agundez2012,Agundez2014a} used a 2D model to address the horizontal quenching that occurs when winds carry hot air to cold places;
and \citet{Benneke2015} combined photochemistry with retrievals from exoplanet transit spectra to mine for C/O ratios in several planets. 

Two recent models do include heavier organic molecules \citep{Rimmer2016, Venot2015}.
\citet{Rimmer2016} compile an extensive reaction network that includes both neutral and ion chemistry;
 they pay particular attention to the formation of prebiotic molecules like glycine, but they do not yet address photochemical hazes.
\citet{Venot2015} have expanded their reaction network to include selected hydrocarbons with as many as eight carbon atoms.
A plus is that their reaction network has been tested against combustion experiments.
On the other hand, it should be borne in mind that complex models of complex systems often achieve empirical agreement by cancellation of errors,
and that things can go awry when the model is applied to new conditions.
 \citet{Venot2015} compute that cyclohexadiene (cC$_6$H$_8$ --- the ``c'' means cyclic --- an obscure but reasonably stable
 molecule that is close kin to benzene, the basic building block of polycyclic aromatic hydrocarbons (PAHs) and soots)
 is a major photochemical product in 500 K stratospheres, exceeding even acetylene (C$_2$H$_2$) and CO in abundance.  
 Although \citet{Venot2015} do not mention photochemical hazes, it is obvious that cyclohexadiene
 is well along the path to building smog.  
 However, the stated pathway for cC$_6$H$_8$ formation goes through  
 \begin{equation}\tag{R60r}
 \mathrm{C}_2\mathrm{H}_2 + \mathrm{C}_2\mathrm{H}_2  \rightarrow \mathrm{nC}_4\mathrm{H}_3 + \mathrm{H},
 \end{equation}
 a very endothermic reaction that we will encounter again in section \ref{Results:carbon} when we discuss its 
 reverse (the ``n'' indicates ``normal'' to indicate that the molecule is linear).
 We estimate that the rate for R60r is $k_{60r}= 3\times 10^{-13} e^{-33000/T}$ cm$^{3}$/s, which at 500 K is very close to never.
 It is hard to imagine how a reaction with such a huge activation energy could actually be a major factor in a planetary atmosphere.
 
We have used our own code to address photochemistry and thermochemistry in giant planets and brown dwarfs
\citet{Zahnle1995, Zahnle2009,Zahnle2014}.
Early versions of this code (2011 and earlier) had some issues with the implementation of thermochemical
equilibrium that were corrected after consultations with Channon Visscher. 
\citet{Miller-Ricci2012} and \citet{Morley2013,Morley2015} used the corrected code 
 to address photochemistry in the warm ($T_{\rm eff}\approx 550$ K) super earth GJ 1214b and similar planets.
They suggested that hazes should form when reduced organic radicals like CH$_3$ (building blocks of
bigger organic molecules) were more abundant than OH.
If so, NMHCs can be abundant enough that organic hazes show potential to provide
a viable alternative to clouds of other condensible substances such as Na$_2$S.     
 However, as with GJ 436b, methane has not been seen in GJ 1214b.

\section{Model Details}

We use a 1D kinetics code to simulate atmospheric photochemistry. 
Such codes parameterize vertical transport as a diffusive process 
with an ``eddy diffusion coefficient,'' denoted $K_{zz}$ [cm$^2$/s].
Volume mixing ratios $f_i$ of species $i$ are obtained by solving continuity
\begin{equation}
\label{eq_one}
N{\partial f_i \over \partial t} = P_i - L_i N f_i - {\partial \phi_i \over \partial z} 
\end{equation}
\noindent and diffusion
\begin{equation}
\label{eq_two}
\phi_i = b_{ia} f_i \left( {m_a g\over kT} - {m_i g\over kT}\right) - \left( b_{ia} + K_{zz}N\right) {\partial f_i \over \partial z}
\end{equation}
\noindent equations for each species.
In these equations $N$ is the total number density (cm$^{-3}$);
$P_i - L_i N f_i$ represent chemical production and loss terms, respectively;
$\phi_i$ is the upward flux (cm$^{-2}$s$^{-1}$); $b_{ia}$, the binary diffusion coefficient (cm$^{-1}$s$^{-1}$) between $i$ and the background atmosphere $a$, describes molecular diffusion ; $m_a$ and $m_i$ are the molecular masses of $a$ and $i$ (grams); and $g$ is surface gravity (units are cm/s$^{2}$ in Eq \ref{eq_two}).
 
For the base model we use 481 forward chemical reactions and 42 photolysis reactions for 78 chemical species made from H, C, O, N, and S.
We supplement these with 12 additional reactions and two additional species for sensitivity tests. 
Every forward chemical reaction (e.g., ${\rm CO} + {\rm OH} \rightarrow {\rm CO}_2 + {\rm H}$) is balanced by the corresponding reverse reaction (e.g., ${\rm CO}_2 + {\rm H} \rightarrow  {\rm CO} + {\rm OH} $) at a rate determined by thermodynamic equilibrium.
 We have not included reverses of the photolysis reactions; that is, we include reactions such as 
 ${\rm H}_2{\rm O} + h\nu \rightarrow  {\rm H} + {\rm OH} $, but we do not include
 ${\rm H} + {\rm OH} \rightarrow {\rm H}_2{\rm O} + h\nu$ because radiative recombination of small
 molecules is typically slow, and our chemical system does not include large molecules for which radiative attachment can
 be important \citep{Vuitton2012}.

Organic photochemistry begins with photolysis of methane.  Methane fragments can react with each other
to make more complicated organic molecules.
Non-methane hydrocarbons (NMHCs) with unsaturated bonds are in turn prone to polymerizing to form chains, rings, 
PAHs, and soots (disorganized agglomerations of PAHs and sheets of PAHs).
In this study we truncate NMHC chemistry at C$_2$H$_{\rm n}$, with the exception of C$_4$H$_2$.
How we handle C$_4$H$_2$ as a proxy for polymerization is discussed in detail in section \ref{Results:carbon} below. 
The more abundant NMHC species in this model are C$_2$H$_2$, C$_2$H$_4$, C$_2$H$_6$, C$_4$H$_2$,
H$_2$CO, CH$_3$OH, and HCN.  
The total NMHC abundance is assessed as the total number of carbon atoms in the NMHCs and reported in several figures below.

Sulfur photochemistry begins with photolysis of, or chemical attack on, H$_2$S.
Sulfur can be successively oxidized by OH (from H$_2$O photolysis) to SO, SO$_2$, and SO$_3$ or H$_2$SO$_4$.
Sulfuric acid (H$_2$SO$_4$) is a major aerosol on Venus and
Earth and therefore to be expected on exoplanets \citep{Hu2013}.  
Sulfur can also react with hydrocarbons to make CS, CS$_2$, and OCS.  All three were abundant in the wake
of the impacts of Comet Shoemaker-Levy 9 into Jupiter in 1994 \citep{Harrington2004}. 
Finally, sulfur can polymerize, condense, and precipitate as the element.
The S$_2$ molecule was seen as a strong signature in the SL9 plumes \citep{Moses1995,Zahnle1995}
and it has been seen in volcanic plumes over Io \citep{Spencer2000}. 
There is strong circumstantial evidence in sulfur's isotopic record in Archean sediments that precipitation of elemental sulfur
was commonplace in the anoxic atmosphere of early Earth \citep{Pavlov2002}.
Here we use a simplified system consisting of S, S$_2$, S$_3$, S$_4$, and S$_8$.
As there is considerable uncertainty in sulfur's reactions, we have listed our choices for key reactions in Table 1.
Most of the key reaction rates will be varied --- and in one case, created --- in sensitivity studies in section \ref{section:sensitivity} below.
All small sulfur-bearing molecules are rather easily photolysed but the sulfur rings
--- here gathered together under the master ring S$_8$ --- are
more stable to UV \citep{Young1983,Kasting1989,Yung2009}.
Thus, as we shall see, there is a strong tendency for sulfur to polymerize to S$_8$ under UV radiation.

The background atmosphere is assumed to be 84\% H$_2$ and 16\% He.
The relative abundances of C, N, O, and S are presumed solar and to scale as a group according to metallicity;
scavenging of O and S by silicates and chalcophiles is taken into account \citep{Lodders2006}.
For simplicity we assume solar metallicity in the base models
(the star 51 Eridani itself is very slightly subsolar, $\left[\mathrm{Fe}/\mathrm{H}\right]=-0.027$).
We consider one set of models with metallicity that is a Jupiter-like $3\times$ solar. 
It is not immediately obvious that higher metallicity always favors haze formation,
despite the greater abundance of haze-forming elements. 
Indeed, in atmospheres where CH$_4$ is less abundant than CO, raising metallicity reduces the CH$_4$/CO
ratio, and hence can make organic haze formation less favorable.
Here we will find that raising the metallicity from solar to $3\times$ solar in 51 Eri b 
has a negative effect on NMHC formation.

51 Eridani is a bright star that was observed decades ago by the {\it International Ultraviolet Explorer} ({\it IUE}).
We use the observed {UV} spectrum for $115 < \lambda < 198$ nm, the range of wavelengths
 for which data are available.   
 This includes Lyman $\alpha$. 
For $\lambda > 198$ nm we use a standard stellar model photosphere for an F0IV star of
 radius $1.6 R_{\odot}$, which makes the star's luminosity appropriate to 51 Eridani itself.
We note in passing that the {UV} irradiation of 51 Eri b is about twice what it is at Earth today,
or about $200\times$ what it is at Titan.  

An important simplification is that we treat vertical mixing by an eddy diffusion parameter $K_{zz}$ that does
not vary with height.
What $K_{zz}$ should be in a stratified atmosphere like that of 51 Eri b is not well-constrained \citep{Freytag2010}. 
Values ranging from $10^3$ cm$^2$/s at the top of the troposphere to $10^6-10^7$ cm$^2$/s at the top of the stratosphere seem to be useful for Jupiter \citep{Moses2005}, and values as high as $10^{10} $ cm$^2$/s have been suggested for hot Jupiters.
Here we consider $10^{5} \leq K_{zz} \leq 10^{10} $ cm$^2$/s.

We set surface gravity to $g=32$ m/s$^2$ in the nominal model.
To test the response of the model to different gravities we consider $g=56$ m/s$^2$ as a variant.
These bracket what is expected for 51 Eri b; $g=32$ m/s$^2$ is not better than $g=56$ m/s$^2$.
The higher gravity models are cooler at a given pressure and thus are more favorable to CH$_4$ and to sulfur condensation.

 The pressure-temperature profile is computed by a radiative-convective equilibrium model assuming a cloud-free atmosphere.
 In the troposphere these assumptions produce a relatively cool model.
 Unlike the thermal structure of the troposphere, which is governed by the planet's own luminosity,
 temperatures at very high altitude depend also on heating by the star.    
 Here we simply extend an isothermal atmosphere to altitudes above the top of the radiative-convective model.
  The isothermal assumption splits the difference between the radiative equilibrium models which get colder at higher
 altitudes and Jupiter itself, which gets hotter for reasons that are poorly understood but may have to do with breaking gravity
 waves propagating up from below. 
 This is an important limitation on our models: we don't know the temperature well enough to categorically state
 that sulfur does or does not condense in 51 Eri b.
 The temperature structure of a sulfurous atmosphere is a big enough topic that it is best  
 deferred to a future study.   

\section{Results}
\label{Results}

We begin with a particular model that illustrates the general features of 51 Eri b photochemistry.
We then look at how the models respond to parameter variations. 

\subsection{Nominal 51 Eri b models: two kinds}

\begin{figure}[!htb]
 \centering
 \begin{minipage}[c]{0.49\textwidth}
   \centering
  \includegraphics[width=1\textwidth]{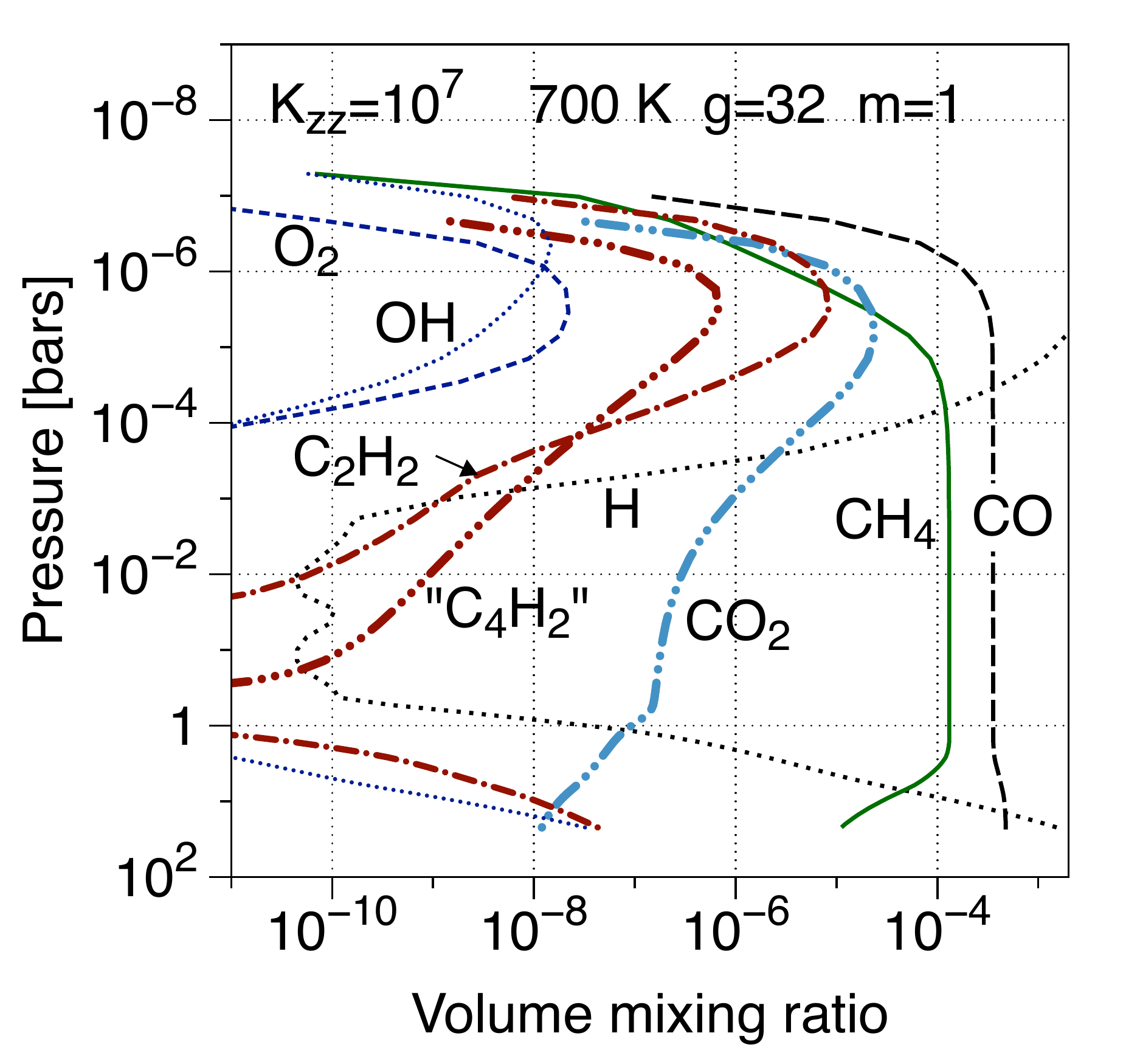} 
 \end{minipage}
 \begin{minipage}[c]{0.49\textwidth}
   \centering
 \includegraphics[width=1\textwidth]{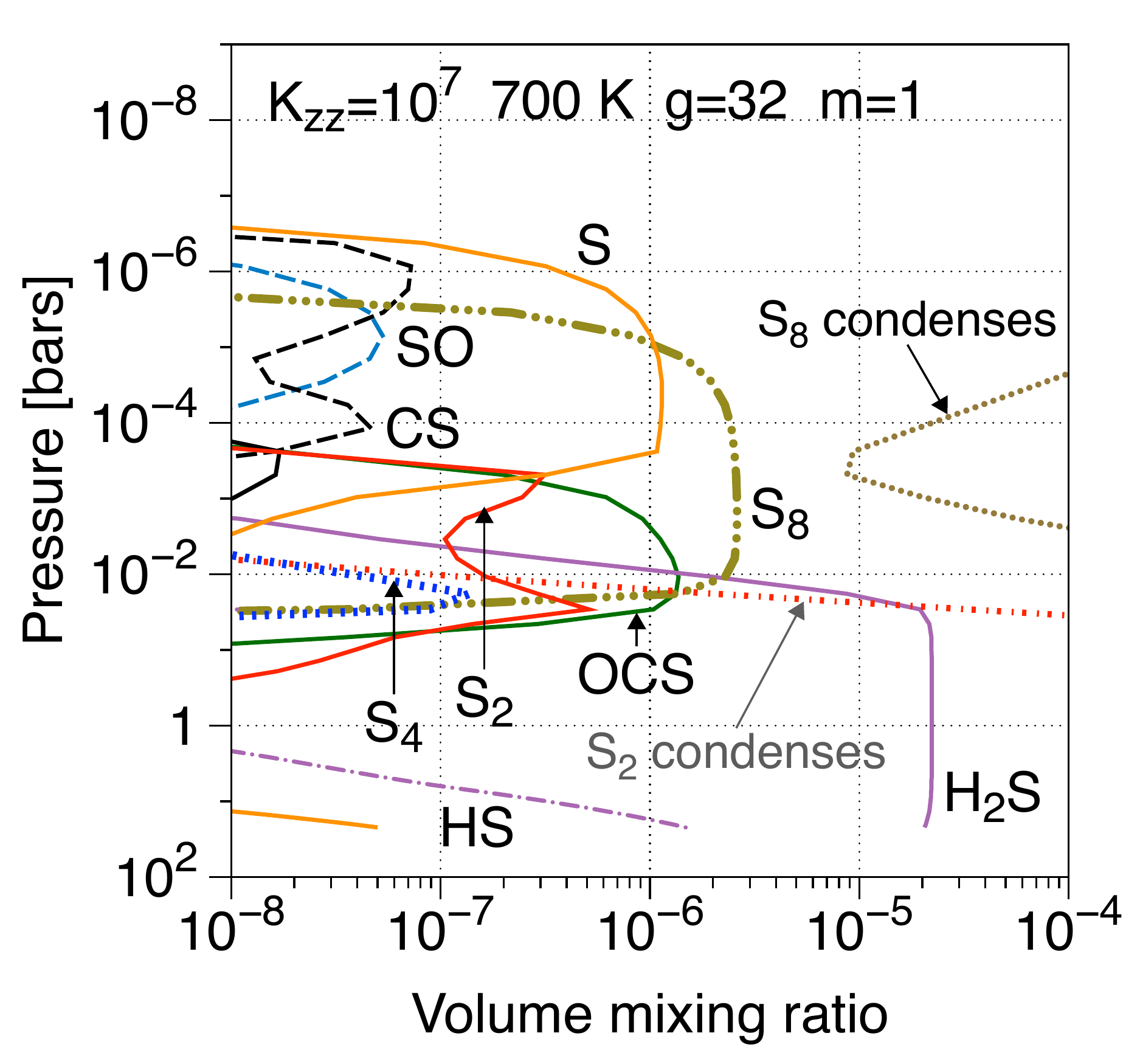} 
 \end{minipage}
 \centering
 \begin{minipage}[c]{0.49\textwidth}
   \centering
 \includegraphics[width=1\textwidth]{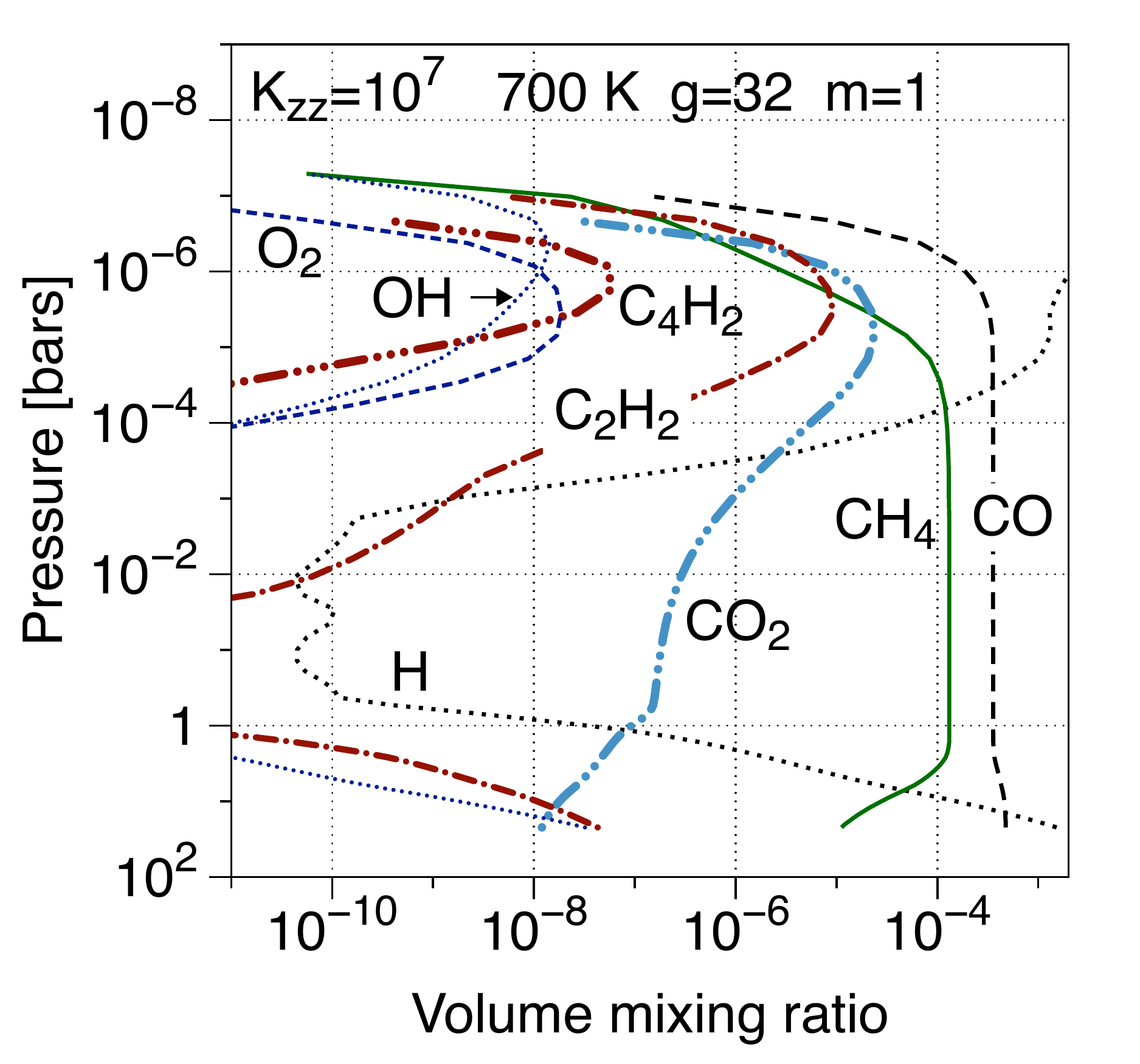} 
 \end{minipage}
 \begin{minipage}[c]{0.49\textwidth}
   \centering
 \includegraphics[width=1\textwidth]{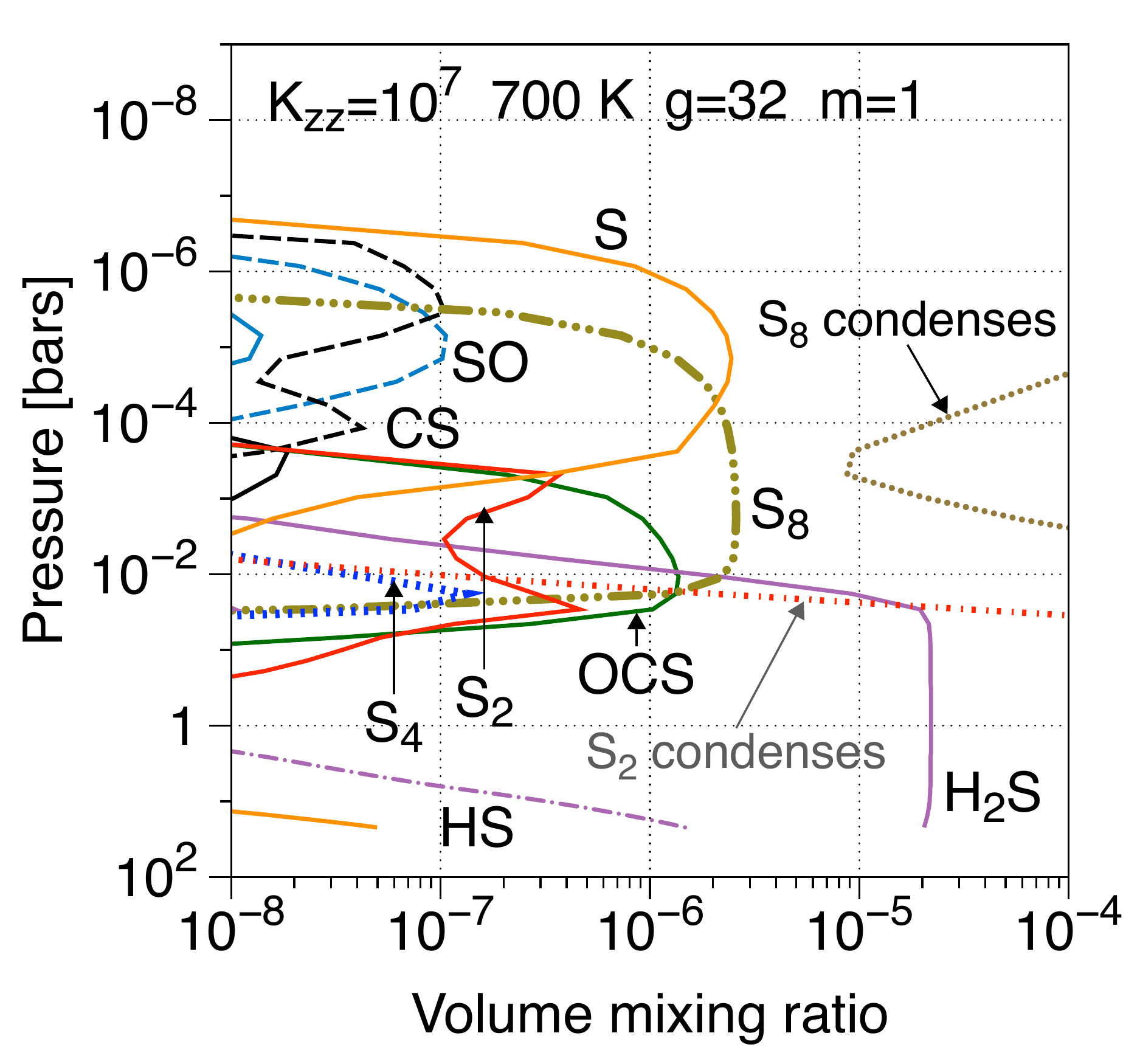} 
 \end{minipage}
 \caption{\small Photochemistry in a nominal 51 Eri b model
  ($T_{\rm eff}=700$ K, $g=32$ m s$^{-2}$, solar metallicity, cloud-free atmosphere, $K_{zz}=10^7$ cm$^2$s$^{-1}$).
  The top and bottom rows differ in how C$_4$H$_2$ is treated. 
  How C$_4$H$_2$ is treated has little effect on the more abundant molecules. 
   {\it Left.} Carbon and oxygen.  In the top panel, ``C$_4$H$_2$'' is treated as the gateway to C$_2$H$_2$ polymerization.
     Where ``C$_4$H$_2$'' is more abundant than acetylene (C$_2$H$_2$), our chemical scheme has broken down.
  In the bottom panel, C$_4$H$_2$ is chemically recycled.     
  {\it Right.} Sulfur shows a rich photochemistry that tends to build toward the relatively photolytically stable S$_8$ molecule.
  This particular model is about 5 K too warm for S$_8$ to condense. 
   Abundances of SO, CS, and S in the upper stratosphere will be smaller
  than shown here if sulfur condenses.
  Note that S$_4$ is abundant at the interface between H$_2$S and S$_8$.  
  }
\label{T700K7}
\end{figure}

The particular model documented in Figure \ref{T700K7}, which we call the nominal model,
 assumes an effective temperature $T_{\rm eff}=700$ K, an eddy diffusivity of $K_{zz}=10^7$ cm$^2$s$^{-1}$,
 constant gravity $g=32$ m/s$^2$, solar metallicity $m=1$, and a cloud-free atmosphere.
Figure \ref{T700K7} plots
volume mixing ratios of selected carbon-, oxygen-, and sulfur- bearing species as a function of altitude (pressure).
For carbon and oxygen we plot CO and CH$_4$,
 the major oxidized photochemical product CO$_2$, 
 the reduced photochemical products acetylene (C$_2$H$_2$) and C$_4$H$_2$,
 the bleaching agents OH and O$_2$, and atomic H.
 For sulfur we plot most of the species that are abundant, although CS$_2$ and SO$_2$ are not
 labeled and S$_3$, which is coincident with S$_4$ but less abundant in these models, is omitted entirely for clarity. 
We do not plot H$_2$O (the most abundant molecule other than H$_2$),
atomic O, other hydrocarbons, nor any N-bearing species.

Figure \ref{T700K7} illustrates the vertical structure of chemical products. 
The top of the atmosphere is relatively oxidized by OH from H$_2$O photolysis, but it is also
where CH$_4$ is photolyzed by Lyman $\alpha$, and so the top is also the primary source of small
hydrocarbon radicals. 
Reactions with OH are the chief competition to hydrocarbon polymerization because the
 CO bond once formed is effectively unbreakable in the haze-forming region.
 Thus NMHC production is possible only when OH is suppressed. 
OH is controlled by reaction with H$_2$ to reconstitute H$_2$O, or with CO to make CO$_2$;
this is why CO$_2$ is always a major photochemical product in all 51 Eri b models.
Conditions are more reduced at greater depth. 

\subsubsection{Alternative carbon polymerizations}
\label{Results:carbon}

It is self-evident that hydrocarbon polymerization can ramify without any known limit,
especially in the presence of nitrogen and a little oxygen.
In the bigger picture this is obviously a good thing, but our modeling effort cannot ramify without limit. 
We must either be able to show that abundances go to zero for molecules with more than a few carbon atoms,
or we must artificially truncate the system.  
If the atmosphere is sufficiently oxidized, the first option is workable.
The system will stop at CO$_2$ without much of interest happening ---
this has historically been the bane of terrestrial prebiotic atmospheric chemistry models \citep[][]{Abelson1966,Pinto1980}.
But here we are dealing with H$_2$-rich atmospheres and it is not obvious {\it a priori} that the chemistry converges.

In this study we truncate the system at C$_4$H$_2$, the first molecule to form as the product of two C$_2$H$_{\rm n}$ molecules.
The state of the art in exoplanets takes the chemistry up to C$_8$H$_{\rm n}$ \citep{Moses2014,Venot2015,Rimmer2016}, but only a tiny fraction of all possible
C$_{\rm m}$H$_{\rm n}$ (${\rm m}\leq 8$) can be taken into account, and the combinatorial nature of the chemistry rapidly
approaches or exceeds the limit of what can be done with a detailed chemical kinetics model.
Further progress requires working with a limited number of generic or representative species.
We consider two extreme assumptions that might bound the problem.

In one set of numerical experiments we treat C$_4$H$_2$ as a bucket in which polymerizing
carbon accumulates, rather than as an actual chemical species.
The only loss is the reverse of the formation reaction,
\begin{equation}\tag{R57}
\mathrm{C}_2\mathrm{H} + \mathrm{C}_2\mathrm{H}_2 \rightarrow \mathrm{C}_4\mathrm{H}_2 + \mathrm{H}.
\end{equation} 
The underlying idea is that C$_4$H$_2$ is destined to grow into ever larger
C$_{\rm m}$H$_{\rm n}$N$_{\rm x}$O$_{\rm y}$S$_{\rm z}$ molecules by
the addition of free radicals.   
When used in this way, we will from here forward put quotes on ``C$_4$H$_2$'' to indicate that 
we are treating it as a representative species rather than as the real C$_4$H$_2$ molecule. 
This is the case documented by the upper left-hand panel of Figure \ref{T700K7} and in most other spaghetti plots in this paper.
 
In the other set of numerical experiments we add three chemical reactions with H to crack C$_4$H$_2$:
first an addition,
\begin{equation}\tag{R58}
  \mathrm{C}_4\mathrm{H}_2 + \mathrm{H} + \mathrm{M} \rightarrow \mathrm{C}_4\mathrm{H}_3  + \mathrm{M}
\end{equation} 
followed either by H-abstraction 
\begin{equation}\tag{R59}
  \mathrm{C}_4\mathrm{H}_3 + \mathrm{H}  \rightarrow \mathrm{C}_4\mathrm{H}_2  + \mathrm{H}_2
\end{equation} 
or by fission
\begin{equation}\tag{R60}
  \mathrm{C}_4\mathrm{H}_3 + \mathrm{H}  \rightarrow \mathrm{C}_2\mathrm{H}_2  + \mathrm{C}_2\mathrm{H}_2.
\end{equation} 
Reaction R58 is a fast reaction that has been studied both theoretically
and experimentally \citep{Eiteneer2003,Klippenstein2005}; we use rates for $k_{58}$ from the latter.
The other two reactions are inventions.
For R59, we assume that $k_{59}=5\times 10^{-11} \exp{\left(-500/T\right)}$ cm$^3$/s, which is not unusual for an H-abstraction,
if perhaps a bit fast.
For R60, the unusual reverse reaction R60r discussed above with respect to cyclohexadiene suggests that there ought to be a considerable
activation barrier and a rather small collision factor to the reverse reaction
to account for the special geometry that would seem required.
We assume that 
\begin{equation}\label{k60}
k_{60}=5\times 10^{-11} \exp{\left(-2000/T\right)} \quad \mathrm{cm}^3/\mathrm{s}.
\end{equation}
 The lower left-hand panel of Figure \ref{T700K7} shows that adding reactions R58-R60 to the network reduces
 the peak abundance of C$_4$H$_2$ and restricts the molecule to the photochemical region.
 Not shown is that if $k_{60}$ is reduced by a factor of 30, the 
 C$_4$H$_2$ altitude profile reverts to the ``C$_4$H$_2$'' profile seen in the upper left panel of Figure \ref{T700K7}.
 
We note that neither R59 nor R60 are likely to be important in reality.   
Much more likely is that the reaction with H will be another addition \citep{Harding2007} and the hydrocarbon will continue to grow,
\begin{equation}\tag{R61}
  \mathrm{C}_4\mathrm{H}_3 + \mathrm{H} + \mathrm{M}  \rightarrow \mathrm{C}_4\mathrm{H}_4  + \mathrm{M}, 
\end{equation}
with no natural truncation point in the photochemical region where C-bearing radicals are also abundant;
that is, additions and ramifications will continue, and there is no obvious end to this.
From this perspective ``C$_4$H$_2$'' is a gateway species. 
At greater depth in a hydrogen-rich atmosphere, hydrogenation will probably focus on the unsaturated carbon
bonds until what is left is an alkane or alkanes, and in the end the alkanes will be hydrogenated to CH$_4$ and H$_2$,
completing the cycle.
 
In most figures that follow we will show ``C$_4$H$_2$'' profiles computed with the high C$_4$H$_2$ 
because these are more interesting to look at.

\subsubsection{Sulfur photochemistry and sulfur condensation}

The righthand panels of Figure \ref{T700K7} line up the sulfur chemistry with the carbon and oxygen chemistry
in the nominal model.
Several things stand out.  The first is that H$_2$S --- sulfur's stable form in the abyss --- barely makes it past the
tropopause.  Although H$_2$S is susceptible to UV photolysis, that is not what is happening here.
Rather, H$_2$S is being destroyed by atomic H flowing down from the high altitude photochemical source region,
\begin{equation}  \tag{R23}
 \mathrm{H}_2\mathrm{S} + \mathrm{H} \rightarrow  \mathrm{HS} + \mathrm{H}_2    .
\end{equation} 
The HS radical reacts quickly with H to free S, 
\begin{equation}  \tag{R9}
 \mathrm{HS} + \mathrm{H} \rightarrow  \mathrm{S} + \mathrm{H}_2    ,
\end{equation} 
and atomic S reacts with HS to make S$_2$,
\begin{equation}  \tag{R8}
 \mathrm{HS} + \mathrm{S} \rightarrow  \mathrm{S}_2 + \mathrm{H}   , 
\end{equation} 
and the polymerization of sulfur has begun, which is the 
second thing to stand out: S$_8$ is very abundant, generally
at a lower altitude than the NMHCs and under more reduced conditions.

The high predicted abundance of S$_8$ suggests that it might condense.
Sulfur vapor is complicated by the presence of several allotropes. 
Our first simplification is to lump S$_6$ and S$_7$ together with the more abundant S$_8$. 
\citet{Lyons2008} gives simple curve fits to many allotropes above the liquid, and then describes a scheme for 
extrapolating these to lower temperatures above solid sulfur.
A complication is that the vapor pressure curves given by \citet{Lyons2008} are discontinuous by
 nearly a factor of two at sulfur's melting point ($T_m=398$ K).
We use a blended approximation in which the vapor pressure over the solid is extended
to higher temperature until it intersects the reported vapor pressure over the liquid,
 \[
 p_v(\mathrm{S}_8) = \exp{\left(20 -11800/T\right)} \qquad  T < 413 \mathrm{~K}
 \]
\begin{equation}
\label{S8_vapor}
 p_v(\mathrm{S}_8) = \exp{\left(9.6 -7510/T\right)} \qquad T > 413 \mathrm{~K}
\end{equation}
where the vapor pressure is in bars.  
In Figure \ref{T700K7}, the S$_8$ mixing ratio is $\sim 2\times 10^{-6}$ for atmospheric pressure levels between 100 $\mu$bars and 10 mbars.
At these partial pressures, $2\times 10^{-10} < p(\mathrm{S}_8) < 2\times 10^{-8}$, sulfur's condensation temperature is between 280 and 310 K.
The uncertainty in Eq \ref{S8_vapor} in this temperature range is probably less than a factor of two (the coldest datum is at $\sim 310$ K),
which is insignificant compared to the uncertainty in the temperature in our models.
For context, the corresponding condensation temperatures for water are between 170 and 200 K at the same altitudes.
At higher metallicity both condensation temperatures are $\sim 20\log_{10}(m)$ K higher. 

The vapor pressure of S$_2$ over solid or liquid sulfur is tiny \citep{Lyons2008}, 
\[
 p_v(\mathrm{S}_2) = \exp{\left(27 -18500/T\right)} \quad\qquad  T < 413 \mathrm{~K}
\]
\begin{equation}
\label{S2_vapor}
 p_v(\mathrm{S}_2) = \exp{\left(16.1 -14000/T\right)} \;\qquad T > 413 \mathrm{~K} .
\end{equation}
All of our models of 51 Eri b predict more S$_2$ than would be consistent
with the presence of condensed sulfur.  Evidently
S$_2$ (and S$_3$ and S$_4$ as well) would be drawn down to negligible amounts where S$_8$ condenses.

Saturation mixing ratios of S$_8$ and S$_2$ over solid sulfur are plotted on Figure \ref{T700K7}. 
A third thing stands out: S$_8$ in 51 Eri b is very close to its condensation point.
In this particular model S$_8$ does not condense, but if the model were a few degrees cooler
it would condense.
If S$_8$ condenses, we can presume that there would be much less S$_2$, SO, CS, and S above the clouds
than is shown here.   

\subsection{Dependence on vertical mixing}

\begin{figure}[!htb]
 \centering
 \begin{minipage}[c]{0.49\textwidth}
   \centering
  \includegraphics[width=1\textwidth]{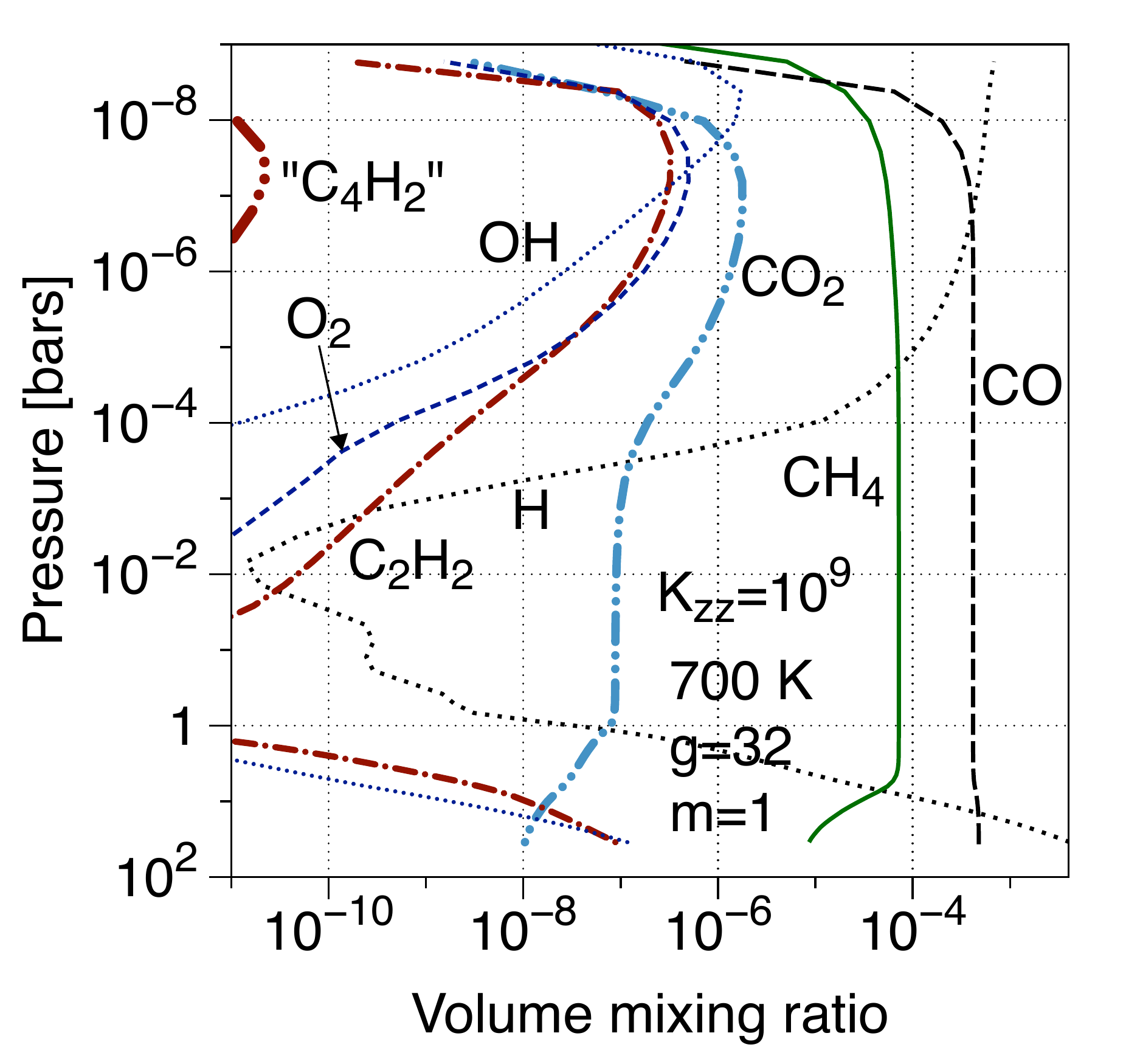} 
 \end{minipage}
 \begin{minipage}[c]{0.49\textwidth}
   \centering
 \includegraphics[width=1\textwidth]{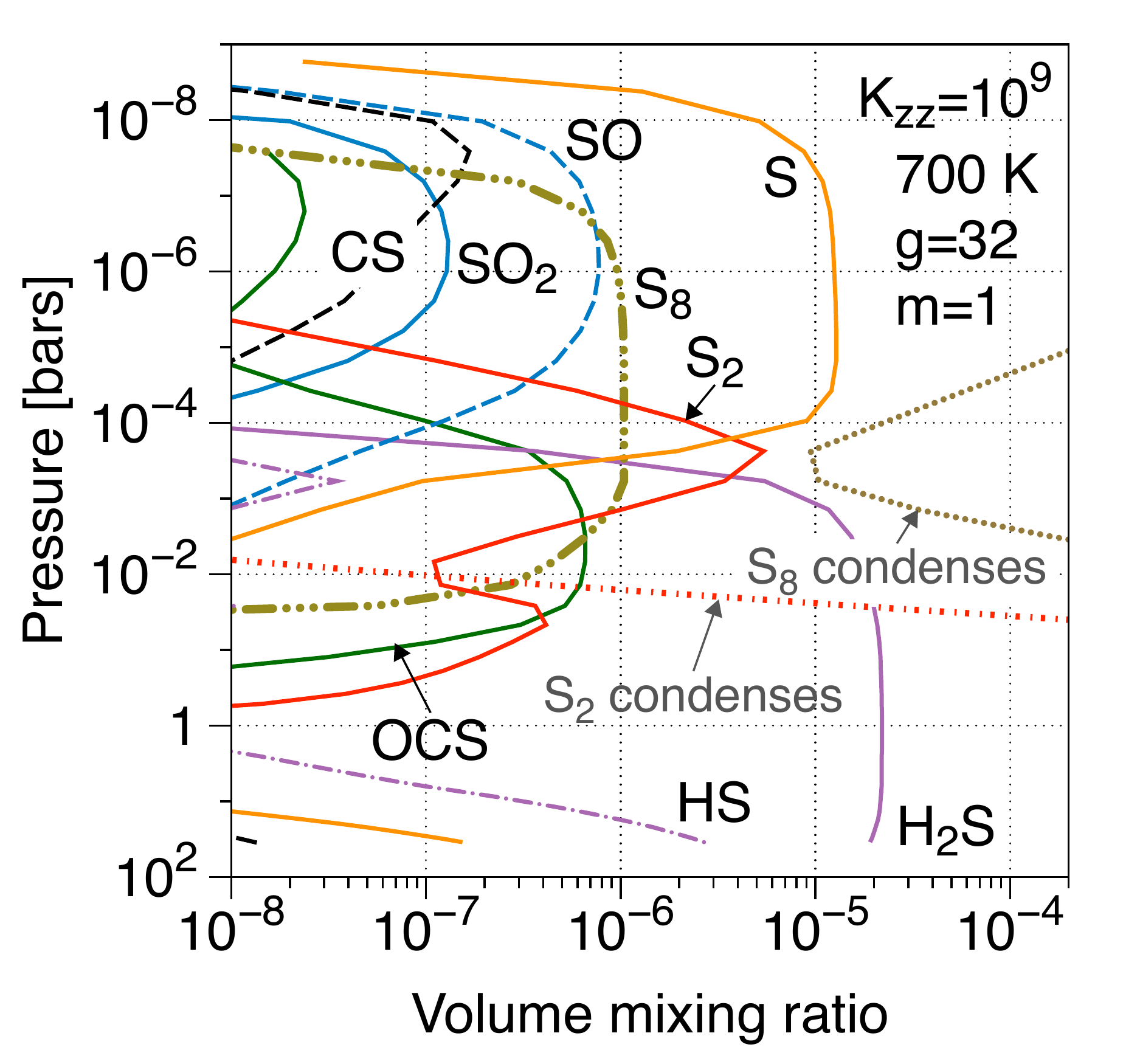} 
 \end{minipage}
 \centering
 \begin{minipage}[c]{0.49\textwidth}
   \centering
 \includegraphics[width=1\textwidth]{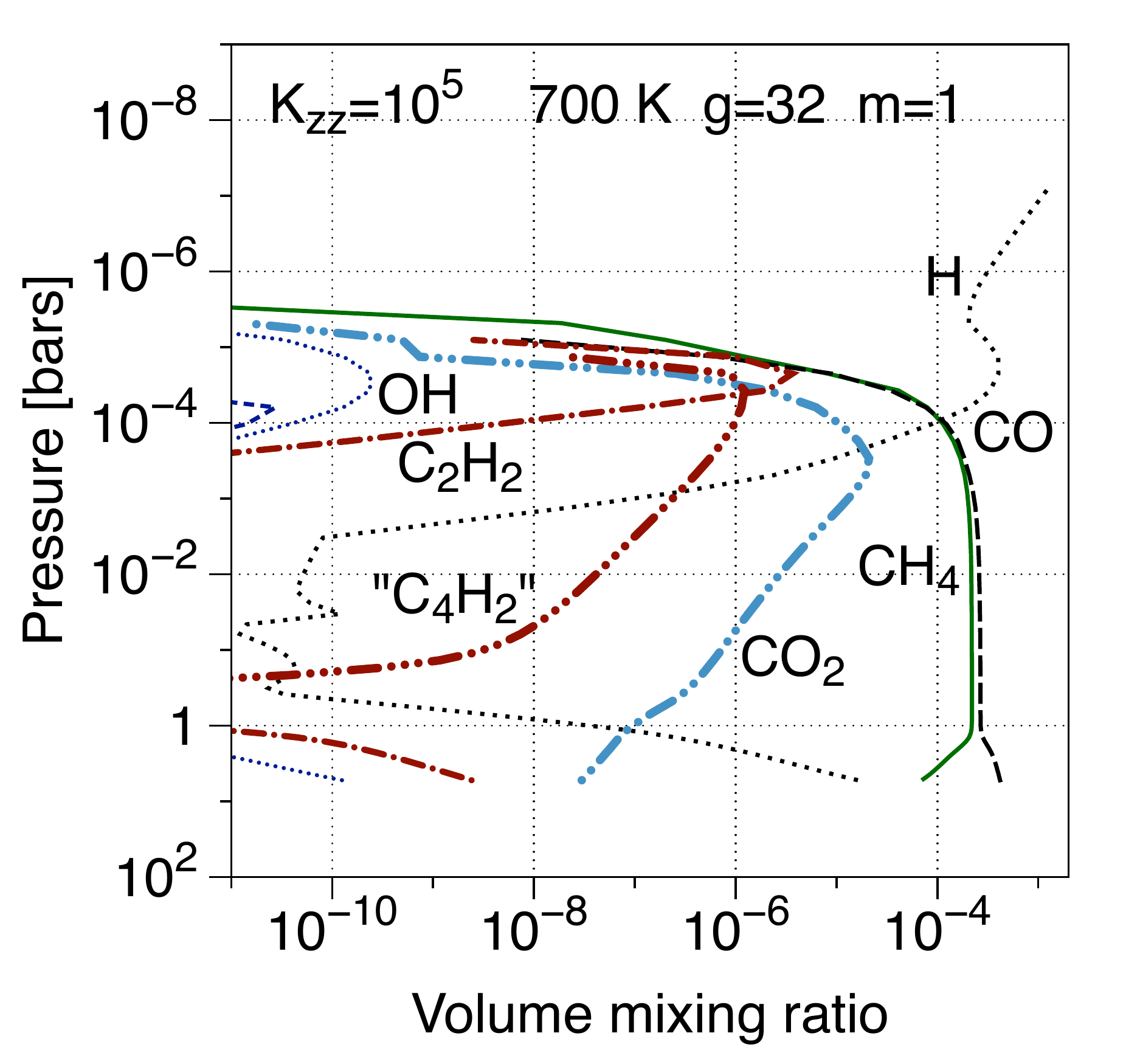} 
 \end{minipage}
 \begin{minipage}[c]{0.49\textwidth}
   \centering
 \includegraphics[width=1\textwidth]{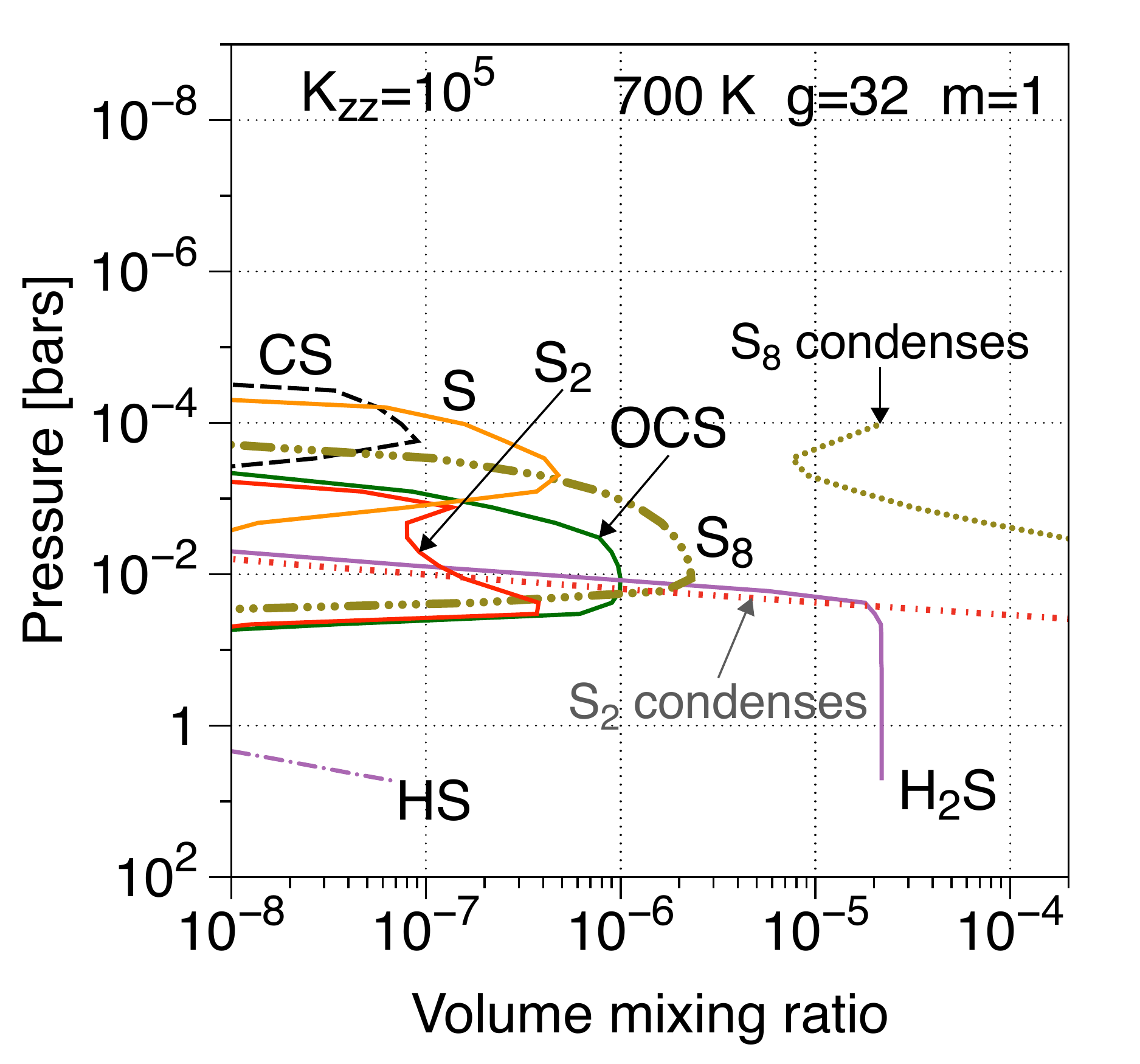} 
 \end{minipage}
 \caption{\small The effect of $K_{zz}$ on carbon and sulfur photochemistry in our nominal 51 Eri b model.
Mixing is $100\times$ stronger (top row) and $100\times$ weaker (bottom row) than in Figure \ref{T700K7}
(these are both high ``C$_4$H$_2$'' models to be compared to the top panels of Figure \ref{T700K7}).
High vertical mixing creates a more oxidized environment at the top of the atmosphere
  that is less favorable to S$_8$ and very unfavorable to NMHC growth.  
   Strong vertical mixing is more favorable to S$_2$, SO, and SO$_2$, less favorable to S$_8$. 
  Weak vertical mixing produces a more reduced atmosphere that 
  is more favorable to NMHCs and to S$_8$, which forms abundantly in deeper, warmer regions.
 }
\label{4Kzz}
\end{figure}

In this study vertical mixing is a free parameter.
Figure \ref{4Kzz} shows what happens when $K_{zz}$ is made much bigger or much smaller.
These are high ``C$_4$H$_2$'' models. 
   Strong vertical mixing ($K_{zz}=10^9$ cm$^2$s$^{-1}$, top panels) 
   creates a more oxidized environment at the top of the atmosphere that is unfavorable to NMHC growth.
   In particular, ``C$_4$H$_2$'' is all but wiped out.
  Weak vertical mixing ($K_{zz}=10^5$ cm$^2$s$^{-1}$, bottom panels) is more favorable to NMHCs,
   especially at lower altitudes that are too deep for oxidants to reach when the mixing is weak.
  This is somewhat obscured by our plotting volume mixing ratios in Figure \ref{4Kzz},
  which exaggerates the apparent importance of trace species at high altitudes,
  and understates the importance of anomalies 
  at $K_{zz}=10^5$ cm$^2$s$^{-1}$. 
  In fact $K_{zz}=10^5$ cm$^2$s$^{-1}$ is more conducive to hydrocarbon polymerization than is $K_{zz}=10^7$ cm$^2$s$^{-1}$. 

The effects of changing $K_{zz}$ on sulfur are parallel to those on carbon but more exaggerated. 
Strong vertical mixing (Figure \ref{4Kzz}, upper right-hand panel) enables H$_2$S to get higher before it gets destroyed,
which creates a more favorable environment for S$_2$, which becomes rather abundant.
If S$_8$ does not condense,
eddy mixing also lifts it to high altitudes where it is photolyzed and oxidized to SO and SO$_2$ or reduced to CS. 
Weak vertical mixing (Figure \ref{4Kzz}, lower right-hand panel) squeezes the sulfur photoproducts into a relatively thin region below
the homopause and above the H$_2$S destruction horizon at 30 mbars; 
the high molecular weight of S$_8$ prevents sulfur getting very high, which markedly depletes the top of the
atmosphere in all sulfur species even if sulfur does not condense.

\subsection{Dependence on effective temperature}

\begin{figure}[!htb]
 \centering
 \begin{minipage}[c]{0.49\textwidth}
   \centering
  \includegraphics[width=1\textwidth]{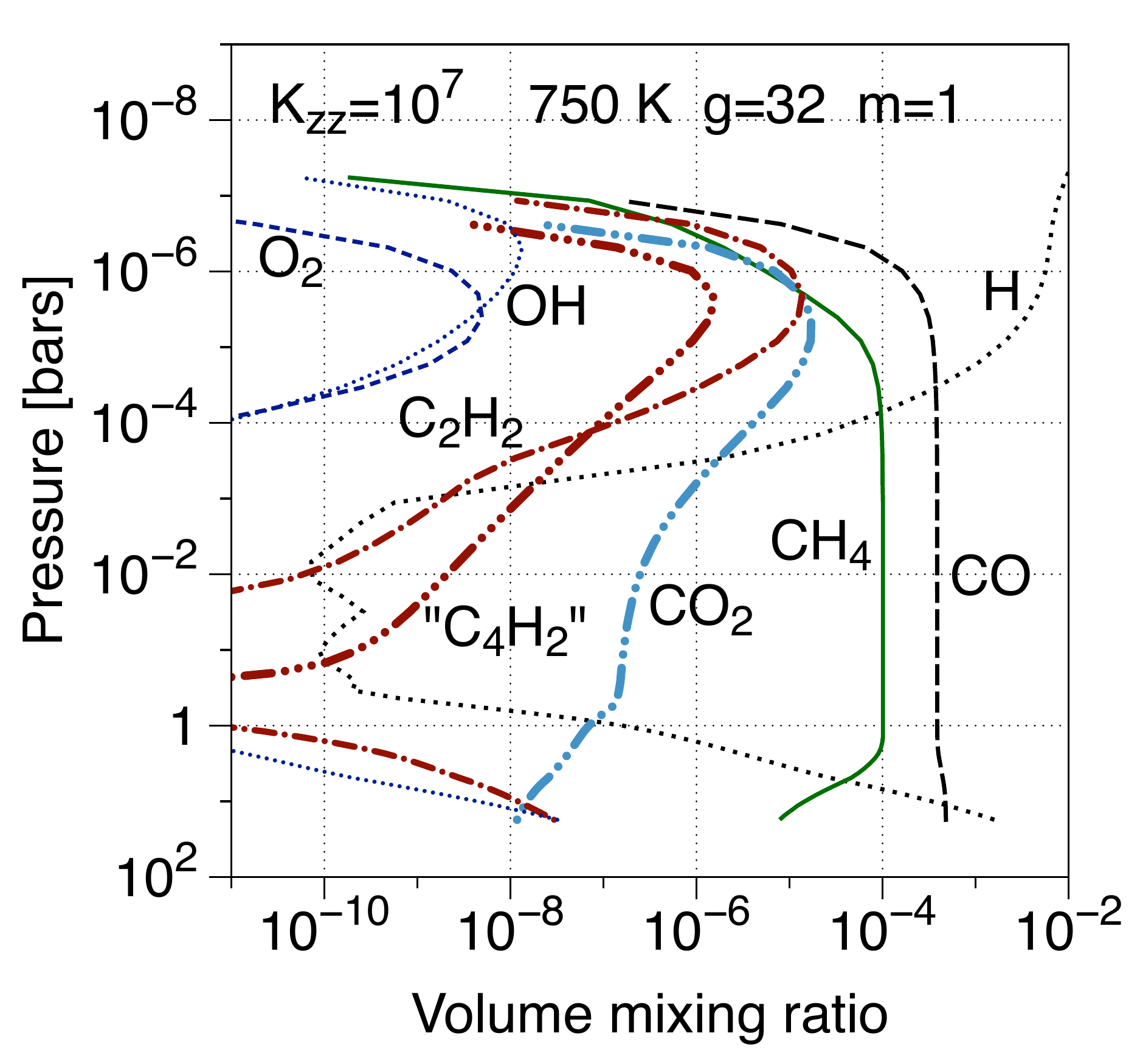} 
 \end{minipage}
 \begin{minipage}[c]{0.49\textwidth}
   \centering
 \includegraphics[width=1\textwidth]{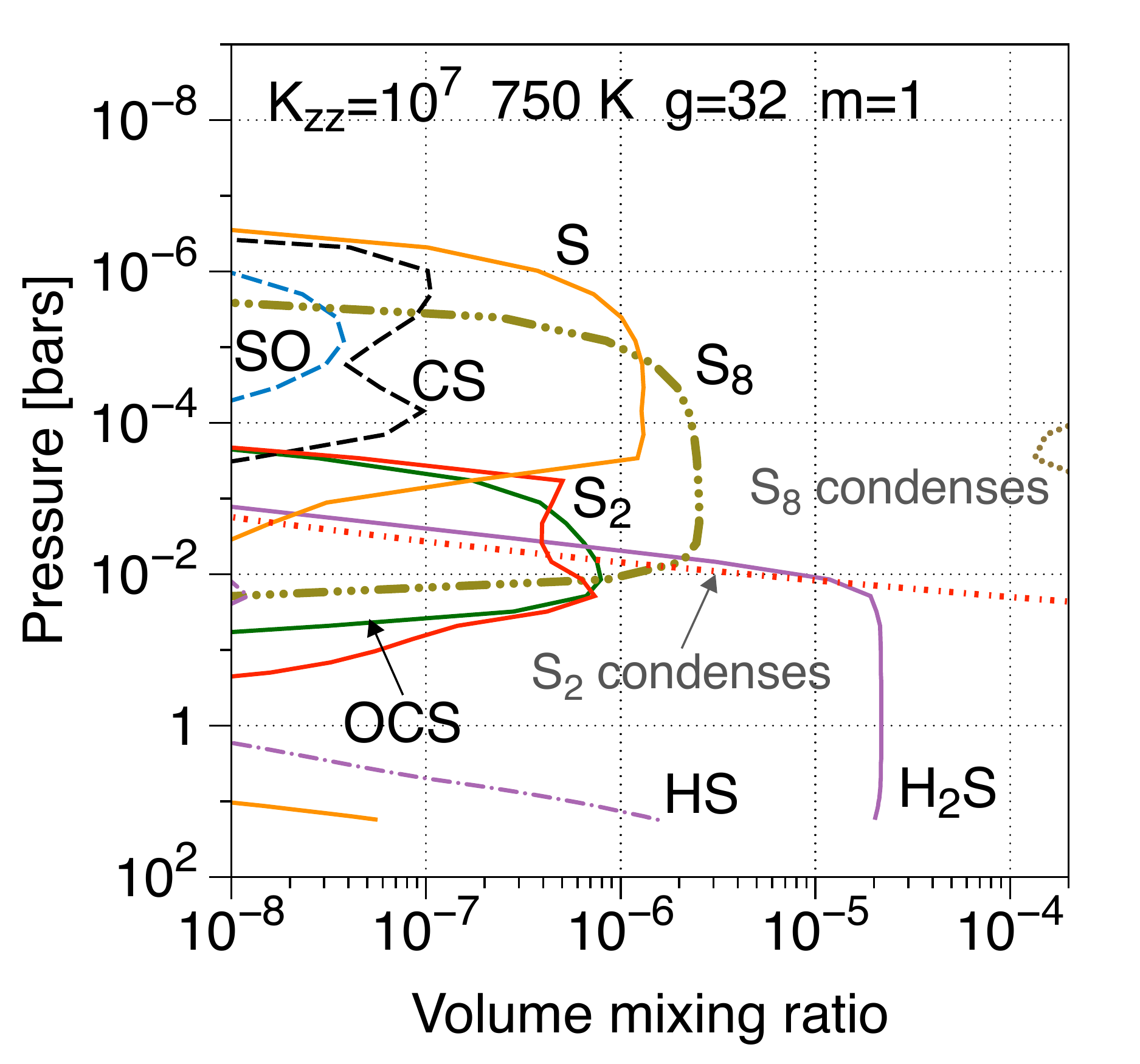} 
 \end{minipage}
 \centering
 \begin{minipage}[c]{0.49\textwidth}
   \centering
 \includegraphics[width=1\textwidth]{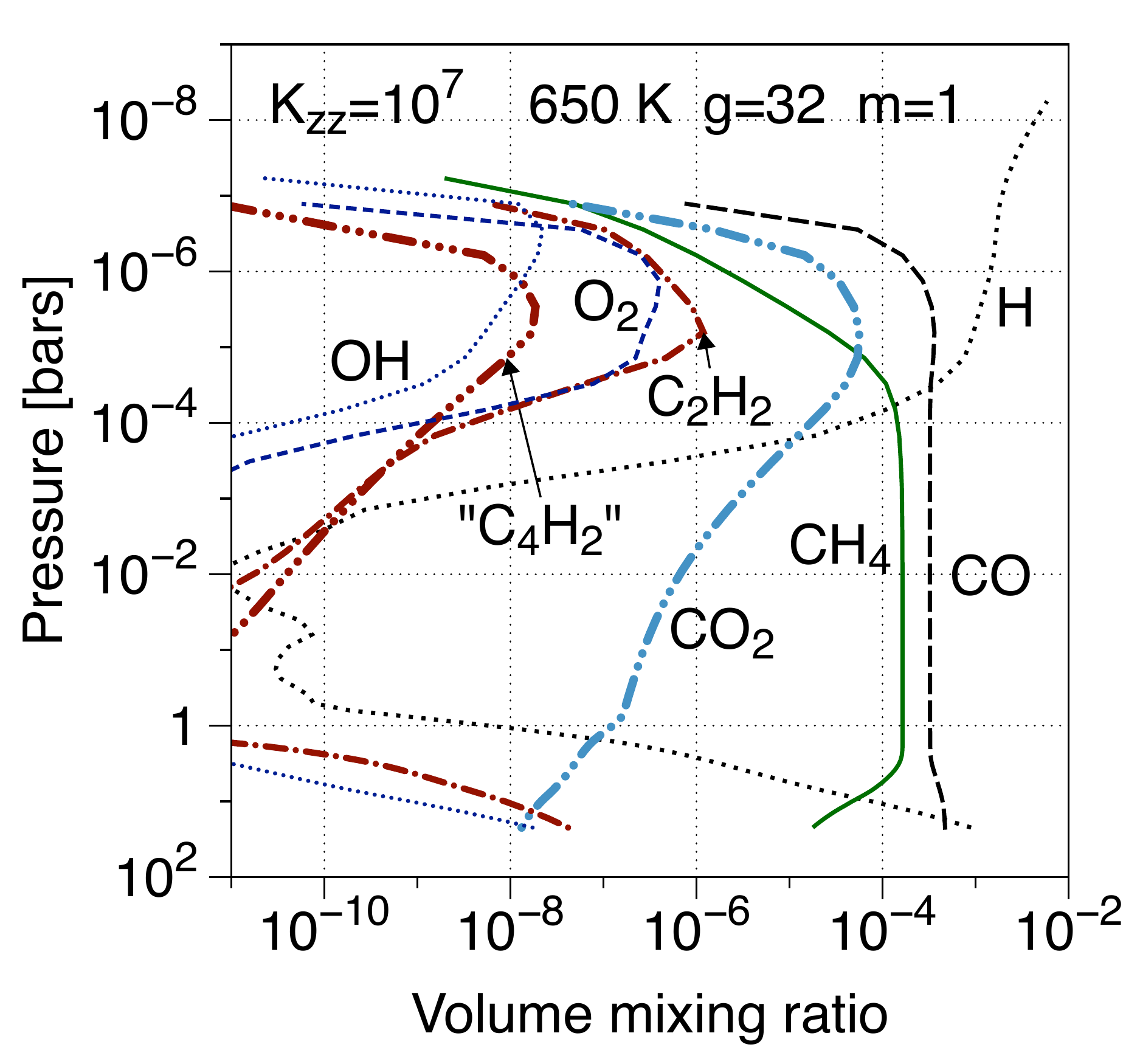} 
 \end{minipage}
 \begin{minipage}[c]{0.49\textwidth}
   \centering
 \includegraphics[width=1\textwidth]{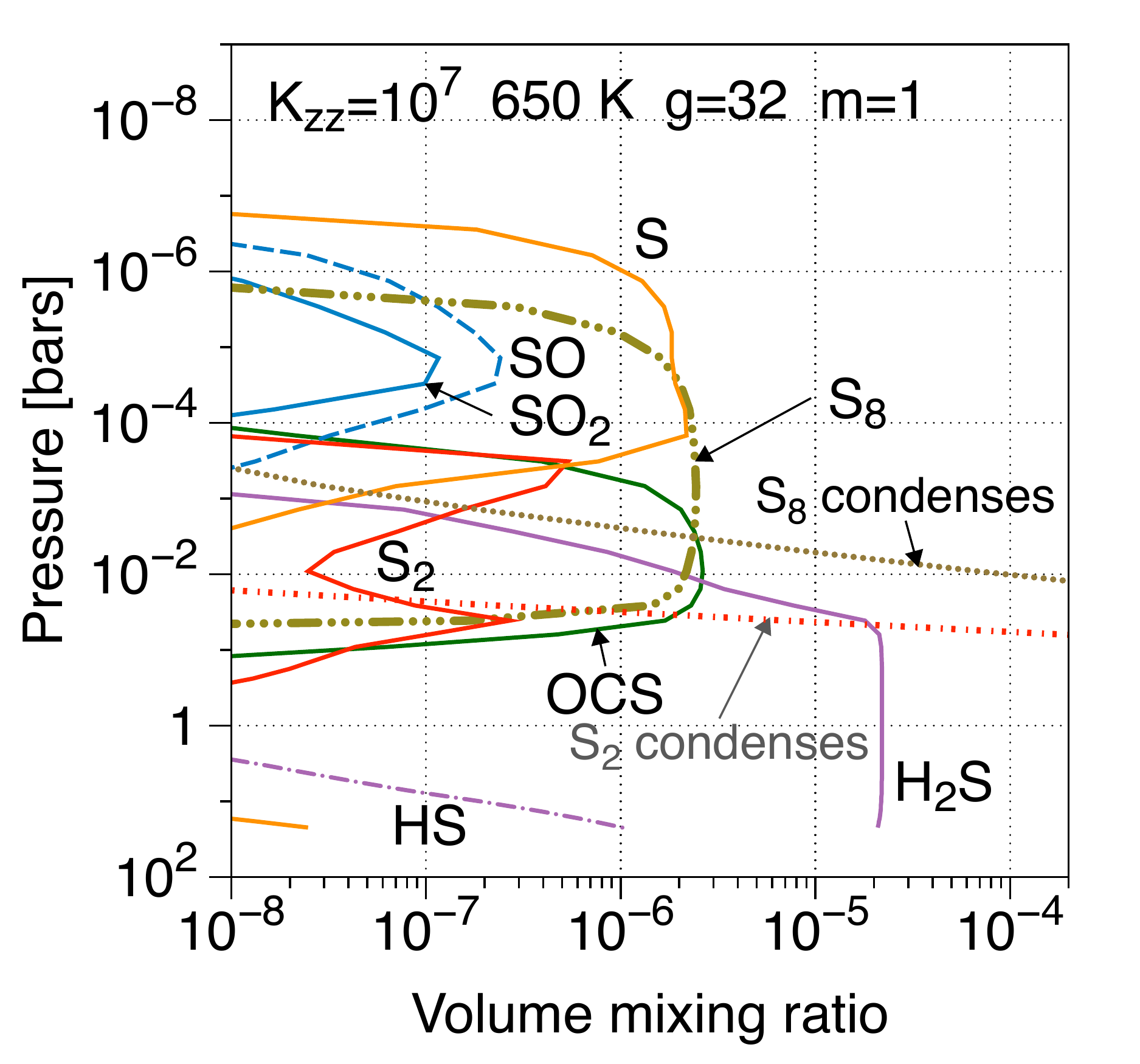} 
 \end{minipage}
 \caption{\small The effect of temperature on carbon and sulfur photochemistries. 
  These are 50 K hotter (top) or colder (bottom) than the corresponding model high C$_4$H$_2$ model in Fig \ref{T700K7}.
  Note that the $T_{\rm eff}=650$ K (lower panels) models are more oxidized and
  less favorable to NMHCs. Note that S$_8$ condenses in the cooler models
  but does not condense in the warmer.}
\label{4Temp}
\end{figure}

Figure \ref{4Temp} shows what happens when the effective temperature of the planet is raised or lowered by 50 K.
These are high ``C$_4$H$_2$'' models. 
The cooler atmosphere is clearly more oxidized.
In carbon this is seen in the higher abundance of CO$_2$ and the lower abundances of C$_2$H$_2$ and ``C$_4$H$_2$,''
in sulfur it is seen in higher abundance of SO and SO$_2$ and the disappearance of CS.
The primary oxidant is OH from H$_2$O photolysis.
The most important sink on OH is the temperature-sensitive reaction with H$_2$
\[
{\rm H}_2 + {\rm OH} \rightarrow {\rm H}_2{\rm O} + {\rm H},
\]
which puts H$_2$O back together.
The high abundance of H$_2$ in a solar composition gas ensures that the 
reaction with H$_2$ is the leading sink on OH for $T>200$.
It is only where $T< 200$ K that the temperature-insensitive reaction with CO,
\[
{\rm CO} + {\rm OH} \rightarrow {\rm CO}_2+ {\rm H} 
\]
becomes more important, but we do not encounter temperatures this low in 51 Eri b models.
The reaction with H$_2$ becomes much slower as the temperature drops and consequently the OH
abundance becomes much larger as the temperature drops.
In turn the higher OH abundance promotes CO$_2$ formation and inhibits NMHC growth.
Both trends are clearly seen in the lower panels of Figure \ref{4Temp}. 

The other effect of temperature on sulfur is the obvious one that condensation becomes more likely in the cooler models.
Sulfur readily condenses in the cooler $T_{\rm eff}=650$ K model at around 3 mbar (Figure \ref{4Temp}, lower right-hand panel). 
This is also the altitude where organic hazes would form if any do,
if the proxy ``C$_4$H$_2$'' is a useful guide. 

\subsection{Overview of carbon chemistry}

\begin{figure}[!htb] 
   \centering
   \includegraphics[width=0.92\textwidth]{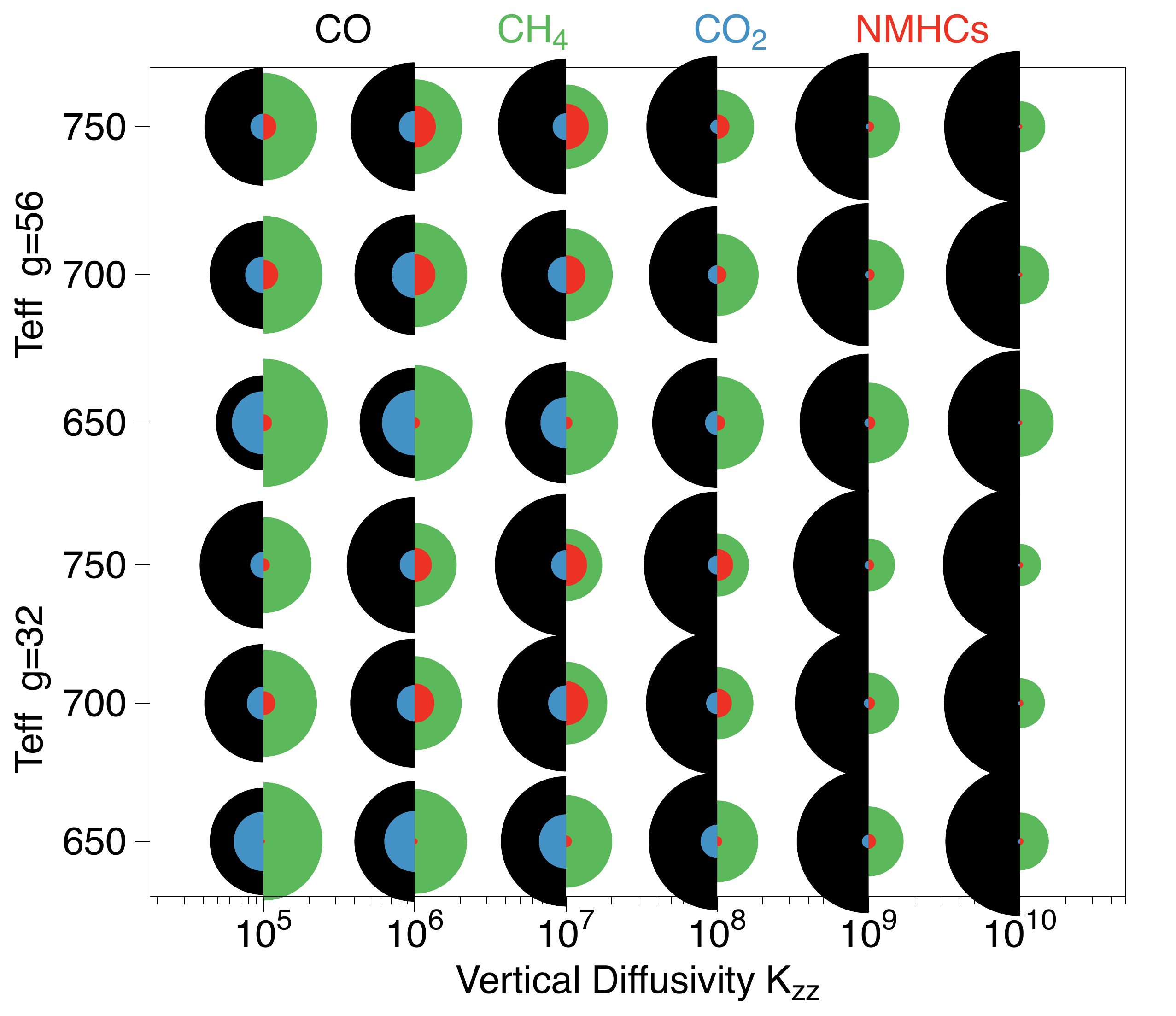} 
   \caption{\small Carbon photochemistry in some possible 51 Eri b's of solar metallicity 
   subject to the observed {\it IUE} {UV} flux from 51 Eri a.
   Peak volume mixing ratios are plotted in proportion to the areas of the disks.
   CO (black) and CH$_4$ (green) are quenched disequilibrium abundances welling up from below. 
  The photochemical products CO$_2$ (blue) and NMHCs (red, chiefly C$_2$H$_2$) are maxima found at higher altitudes where
   photochemistry is king. 
 }
\label{test95}
\end{figure}

Figure \ref{test95} gives an overview of carbon photochemistry for solar composition
models over the phase space of different $T_{\rm eff}$, $g$, and $K_{zz}$ pertinent to 51 Eri b.  
We consider temperatures of $T_{\rm eff}=750$ K and $T_{\rm eff}=650$ K
in addition to the nominal model with $T_{\rm eff}=700$ K,
and we consider a gravity of $g=56$ m/s$^2$ in addition to the nominal $g=32$ m/s$^2$.
Figure \ref{test95} is restricted to solar metallicity and the {UV} radiation observed by {\it IUE}.
We vary $K_{zz}$ between $10^5$ cm$^2$/s and $10^{10}$ cm$^2$/s for all variants of $T_{\rm eff}$ and $g$. 

Figure \ref{test95} plots the quenched disequilibrium CO and CH$_4$ mixing ratios
and it plots the peak mixing ratios reached by the major photochemical products. 
 51 Eri b is near the boundary between CO-dominated and CH$_4$-dominated atmospheres (in equilibrium, the carbon
 would almost entirely be in ${\rm CH}_4$).
Both gases are abundant in all models, although CO is more abundant in most of them.
In general, CH$_4$ is most abundant when $K_{zz}$ is small, or the gas cooler, or the gravity higher \citep{Zahnle2014}.
None of the cases are truly methane-rich.

Smaller values of $K_{zz}$ are more favorable to photochemical NMHC formation
and high values of $K_{zz}$ are very unfavorable.
 The apparently lower NMHC production at low values of $K_{zz}$ is illusory,
   a consequence of plotting peak mixing ratios in Figure \ref{test95}.
   The peak occurs at higher pressure at $K_{zz}=10^5$ cm$^2$s$^{-1}$ than
   at $K_{zz}=10^7$ cm$^2$s$^{-1}$, so that NMHC densities at $K_{zz}=10^5$ cm$^2$s$^{-1}$ are actually higher. 
   Some of the trend with $K_{zz}$ can be ascribed to the CH$_4$/CO ratio, but the trend is even stronger in CO$_2$,
   which suggests that the weaker mixing is also acting to isolate and preserve the photochemical products.
   On the other hand, the relative dearth of NMHCs in the cooler $T_{\rm eff}=650$ K models is a real feature
   caused by the strong temperature dependence of the 
   $\mathrm{H}_2 + \mathrm{OH} \rightarrow \mathrm{H}_2\mathrm{O}+\mathrm{H}$ reaction that holds OH in check.

\subsection{Overview of sulfur chemistry}

\begin{figure}[!htb] 
   \centering
   \includegraphics[width=0.92\textwidth]{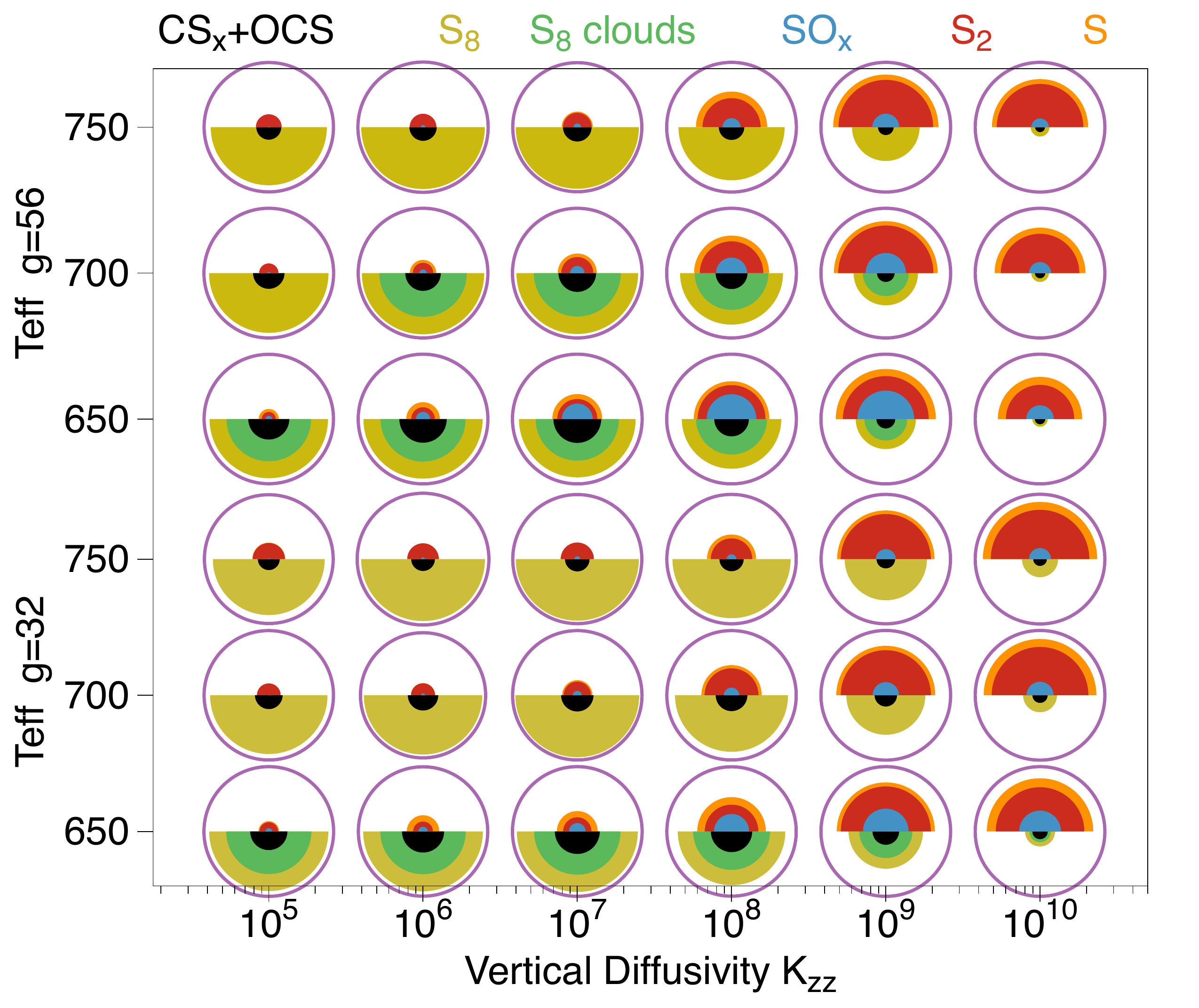} 
   \caption{\small Overview of sulfur photochemical products. 
   Sulfur is grouped into relatively oxidized species (SO and SO$_2$), 
   carbonized species (mostly OCS), and two allotropes of elemental sulfur (S$_2$ and S$_8$).
   The circles and semicircles represent maximum mixing ratios as the areas of the implicit disks. 
   The outer ring is the mixing ratio of H$_2$S in the deep atmosphere.
   In cooler models S$_8$ is predicted to condense; in these half of the S$_8$ is colored green.
 }
\label{test85}
\end{figure}

Figure \ref{test85} presents the corresponding overview of sulfur photochemistry.
Here we count sulfur atoms, so that S$_2$ is counted doubly and
    S$_8$ is counted eight-fold.  The symbols are not like pie charts.
    They do not show how sulfur is apportioned at any one height.
    Rather, they show each category at its peak abundance, which in most cases are at different heights.
    In this way we see that, for example at $K_{zz}=10^7$, almost all the sulfur transitions from H$_2$S to S$_8$ at higher altitudes,
    or that at $K_{zz}=10^{10}$, almost all the sulfur that started in H$_2$S is found in S$_2$ higher up
    and then still higher up it is found as S. 

   Several trends are evident in Figure \ref{test85}. 
   One is that H$_2$S is quantitatively converted to elemental sulfur.
  For weaker vertical mixing the sulfur will pool in S$_8$. 
   The OCS molecule will be abundant.
  Strong vertical mixing favors S$_2$ and S.
   As with carbon, the cooler atmospheres are more strongly oxidized, but with sulfur
   the more strongly oxidized species are more prevalent when the vertical mixing is stronger
   because when mixing is weak S$_8$ settles out, as was seen in the bottom-right panel of Figure \ref{4Kzz}.
   About half the models predict that sulfur condenses in clouds.

\subsection{Metallicity}

\begin{figure}[!htb] 
   \centering
   \includegraphics[width=1.0\textwidth]{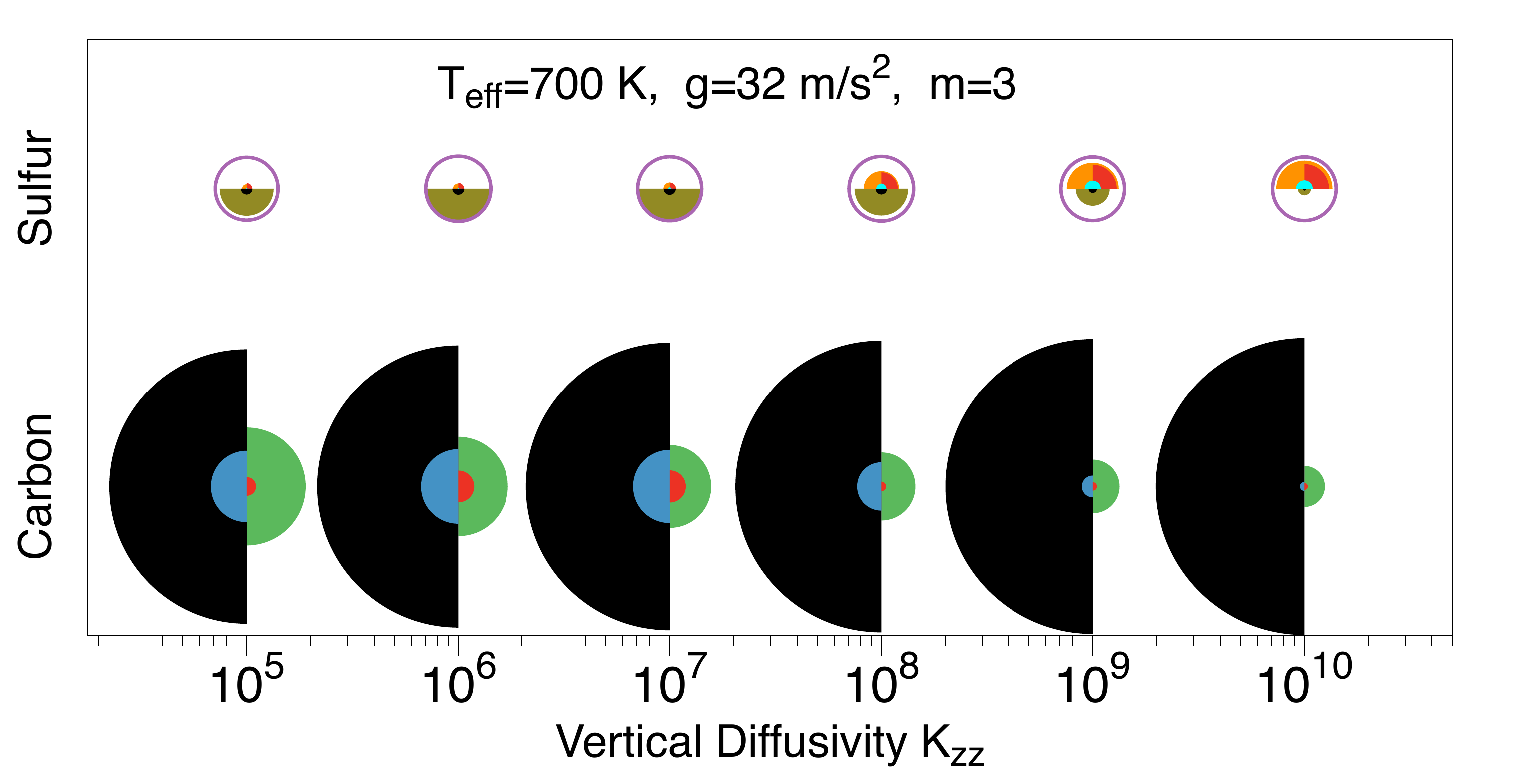} 
   \caption{\small Higher metallicity ($m=3$). 
   Symbols and colors have the same meaning as in Figures \ref{test95} and \ref{test85} above.
   On this figure we plot sulfur and NMHC mixing ratios to the same scale to facilitate direct comparison.
}
\label{test93}
\end{figure}

Figure \ref{test93} illustrates the effect of higher metallicity $m=3$ in the $g=32$ m/s$^2$ models
as a function of $K_{zz}$ for both carbon and sulfur chemistry.
In this figure the carbon and sulfur mixing ratios are plotted to the same scale to facilitate cross-comparison,
but as a consequence sulfur's circles are rather small.
In order to see both S$_2$ and S, these are plotted as quarter circles.
   As is well-known, higher metallicity strongly favors CO and CO$_2$ over CH$_4$.  
   Higher metallicity has little effect on sulfur speciation because
   (i) all of its major products are metal-rich and (ii) its most abundant product, S$_8$, is the metal-richest.
   
\subsection{Optical depths}

It is possible that sulfur will be optically thick when it condenses,
and it is possible that several of its optically-active allotropes will be visible when it does not.
{ To quantify these notions, we estimate the possible optical depths of sulfur clouds and of sulfur vapor through the S$_4$ di-radical,
and we give an upper bound on the opacity from organic hazes.}

  \begin{figure}[!htb] 
   \centering
   \includegraphics[width=1.0\textwidth]{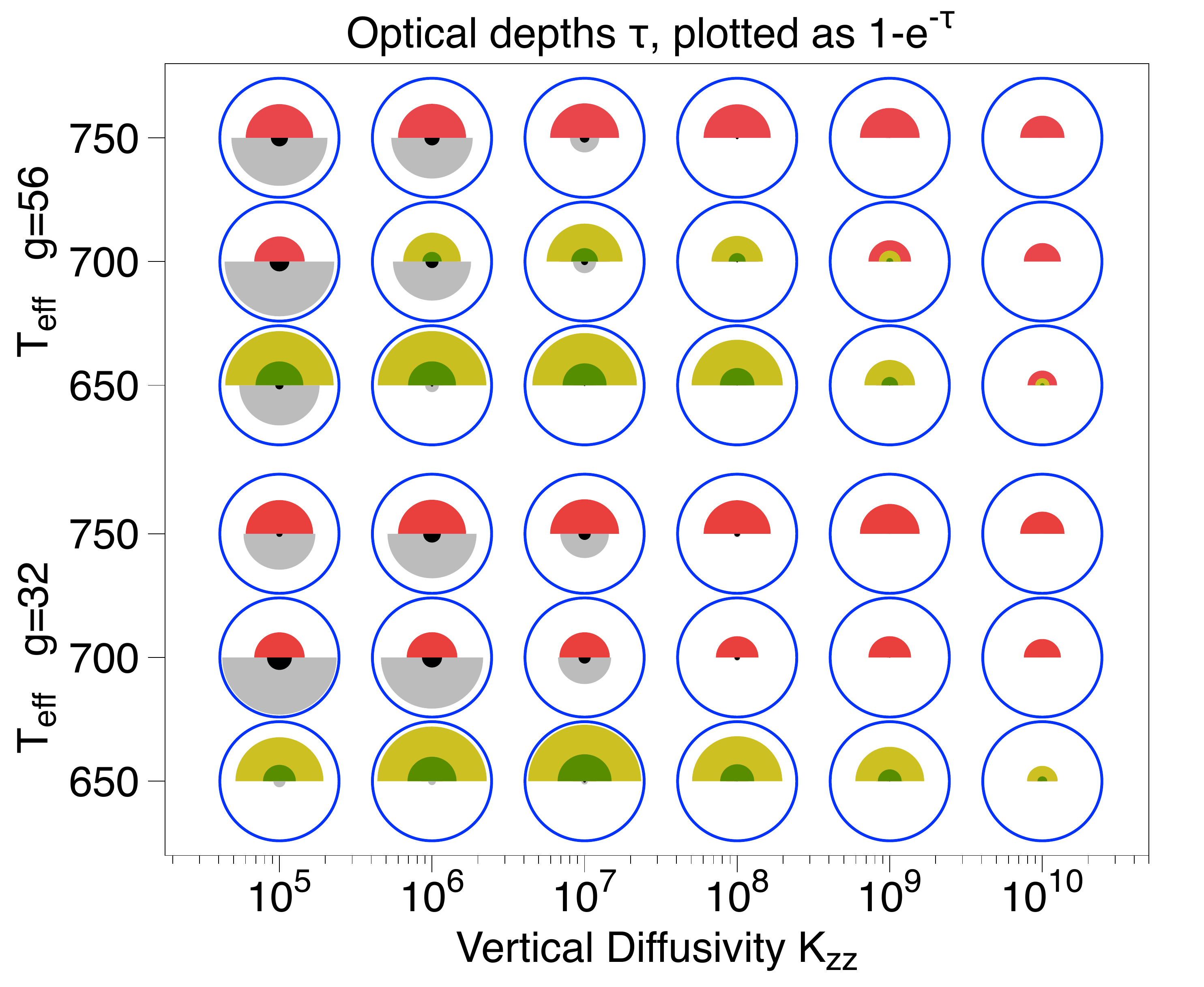} 
   \caption{\small { Geometrical }optical depths at a glance. 
   We plot $1-e^{-\tau}$ rather than $\tau$ itself to give a better graphical
   sense of how much light { might be} blocked.  The outer circle represents the incident light.
   The black and gray disks show two upper bounds on { hazes} of 100 nm diameter organic particles.
   Black is deduced from mixing ratios in the photochemical
   zone where acetylene peaks, while gray refers to number densities of ``C$_4$H$_2$,'' which typically peaks deeper in the atmosphere. 
   Gold shows clouds of 1 micron diameter sulfur particles and green
   shows clouds of 10 micron diameter sulfur particles, presuming in both cases that the particles are opaque.
   Red disks show the optical depth of S$_4$ vapor at 500 nm.
   Where sulfur condenses, S$_4$ should condense too, and so it is not shown.  
   Results are for solar composition; $\tau$ for sulfur scales with metallicity but $\tau$ of organic hazes may not.
    }
\label{optical depth}
\end{figure}

For the sulfur clouds we gather all the S$_8$ above the condensation height into either 1 or 10 $\mu$m diameter particles,
a size range that seems appropriate for condensation clouds.
We assume effective particle densities of 1.5 g/cm$^3$. 
{We then compute geometrical optical depths assuming that the particles are opaque.
An implicit and imperfect assumption is that the particles are large compared to the wavelength of light.
Geometrical optical depths for the two cases are shown} in Figure \ref{optical depth} as gold and green disks, respectively.
For S$_4$ we show the optical depth at 500 nm (red disks), which is near the center of its strong broadband visible light absorption \citep{Meyer1976}.

For the organic hazes, we consider two cases.
In both we assume that organic particles are 100 nm diameter and of effective density 0.6 g/cm$^3$. 
{As with sulfur, we compute the geometrical optical depths of opaque particles.
At 100 nm the particles would be too small to block much visible or infrared light; these would be more likely
to form a blue haze.
Nevertheless geometrical optical depth is a convenient way to quantify the amount of material that might be in a haze. } 
The first (black disks) is based on NMHC mixing ratios:
we presume that 10\% of the NMHCs (chiefly C$_2$H$_2$) go into haze particles at altitudes where
the total mixing ratio of NMHC's exceeds 1 ppmv.
The second case (gray disks) is based on the computed number densities
of ``C$_4$H$_2$'' when treated as a portal through which every carbon that
passes ultimately gets incorporated in a haze particle. 

The results of the exercise are presented in Figure \ref{optical depth} for the same range of 
solar composition models discussed above. 
Sulfur is in the top hemisphere of each circle and carbon in the bottom half.
There appears to be considerable potential for sulfur to be optically significant in 51 Eri b.
This can be as sulfur clouds if 51 Eri b is a cool object, or as vapor  
if sulfur does not condense.  
The sulfur clouds can be optically thick at solar metallicity, and they could be significantly thicker on planets
 because sulfur optical depths will scale linearly with metallicity. 
The sulfur vapors can also be important, especially the chains.
E.g., S$_4$ absorbs strongly at 500 nm, and longer chains absorb 
 to 750-850 nm \citep{Meyer1976} (the rings, which confer invisibility, typically absorb $\lambda < 330$ nm).
 It is well known that liquid sulfur when heated turns from light yellow to dark red as S$_8$ rings
  decompose into a soup of chains and rings (the depth of red depends on impurities, especially hydrocarbons \citep{Moses1991}).   
 We might expect similar behavior in 51 Eri b as S$_8$ rings thermochemically decompose between 10 and 100 mbars.
 On the other hand sulfanes, alkane analogs with the general formula HS$_{\rm n}$H, may be the
 intermediaries between S$_8$ and H$_2$S; like the rings, sulfanes typically absorb $\lambda < 330$ nm \citep{Meyer1976}.
 
There is also some potential for organic hazes to be important, especially where $K_{zz}$ is small,
but this potential is model dependent.
At high altitudes where CH$_4$ and H$_2$O are photolyzed,
optical depths near unity (black disks) are achievable only if conversion of acetylene into PAHs is highly efficient,
which seems unlikely.  
Lower altitudes that coincide with the more reduced S$_{\rm n}$-H$_2$S photochemistry are more promising,
but interpreting ``C$_4$H$_2$'' as a bucket full of particles is a leap that future work could prove baseless.
A difference from sulfur is that we do not expect that modestly higher metallicity will lead to more organic haze. 
   
\section{Sensitivity of the results to model uncertainties}
\label{section:sensitivity}

We have found that most of our models predict
that S$_8$ is a major product of sulfur photolysis (Figure \ref{test85}).
We have also found that NMHC formation is sensitive to sulfur photochemistry.
We have discussed truncation of hydrocarbon chemistry at C$_4$H$_2$ above.
 Here we perform a series of tests to determine how sensitive the model is to other uncertain or unknown factors.
These are (i) different amounts of stellar ultraviolet radiation;
(ii) different rates of S$_8$ photolysis;  
(iii) different estimates of H$_2$S thermolysis and recombination;
(iv) different rates of sulfur polymerization;
and (v) unknown chemical reactions that would compromise S$_8$'s stability.
The latter proves the matter of most concern. 

\subsection{Sensitivity to UV}

\begin{figure}[!htb]
 \centering
 \begin{minipage}[c]{0.49\textwidth}
   \centering
  \includegraphics[width=1\textwidth]{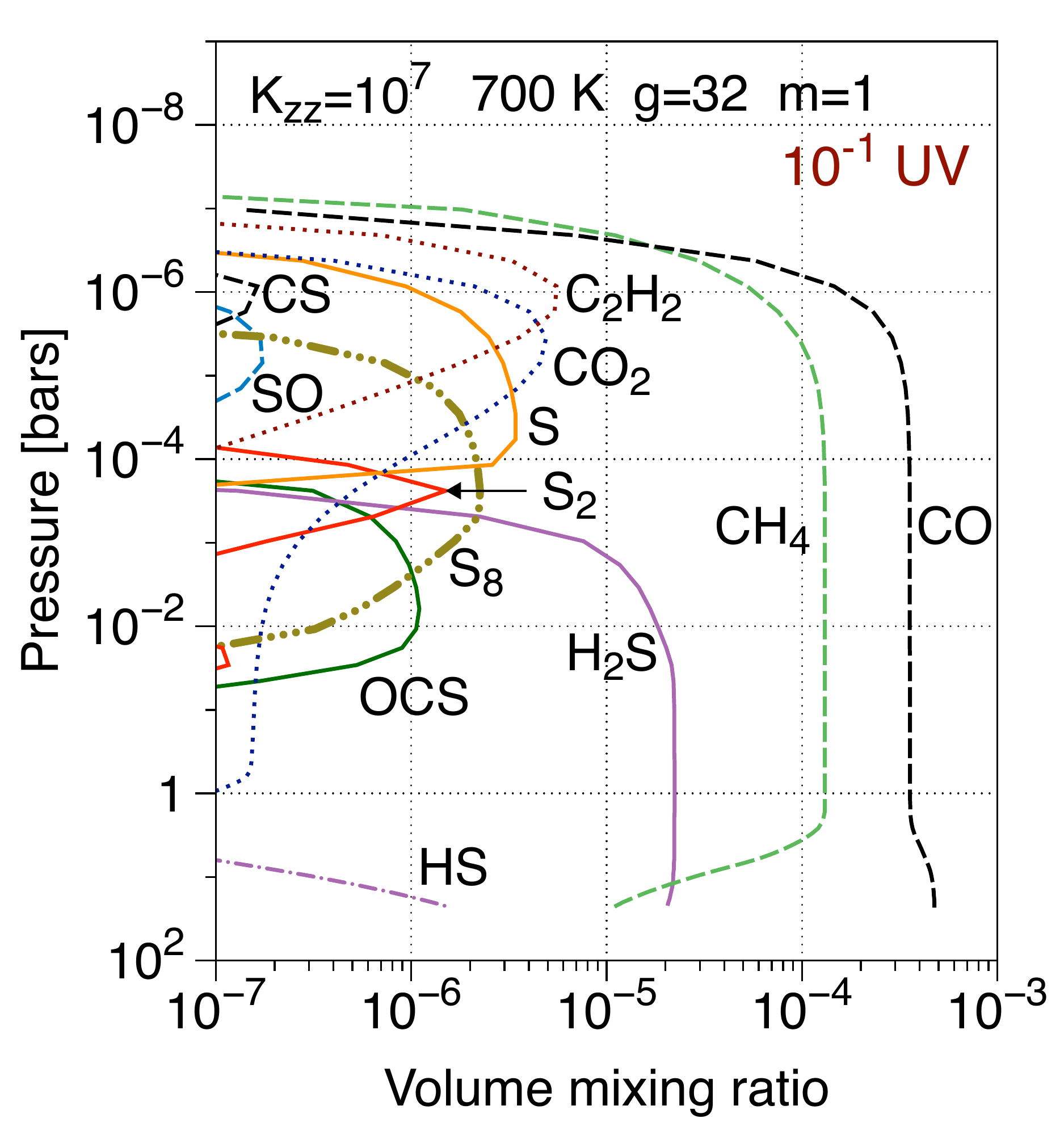} 
 \end{minipage}
 \begin{minipage}[c]{0.49\textwidth}
   \centering
 \includegraphics[width=1\textwidth]{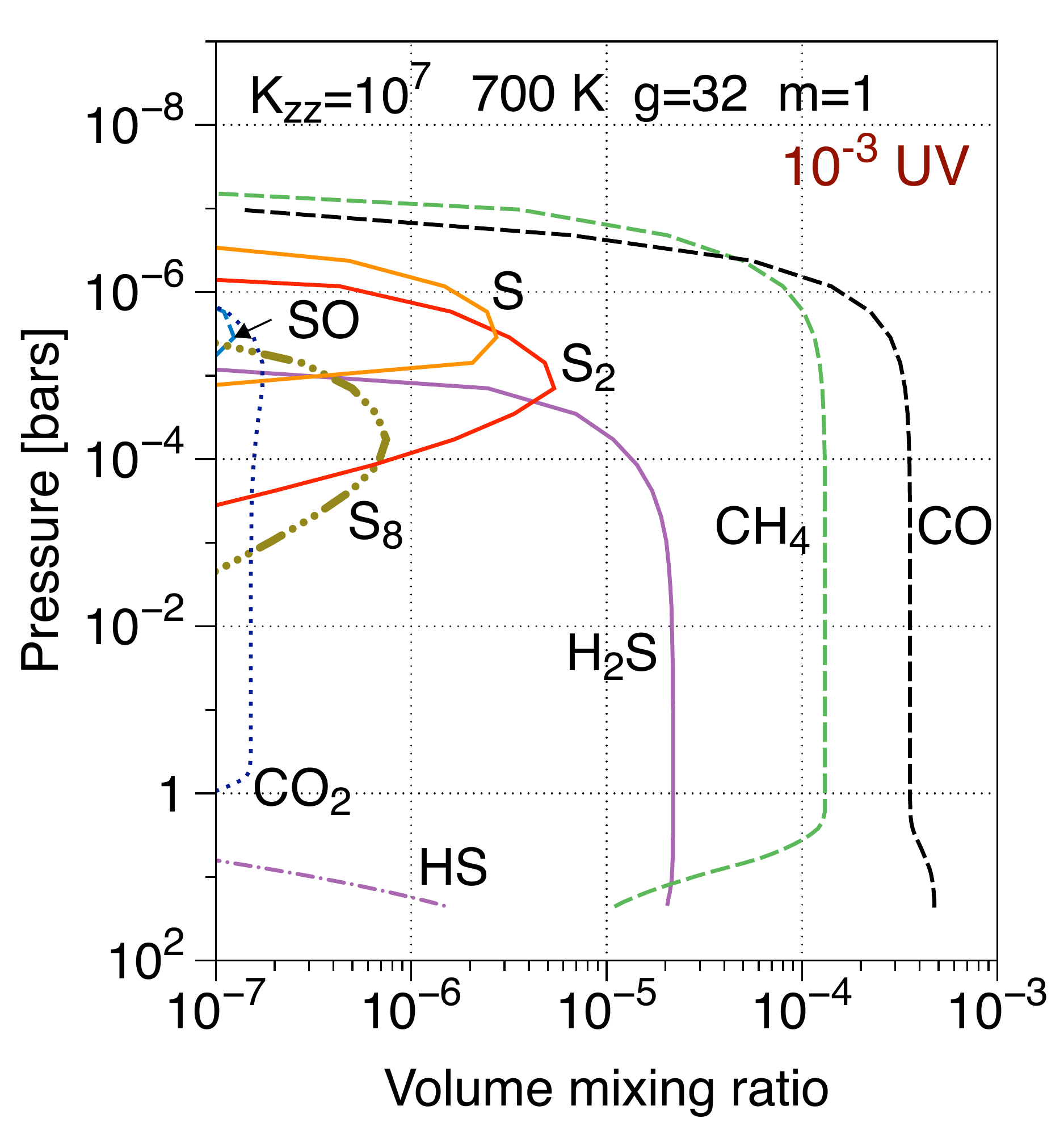} 
 \end{minipage}
  \caption{\small 51 Eridani b models with reduced and greatly reduced UV irradiation. 
   {\it Left.} UV irradiation is 10\% that in the nominal model.
  {\it Right.} UV irradiation is reduced to 0.1\% that in the nominal model. 
  Photochemical CO$_2$ and C$_2$H$_2$ nearly disappear, but a rich sulfur photochemistry remains.
  }
\label{No_UV}
\end{figure}

In Figure \ref{No_UV} we have explored the sensitivity of the nominal model to reduced levels of UV radiation.
With UV irradiation at 10\% that in the nominal model, the general pattern of the photochemistry
   is similar to that in the nominal model.  Chief differences are that there is less CO$_2$ and 
   C$_2$H$_2$, H$_2$S reaches higher altitudes before it
   is destroyed, and there is a modest shift away from S$_8$ as the chief product.
   Even when the UV is reduced to 0.1\% that of the nominal model, there are enough photons
   for H$_2$S to be fully consumed and a complete suite of sulfur photochemical products is generated.
  It is only when UV irradiation is reduced by another factor of ten that most of the H$_2$S survives
  and the sulfur photochemistry becomes photon-limited.

\subsection{Sensitivity to S$_8$ photolysis}

\begin{figure}[!htb]
 \centering
  \begin{minipage}[c]{0.49\textwidth}
   \centering
 \includegraphics[width=1\textwidth]{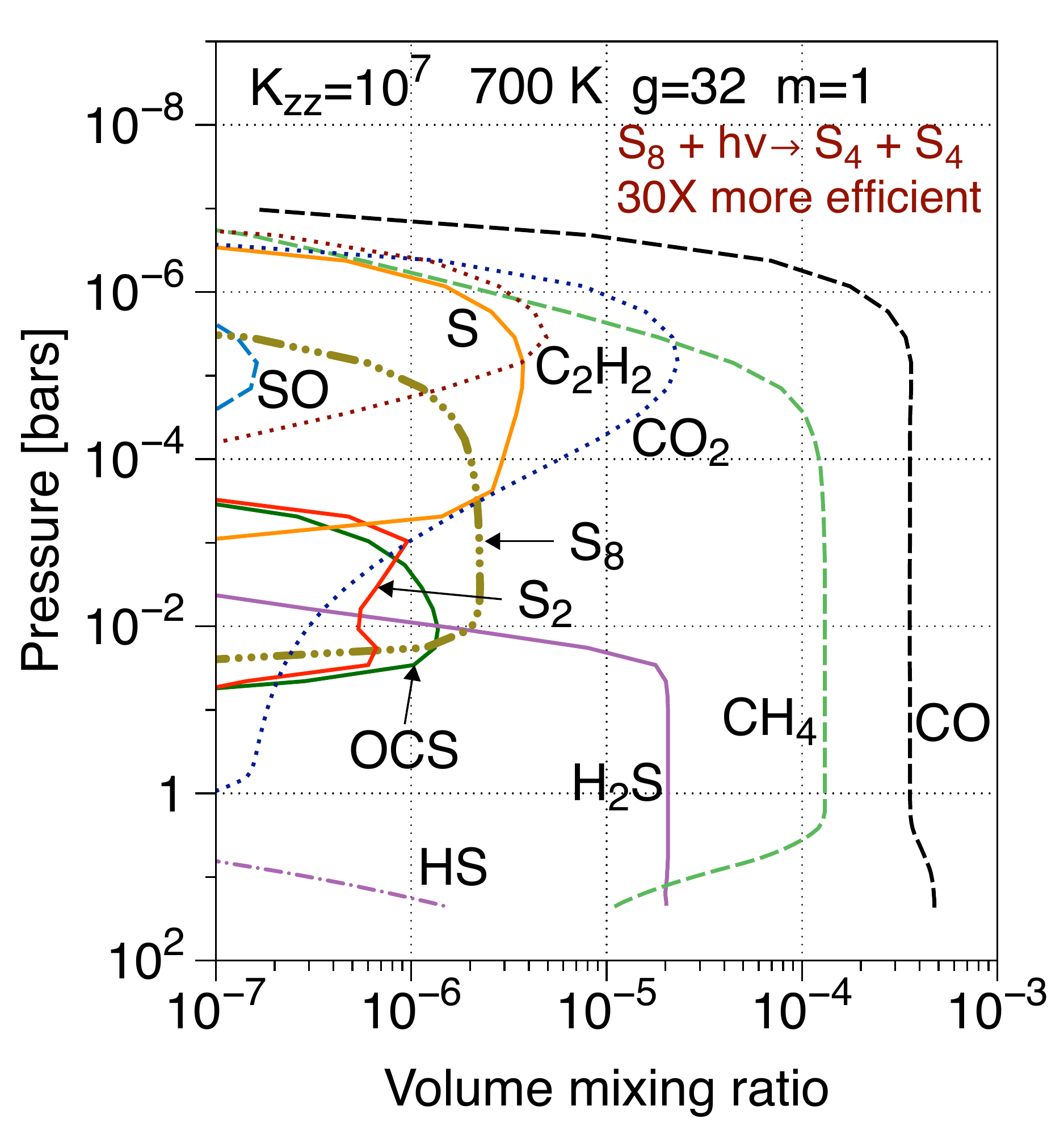} 
 \end{minipage}
\begin{minipage}[c]{0.49\textwidth}
   \centering
  \includegraphics[width=1\textwidth]{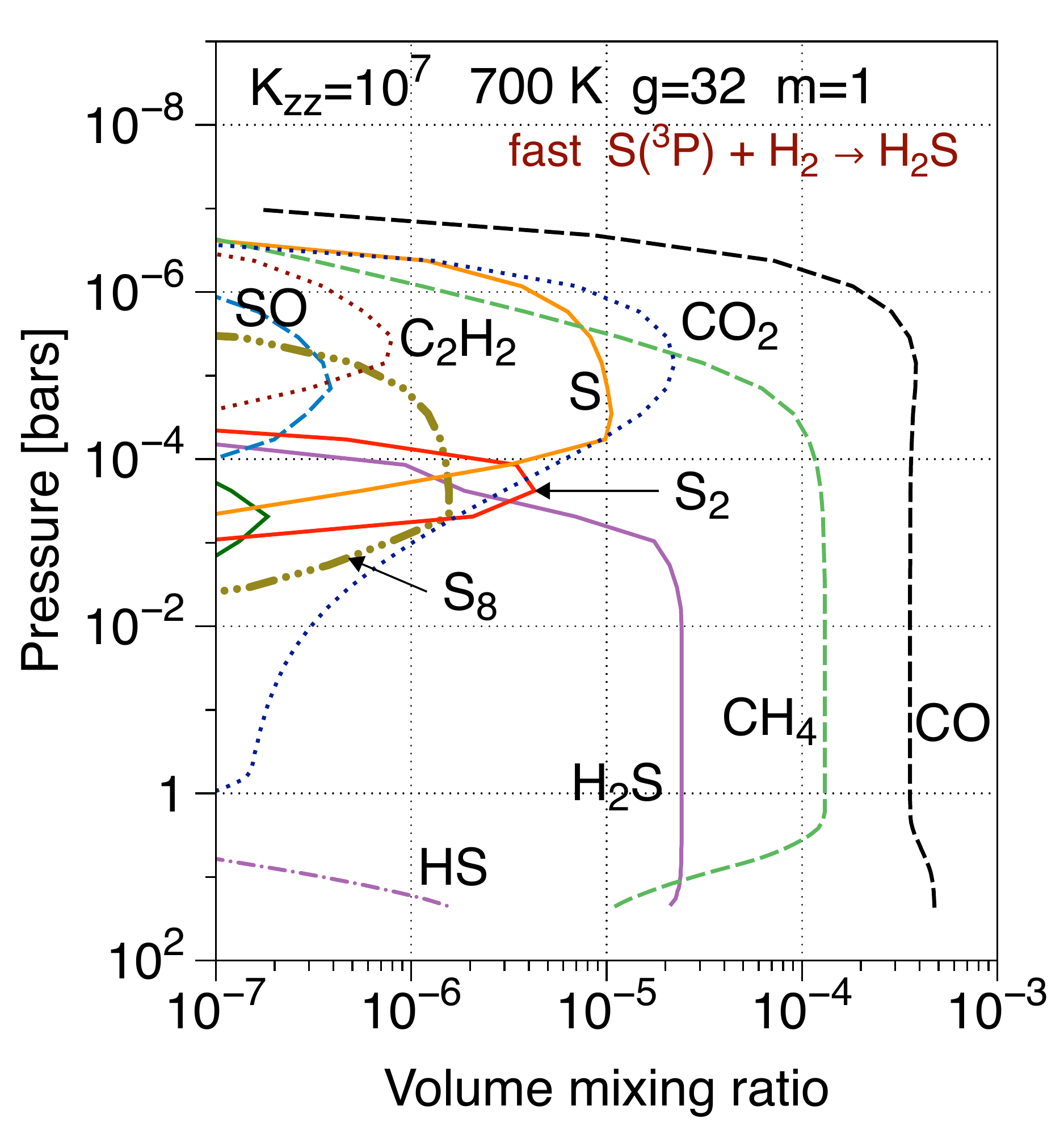} 
 \end{minipage}
  \caption{\small Two sensitivity tests to compare to Figure 1.
   {\it Left.} Enhancing the efficiency of S$_8$ photolysis at low pressures by raising $N_c$ in Eq \ref{Young}
   also generates more S-bearing free radicals that reduce the yield of C$_2$H$_2$ and other NMHCs at the top of the atmosphere.
   {\it Right.} This model uses the faster rate $k'_{22r}$ for the spin-forbidden insertion reaction
   $\mathrm{H}_2 + \mathrm{S}(^3\mathrm{P}) + \mathrm{M} \rightarrow \mathrm{H}_2\mathrm{S} + \mathrm{M} $. 
   The model is quite sensitive to this.
   With the faster rate H$_2$S survives to much higher altitudes than in the nominal model (Figure 1) and
   S$_8$ is less abundant and restricted to higher levels in the atmosphere.  Photolysis of
   other sulfur-containing small molecules generates S-bearing free radicals that reduce the yield of C$_2$H$_2$ and other NMHCs. 
  }
\label{sensitivity}
\end{figure}

We have used \citet{Young1983}'s method for estimating S$_8$'s photolysis rate.
Young suggested that the first UV photon absorbed cleaves the ring.
The resulting linear S$_8$ molecule can either be put back into the form of a ring by a collision,
or it can be broken into two pieces (here both S$_4$) by absorbing a visible light photon.
We assume an absorption cross section of $3\times 10^{-18}$ cm$^2$ to visible light \citep[$\lambda < 850$ nm,][]{Meyer1976}.
 The effective photolysis rate is 
\begin{equation}
\label{Young}
 P(\mathrm{S}_{8}) = P(\mathrm{S}_{8,r}) {N_c P(\mathrm{S}_{8,l})\over N_c P(\mathrm{S}_{8,l}) + \nu_{c} },
\end{equation} 
 where $P(\mathrm{S}_{8,r})$ and $P(\mathrm{S}_{8,l})$ 
 are the photolysis rates of the ring and linear S$_8$ molecules, respectively;
 $N_{c}$ is the number of collisions required to close the ring;
$\sigma_{c}=3\times 10^{-15}$ cm$^2$ is the collision cross section of a molecule;
and $\nu_{c} = N\sigma_{c}{\bar v}$ is the collision frequency in terms of the mean thermal speed ${\bar v}$. 
In the nominal model we take $N_{c}=1$.  For the sensitivity test (Figure \ref{sensitivity}) we take $N_c=30$.
The chief consequence of higher S$_8$ photolysis is that catalytic sulfur is more abundant
 and NMHC yield is reduced. 

\subsection{Sensitivity to H$_2$S recombination}

Rates of many of the chemical reactions that involve sulfur are poorly known.
In particular, a major source of model pathology 
is the 3-body recombination of H$_2$S, either from HS and H, or from H$_2$ and S. 
There is limited information on H$_2$S recombination,
but the reverse process, thermolysis of H$_2$S, is industrially important and
 has been the subject of several experiments that elude easy consensus
 \citep{Bowman1977,Roth1982,Tesner1990,Woiki1994,Woiki1995a,Olschewski1994,Shiina1996,Shiina1998, 
 Karan1999}.
 Measured rates from high temperature ($1800<T<3500$ K) shock tube experiments in Ar
 \citep{Bowman1977,Woiki1994,Woiki1995a,Olschewski1994,Shiina1996,Shiina1998}
 differ among themselves by an order of magnitude; it is not clear why.
 Moreover, lower temperature ($800 < T < 1400$ K) flow reactor experiments in N$_2$ \citep{Tesner1990,Karan1999}
 imply rates that are 100-300 times higher than extrapolation of the shock tube data predict.
  
 It was at first presumed that the dominant decomposition channel was 
\begin{equation}  \tag{R21}
 \mathrm{H}_2\mathrm{S} + \mathrm{M} \rightarrow  \mathrm{HS} + \mathrm{H} + \mathrm{M}   
\end{equation} 
as with H$_2$O, because the alternative
 \begin{equation}\tag{R22}
 \mathrm{H}_2\mathrm{S} + \mathrm{M} \rightarrow  \mathrm{H}_2 + \mathrm{S}(^3\mathrm{P}) + \mathrm{M} , 
 \end{equation}
although much less endothermic, is spin forbidden \citep{Roth1982}.
But parallel shock tube experiments by \citet{Woiki1994}, who monitored S($^3$P) production, 
and \citet{Olschewski1994}, who monitored H$_2$S disappearance, gave a consistent picture
of the spin-forbidden path being dominant.
The straightforward, thermodynamically self-consistent reverse reaction 
 \begin{equation}\tag{R22r}
 \mathrm{H}_2 + \mathrm{S}(^3\mathrm{P}) + \mathrm{M} \rightarrow  \mathrm{H}_2\mathrm{S} + \mathrm{M}  
 \end{equation}
 was therefore predicted to be fast at low temperatures. 
The possibility that the interesting reaction R22r might be fast 
motivated follow-up experiments by \citet{Woiki1995a} and \citet{Shiina1996,Shiina1998} to directly determine the reaction rate
between S($^3$P) and H$_2$. 
\citet{Shiina1998} found that, for $T>900$ K, R22r is negligible compared to the competing abstraction
reaction 
 \begin{equation}\tag{R9r}
 \mathrm{H}_2 + \mathrm{S}(^3\mathrm{P}) \rightarrow  \mathrm{HS} + \mathrm{H}.
 \end{equation}
\citet{Shiina1998} do not dispute that R22 is the more important thermolysis channel for H$_2$S,
but they change the extrapolation to low
temperatures to take into account the considerable energy barrier that they computed,
\begin{equation}
\label{k22}
k_{22} = 8.9\times 10^{-7} \left(T/300\right)^{-2.61} \exp{\left(-44640/T\right)} .
\end{equation}   
The rate we use for R22r in our standard models is the reverse of $k_{22}$,  
\begin{equation}
\label{k22r}
k_{22r} = 1.4\times 10^{-31} \left(T/300\right)^{-1.9} \exp{\left(-8140/T\right)},
\end{equation}   
which is far below the upper bound determined by \citet{Shiina1998}
and very slow (but not negligible) at low temperatures.  

For the sensitivity test (Figure \ref{sensitivity}, right-hand panel) we use a parallel pair of rates that are consistent 
both with the higher thermolysis rates reported by \citet{Olschewski1994} and \citet{Woiki1994}
and with the lower activation energy estimated by \citet{Olschewski1994}:
\begin{equation}
\label{kprime22}
k'_{22} = 8.9\times 10^{-7} \left(T/300\right)^{-2.61} \exp{\left(-38800/T\right)} 
\end{equation}   
with reverse  
\begin{equation}
\label{kprime22r}
k'_{22r} = 1.4\times 10^{-31} \left(T/300\right)^{-1.9} \exp{\left(-2300/T\right)} .
\end{equation}   
The rate $k'_{22r}$ is comparable to the upper bound reported by \citet{Shiina1998}.
Although much slower than the rate that \citet{Shiina1998} had hoped to see,
$k'_{22r}$ is fast enough to affect our results significantly (Figure \ref{sensitivity}). 
With $k'_{22r}$, H$_2$S reaches altitudes 3 scale heights above where it gets to with $k_{22r}$. 

We do not attempt to take into account the flow reactor data.
These experiments suggest thermolysis rates that are orders of magnitude faster than either $k_{22}$ or $k'_{22}$ at 1000 K,
and therefore the recombination reactions must also be.
However, we were unable to reproduce \citet{Karan1999}'s argument that the different reported rates can be brought into agreement.
We favor the shock tube data because the flow reactor system is more complicated (more reactions need to be taken into account)
and less straightforwardly interpreted.
E.g., what \citet{Karan1999} actually measured is whether the system has had time enough to reach thermochemical equilibrium,
which isn't quite the same thing as determining a particular reaction rate.
In the end, we think that the slower rates $k_{22r}$ and $k'_{22r}$ are more plausible given the extensive
molecular rearrangements that must occur if an unlikely-looking reaction like R22r
is to take place

\subsection{Sensitivity to S$_{\rm n}$ polymerization}

\begin{figure}[!htb]
 \centering
 \begin{minipage}[c]{0.49\textwidth}
   \centering
  \includegraphics[width=1\textwidth]{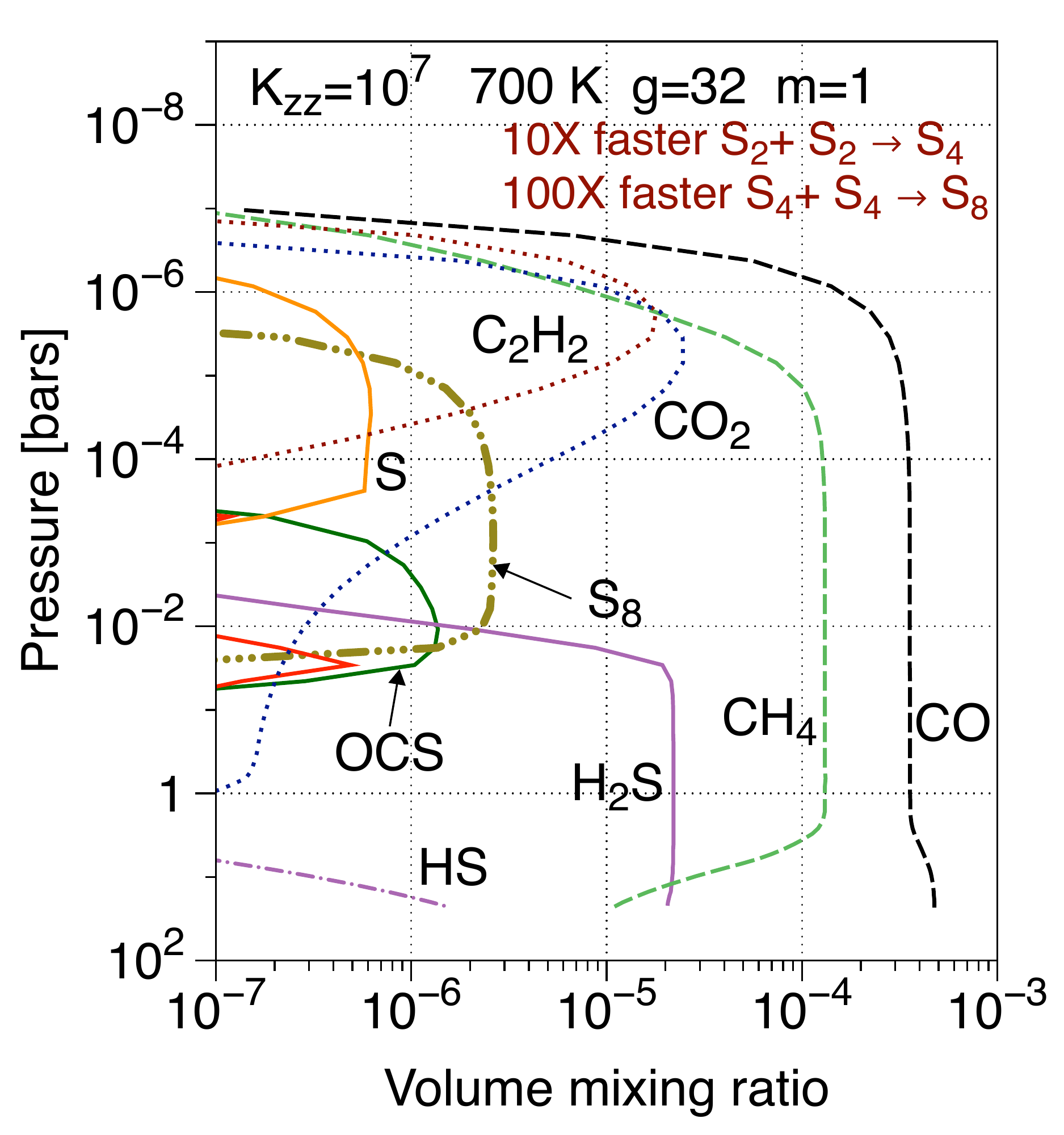} 
 \end{minipage}
 \begin{minipage}[c]{0.49\textwidth}
   \centering
 \includegraphics[width=1\textwidth]{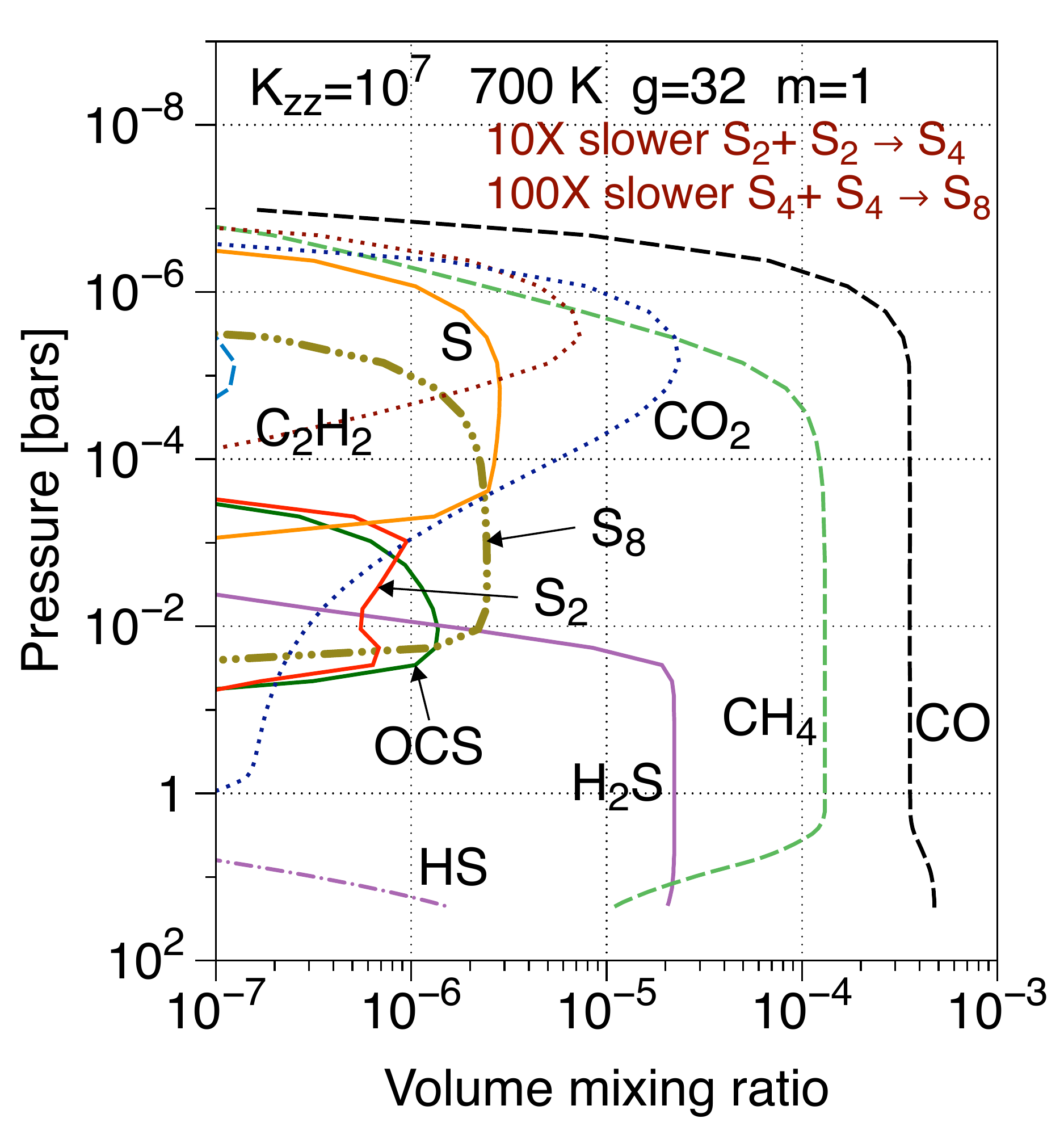} 
 \end{minipage}
  \caption{\small Sensitivity of results to rates of S$_n$ polymerization (to be compared to Figure 1).
   {\it Left.} Faster polymerization decreases the abundances of S and S$_2$ without noticeably changing
   S$_8$, because in the nominal case most of the sulfur was already pooling in S$_8$.
  {\it Right.} Slower polymerization increases the abundances of S and S$_2$, but not enough in this model
  to noticeably affect the S$_8$ abundance.  The greater abundance of S-bearing radicals 
  causes the acetylene (C$_2$H$_2$) yield to shrink.
  }
\label{polymerize}
\end{figure}

Our nominal sulfur polymerization scheme is mostly encompassed by reactions R2-R8 in Table 1.
We have kept the system simple because the reactions and rates are very uncertain.
Our rates are similar to those used elsewhere \citep[e.g.,][]{Moses2002,Yung2009} and are not inconsistent
with the few experimental reports \citep{Fair1969,Langford1972,Langford1973,Nicholas1979}.
For the sensitivity tests we raise [lower] the rates of
\begin{equation}\tag{R4}
\mathrm{S}+\mathrm{S}_3+\mathrm{M} \rightarrow \mathrm{S}_4 +\mathrm{M}
\end{equation}
and 
\begin{equation}\tag{R5}
\mathrm{S}_2+\mathrm{S}_2+\mathrm{M} \rightarrow \mathrm{S}_4 +\mathrm{M}
\end{equation}
by a factor of 10, and raise [lower] the rate of
\begin{equation}\tag{R7}
 \mathrm{S}_4+\mathrm{S}_4+\mathrm{M} \rightarrow \mathrm{S}_8 +\mathrm{M}
\end{equation}
 by a factor of 100. 
 These two cases of faster and slower polymerization are illustrated in left- and right-hand
 panels of Figure \ref{polymerize}, respectively.
 The figure shows that our model is not very sensitive to reasonable uncertainties in the sulfur polymerization rate.

\subsection{Sensitivity to unknown mechanisms of S$_8$ chemical destruction}

Other than photolysis, the main sink of S$_8$ in our basic model is thermal destruction
\begin{equation}\tag{R7r}
\mathrm{S}_8 + \mathrm{M} \rightarrow \mathrm{S}_4 + \mathrm{S}_4 + \mathrm{M} .
\end{equation}
This is predicted to be rather fast, because $\Delta H$ is a relatively modest 150 kJ/mol,
there is a considerable gain in entropy, and the rate for the forward reaction R7 is probably fast.
We know of no reported kinetic data regarding S$_8$'s reactions.
Yet S$_8$ should be reactive because reactions with important free radicals should be significantly exothermic.
To test the sensitivity of our model to these unknown reactions, we need to invent both the reactions and the products.
The most abundant free radical by far is H, which makes reactions with H the likeliest to be important.

At high pressures we might expect a 3-body reaction to unmake the ring into a quasi-linear HS$_8$ radical,
which would then be followed by reactions that either return S$_8$ or cleave the S$_8$ chain.
We have not pursued this strategy here because (i) we would have to invent many species and many rates
and (ii) our rate for R7r is pretty fast at high pressures.

At low pressures any plausible reaction would have to cleave the chain in two places.
The invented reaction that adds the least new complexity to our model is
\begin{equation}\tag{R51}
\mathrm{S}_8 + \mathrm{H} \rightarrow \mathrm{HS}_4 + \mathrm{S}_4,
\end{equation}
because we need to add only one invented species, HS$_4$.
We estimate a standard heat of formation of 110 kJ/mol and standard entropy of 330 J/mol/K
 by analogy to HS$_2$ \citep{Benson1978}.
The invented reaction R51 is therefore substantially exothermic but undoubtedly faces a considerable energy barrier.
We consider a slow rate   
\begin{equation}\label{slow}
k_{\rm slow} = 3\times 10^{-12} \exp{\left(-5000/T\right)}
\end{equation}
and a fast rate   
\begin{equation}\label{fast}
k_{\rm fast} = 1\times 10^{-11} \exp{\left(-2500/T\right)} .
\end{equation}
For the temperature dependence we use Pauling's rule of thumb \citep[][p.\ 568]{Pauling1970} that the
activation barrier of a radical-molecule reaction is about 8\% of the bond energy of the bond to be broken.
The slow rate presumes that the two S-S bonds are additive.

We then need a set of reactions for HS$_4$.
One category of reaction will be the H-abstraction reactions with H, OH, and some other radicals, such as
\begin{equation}
\mathrm{H} + \mathrm{HS}_4 \rightarrow \mathrm{H}_2 + \mathrm{S}_4 .
\end{equation}
These will probably have small activation barriers.
At high altitudes S$_4$ will be promptly photolysed by visible light, so that the sulfur chain is quickly broken down. 
 The other representative category will be molecular rearrangements that reconstitute H$_2$S, such as
 \begin{equation}
\mathrm{H} + \mathrm{HS}_4 \rightarrow \mathrm{H}_2\mathrm{S} + \mathrm{S}_3 .
\end{equation}
 Reactions of this type face considerable activation barriers but can be important at depths
 where downwelling S$_8$ is converted back to H$_2$S.
 For these we use the same 2500 K activation barrier that we used for breaking the S-S bond. 
   
\begin{figure}[!htb]
 \centering
 \begin{minipage}[c]{0.49\textwidth}
   \centering
  \includegraphics[width=1\textwidth]{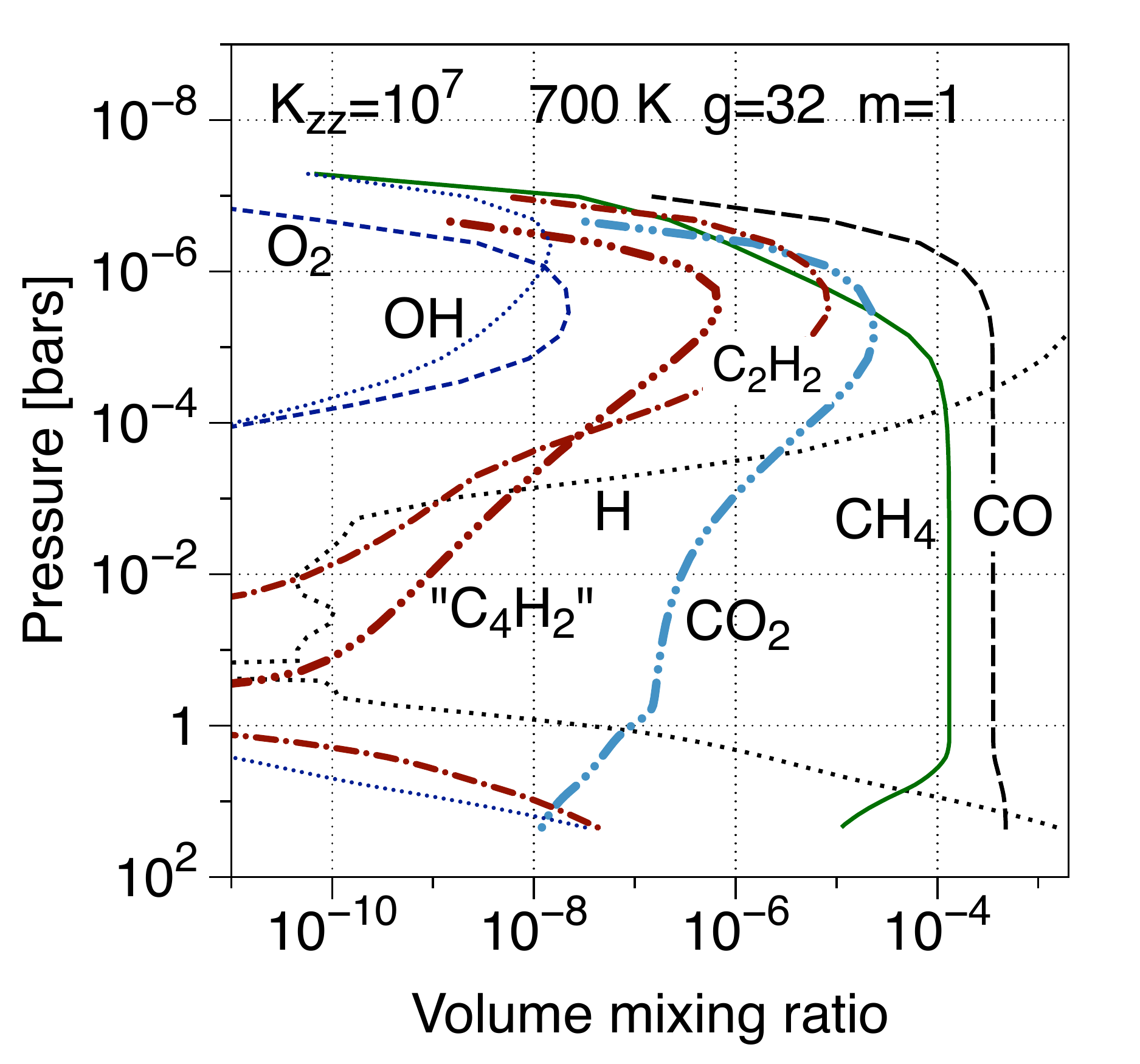} 
 \end{minipage}
 \begin{minipage}[c]{0.49\textwidth}
   \centering
 \includegraphics[width=1\textwidth]{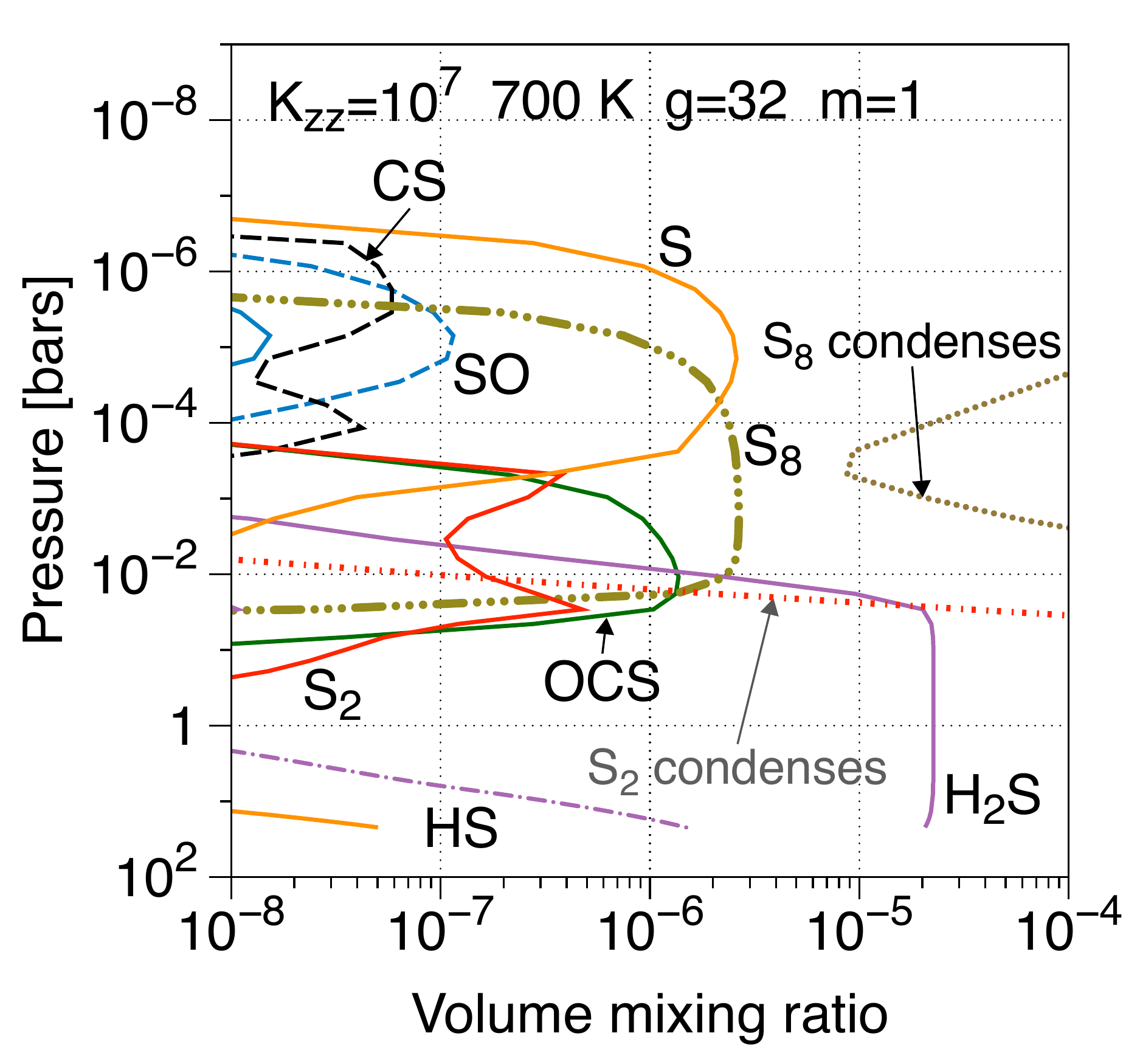} 
 \end{minipage}
 \centering
 \begin{minipage}[c]{0.49\textwidth}
   \centering
 \includegraphics[width=1\textwidth]{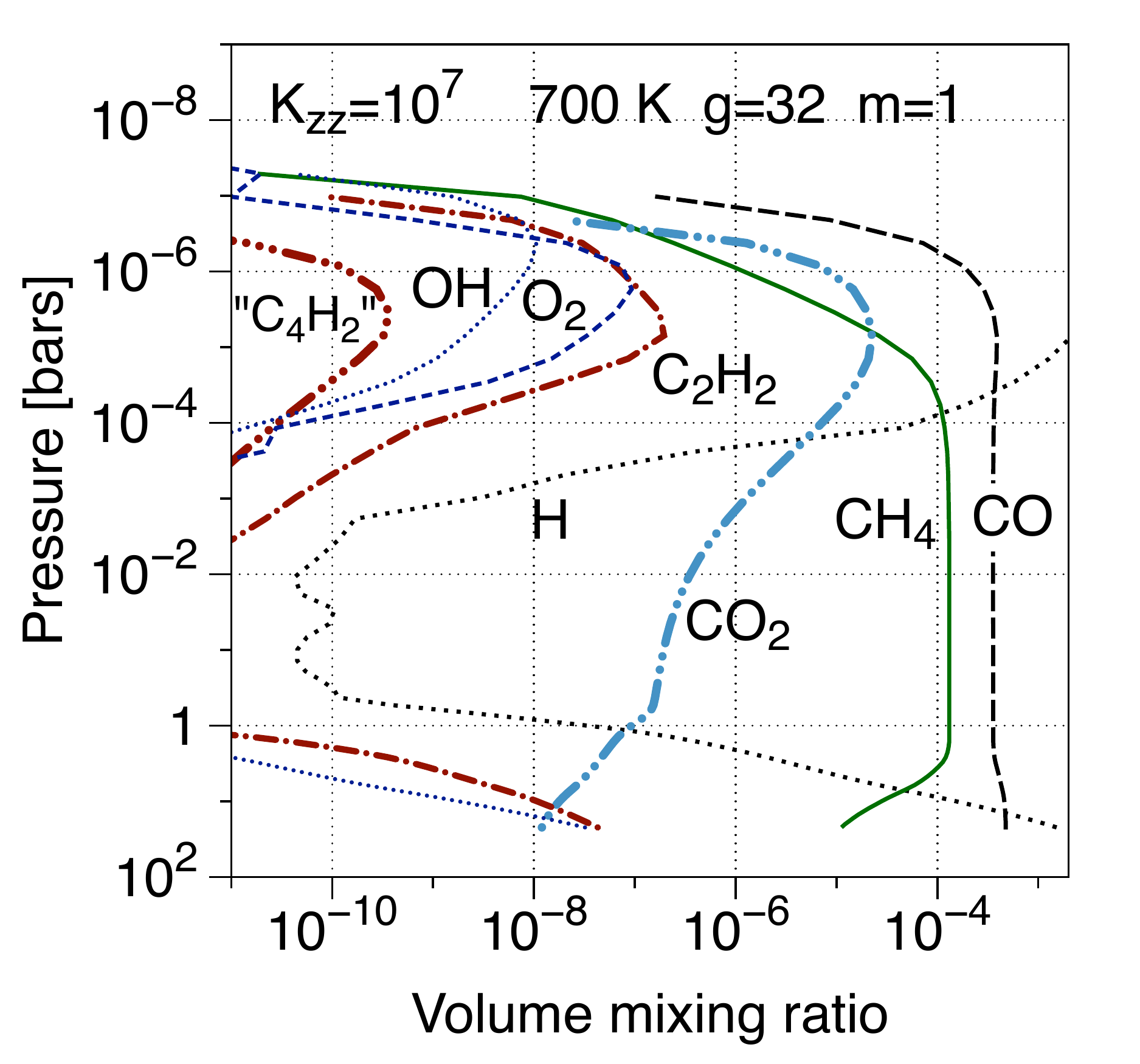} 
 \end{minipage}
 \begin{minipage}[c]{0.49\textwidth}
   \centering
 \includegraphics[width=1\textwidth]{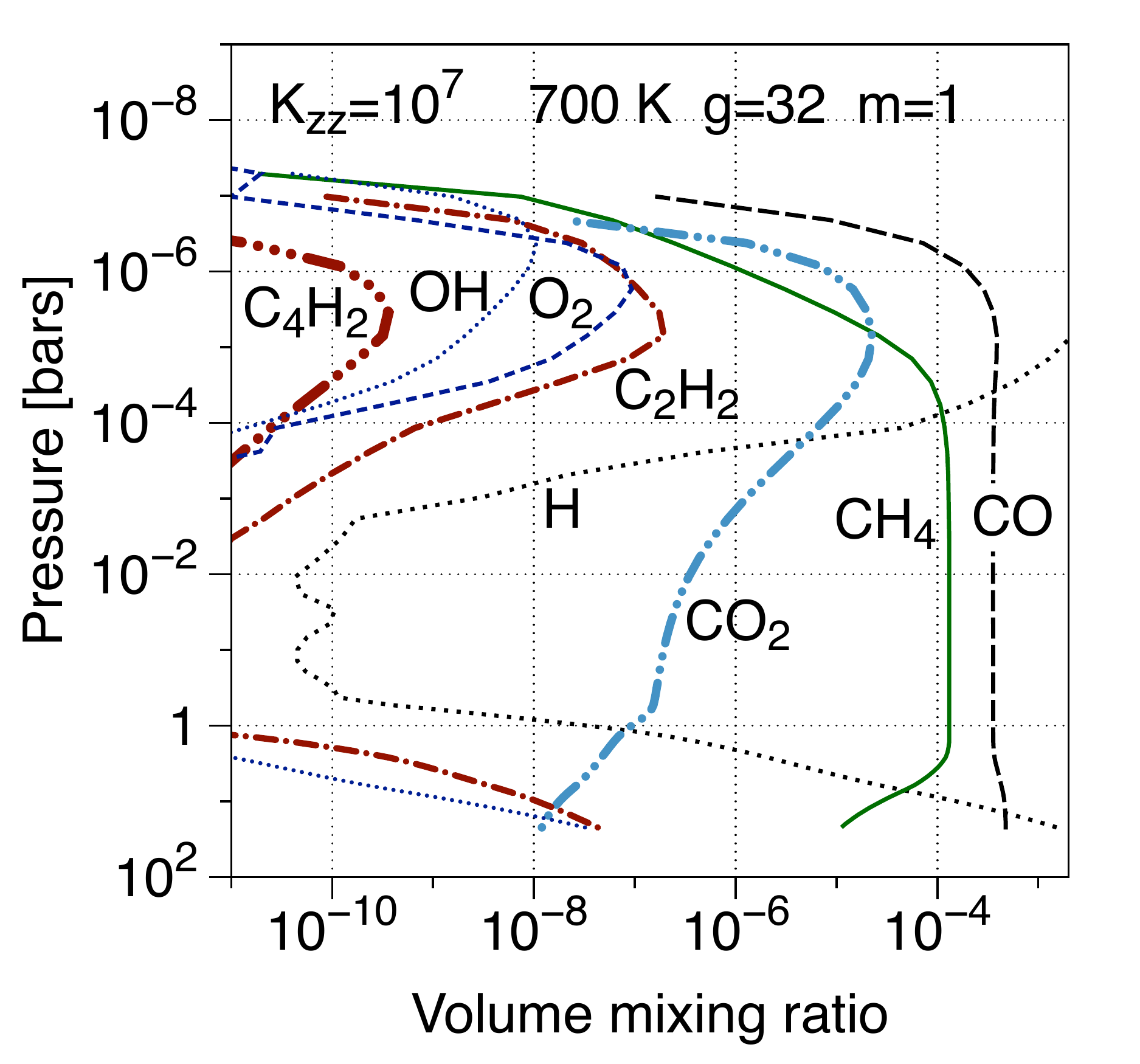} 
 \end{minipage}
 \caption{\small Chemical sensitivity tests.  {\it Top.} With the slow rate $k_{\rm slow}$ (Eq \ref{slow}) for 
$\mathrm{H} + \mathrm{S}_8$, the nominal model is little altered, although there is notably more atomic S at high altitudes.
  {\it Bottom.} With the fast rate $k_{\rm fast}$ (Eq \ref{fast}), the model looks rather different, with S$_8$ eliminated above 100 $\mu$bars
  and much increased abundances of photochemically active S-containing radicals and molecules.  The overall character
  of the upper atmosphere is more oxidized, acetylene is much reduced, and the proxy ``C$_4$H$_2$'' is nearly wiped out. }
\label{4Sensitivity}
\end{figure}

Our expectation had been that deep thermal recycling of S$_8$ would be much sped up by the new chemistry, but this is
not really evident in Figure \ref{4Sensitivity}.  Rather, the greater impact of the new chemistry is
to convert S$_8$ in the upper atmosphere into other more active species, and finally to atomize it.
The more abundant S-containing radicals catalyze the oxidation of organics.
On the other hand our chemical schemes do not encompass the speculative possibility that sulfur
might also catalyze carbon polymerization. 
 In summary, what we don't know about sulfur chemistry appears to have relatively little impact on 
 whether S$_8$ forms, but there appears to be a strong impact on carbon chemistry. 
 If sulfur does not condense, the fast rate $k_{\rm fast}$ for S$_8$ destruction does not bode well for organic hazes.
 Prospects for organics then become better in cooler atmospheres
  because sulfur condensation would deplete S-containing radicals above the cloudtops.

\section{Discussion}

Photochemical hazes are widespread in the solar system 
but they are not yet established as fact on any actual exoplanet.
Observations do not go much beyond showing that many exoplanetary spectra require a broad-band opacity
resembling that of clouds.  
What these clouds might be made of has been a problem for theory \citep{Helling2014,Morley2015},
but as many substances can condense, it is reasonable to expect that there are many kinds of cloud.

Our purpose when we began this study was to make a case for organic hazes on the particular planet 51 Eri b.
The idea was to use C$_4$H$_2$, the first product of acetylene polymerization, as a
proxy for further polymerization: every ``C$_4$H$_2$'' formed was assumed to eventually become incorporated into a particle.
The quotes on ``C$_4$H$_2$'' indicate that we are no longer
talking about C$_4$H$_2$ the molecule but instead about everything downstream from it. 
But even in those cases where we have clearly tipped the scales to favor ``C$_4$H$_2$,''
 it only becomes more abundant than C$_2$H$_2$ at depths well below the primary photochemical region, 
 where conditions are more reducing.
It doesn't help our case that when we treat C$_4$H$_2$ as an actual molecule subject to cracking,
we find that there isn't all that much of it there.
In summary, the case for soots is intriguing but falls short of being compelling. 

On the other hand we have rediscovered the importance of sulfur. 
Sulfur can have a disproportionate influence on photochemistry
because most S-bearing species are relatively easily photolyzed by 
UV photons with $\lambda > 200$ nm, which 51 Eri a, an F star, emits copiously.
Thus sulfur photochemistry becomes a major source of free radicals that can catalyze other chemistries.
One consequence of sulfur catalysis is a tendency to drive carbon 
away from the disequilibrium NMHCs and toward the stronger bonds of CO and CO$_2$.

More interesting is that sulfur itself can be the photochemical cloud that we are looking for.
We find that for a wide range of conditions the major photochemical product of sulfur in 
a planet like 51 Eri b is the ring molecule S$_8$, which typically forms at $\sim 10$ mbars
and extends up to 100 $\mu$bars.
The overall sulfur cycle is simple: H$_2$S flows up and S$_8$ and H$_2$ flow down.
In the cooler half of our models sulfur condenses to make a photochemical haze that, depending on particle size,
can be optically thick, while in the warmer half of the models sulfur remains in the vapor phase.
The sulfur vapor itself might also be optically important, especially at the interface between
abyssal H$_2$S and S$_8$, where the latter thermally decomposes into a wide range of optically active
molecules that are eventually hydrogenated to recombine H$_2$S. 
The sulfur photochemistry we have discussed in this paper is quite general and ought to be found
in a wide variety of worlds over a broad temperature range, both much cooler and much hotter
than the 650-750 K range studied here, and will be present on planets where the UV irradiation is very weak.  
Sulfur clouds should be found in many of these.  
Whether 51 Eri b itself is cold enough for sulfur to condense cannot be answered until
radiative transfer models incorporate sulfur vapors and sulfur clouds,
which is a project beyond the scope of this paper, or until the yellow clouds are seen.

\section{Acknowledgements}
We thank Channon Visscher and Michael Line for insightful and incisive commentary, and
we thank an anonymous reviewer for hinting that a previous draft of this paper
 that only addressed organic hazes was not really very interesting.     
M.S.M. and C.M. gratefully acknowledge the support of the NASA Origins Program.
J.I.M gratefully acknowledges the support of the NASA Planetary Atmospheres Program.
K.J.Z. acknowledge the support of the Virtual Planet Laboratory of the National Astrobiology Institute.



\small

\clearpage

\newpage

\setlongtables 
\begin{longtable}{l lcl l p{3.5cm} } 
 & {\bf Table 1}  & &  & & \\
\hline
 & {\strut Reactants}  &  & {Products} & \!\!{Rate [cm$^{3}$s$^{-1}$] or [cm$^{6}$s$^{-1}$]} & {Reference} \\
\hline 
\endfirsthead
\hline
 & {\strut Reactants}  &  & {Products} & \!\!{Rate [cm$^{3}$s$^{-1}$] or [cm$^{6}$s$^{-1}$]} & {Reference} \\
\hline 
\endhead 

 \refstepcounter{reaction}R\arabic{reaction}   & S  + S   + M$^{\ast}$ & $\!\!\!\rightarrow\!\!$ &  S$_2$ + M &$\!\!  2.0\!\times\! 10^{-33} e^{ 206/T}$ &  \citet{Du2008}\\
             & S + S    &$\!\!\!\rightarrow\!\!$&  S$_2$    &$\!\!  2.3\!\times\! 10^{-14} e^{ 415/T}$ &  \citet{Du2008}\\
 \refstepcounter{reaction}R\arabic{reaction}   & S  + S$_2$  + M & $\!\!\!\rightarrow\!\!$ &  S$_3$  + M & $\!\!  1.0\!\times\! 10^{-30} \left(T/298 \right)^{-2.0}$ & assumed\\
             & S   + S$_2$       &$\!\!\!\rightarrow\!\!$&  S$_3$      &$\!\!  5.0\!\times\! 10^{-11}$ & assumed \\
\refstepcounter{reaction}R\arabic{reaction}  & S   + S$_3$   &$\!\!\!\rightarrow\!\!$ &  S$_2$  + S$_2$      & $\!\!  4.0\!\times\! 10^{-11}$ & assumed\\
 \refstepcounter{reaction}R\arabic{reaction}   & S   + S$_3$  + M & $\!\!\!\rightarrow\!\!$ &  S$_4$  + M &$\!\!  1.0\!\times\! 10^{-30} \left(T/298 \right)^{-2.00}$ & assumed, varied $10\times$ \\
           & S  + S$_3$   &$\!\!\!\rightarrow\!\!$&  S$_4$   &$\!\!  5.0\!\times\! 10^{-11}$ & assumed \\
\refstepcounter{reaction}R\arabic{reaction}  & S$_2$  + S$_2$   + M & $\!\!\!\rightarrow\!\!$ &  S$_4$  + M &$\!\!  1.0\!\times\! 10^{-30} \left(T/298 \right)^{-2.00}$ & varied $10\times$, see note\\
            & S$_2$  + S$_2$ &$\!\!\!\rightarrow\!\!$&  S$_4$     &$\!\! 3.0\!\times\! 10^{-11}$ & assumed \\
 \refstepcounter{reaction}R\arabic{reaction}   & S + S$_4$ &$\!\!\!\rightarrow\!\!$ &  S$_2$  + S$_3$ & $\!\!  4.0\!\times\! 10^{-11} e^{  -500/T}$ & \citet{Moses1995}\\
\refstepcounter{reaction}R\arabic{reaction}  & S$_4$ + S$_4$   + M & $\!\!\!\rightarrow\!\!$ &  S$_8$ + M &$\!\!  7.0\!\times\! 10^{-30} \left(T/298 \right)^{-2.00}$ & assumed, varied $100\times$ \\
            & S$_4$  + S$_4$ &$\!\!\!\rightarrow\!\!$&  S$_8$   &$\!\!  7.0\!\times\! 10^{-11}$ & assumed \\
 \refstepcounter{reaction}R\arabic{reaction}  & S  + HS   &$\!\!\!\rightarrow\!\!$ &  S$_2$  + H   & $\!\!  1.0\!\times\! 10^{-11}$ &  assumed, see note \\
 \refstepcounter{reaction}R\arabic{reaction}  &  H  + HS    &$\!\!\!\rightarrow\!\!$ &   S   + H$_2$  & $\!\!  3.0\!\times\! 10^{-11}\left(T/298 \right)^{0.7}$  & reverse of R\arabic{reaction}r \\
  R\arabic{reaction}r  & S  + H$_2$   &$\!\!\!\rightarrow\!\!$ &  H  + HS  & $\!\!  5.3\!\times\! 10^{-10}\left(T/298 \right)^{0.95} e^{-9920/T}$ & see note\\

 \refstepcounter{reaction}R\arabic{reaction}   & HS  + HS    &$\!\!\!\rightarrow\!\!$ &  S$_2$ + H$_2$  & $\!\!  1.3\!\times\! 10^{-11} e^{-20600/T}$ & like $2\mathrm{OH} \rightarrow \mathrm{H}_2+\mathrm{O}_2$ \\
 \refstepcounter{reaction}R\arabic{reaction}   & H  + S$_3$  &$\!\!\!\rightarrow\!\!$ &  HS + S$_2$ & $\!\!  5.0\!\times\! 10^{-11} e^{  -500/T}$ & like $\mathrm{H}+\mathrm{O}_3 \rightarrow \mathrm{OH}+\mathrm{O}_2$\\
 \refstepcounter{reaction}R\arabic{reaction}   & H  + S$_4$  &$\!\!\!\rightarrow\!\!$ &  HS + S$_3$  & $\!\!  5.0\!\times\! 10^{-11} e^{  -500/T}$ & like H  + S$_3$\\
 \refstepcounter{reaction}R\arabic{reaction}   & O + HS  & $\!\!\!\rightarrow\!\!$ &  OH  + S & $ \!\! 1.7\!\times\! 10^{-11} \left(T/298\right)^{ 0.67}e^{  -956/T}$ & \citet{Schofield1973}\\
 \refstepcounter{reaction}R\arabic{reaction}   & HS + OH  &$\!\!\!\rightarrow\!\!$ &  H$_2$O + S  & $\!\! 4.0\!\times\! 10^{-12} e^{  -240/T}$ & inspired by R23\\
 \refstepcounter{reaction}R\arabic{reaction}   & S + CH  & $\!\!\!\rightarrow\!\!$ &  HS  + C   & $\!\! 1.7\!\times\! 10^{-11} \left(T/298\right)^{ 0.50}e^{ -4000/T}$ & \citet{Millar1997}\\
 \refstepcounter{reaction}R\arabic{reaction}   & S + NH   & $\!\!\!\rightarrow\!\!$ &  HS  + N  & $\!\! 1.7\!\times\! 10^{-11} \left(T/298\right)^{ 0.50}e^{ -4000/T}$ & \citet{Millar1997}\\
 \refstepcounter{reaction}R\arabic{reaction}  & NH$_2$  + HS   &$\!\!\!\rightarrow\!\!$ &  NH$_3$ + S & $\!\! 5.0\!\times\! 10^{-12}e^{ -500/T}$ & \citet{Moses1995}\\
 \refstepcounter{reaction}R\arabic{reaction}  & HS  + CH$_2$ &$\!\!\!\rightarrow\!\!$ &  S  + CH$_3$ & $\!\! 4.0\!\times\! 10^{-12}e^{ -500/T}$ & \citet{Moses1995}\\ 
 \refstepcounter{reaction}R\arabic{reaction}  & HS   + CH$_3$  &$\!\!\!\rightarrow\!\!$ &  S+ CH$_4$  & $\!\! 4.0\!\times\! 10^{-11}e^{ -500/T}$ & \cite{Shum1985}\\ 
 \refstepcounter{reaction}R\arabic{reaction}  & S  + HCO   &$\!\!\!\rightarrow\!\!$ &  HS  + CO  & $\!\! 6.0\!\times\! 10^{-11}$ & \citet{Moses1995}\\

\refstepcounter{reaction}R\arabic{reaction}   & H  + HS + M & $\!\!\!\rightarrow\!\!$ &  H$_2$S + M &$\!\! 1.4\!\times\! 10^{-31} \left(T/298 \right)^{-2.5}e^{ +500/T}$ &  see note \\
             & H  + HS  &$\!\!\!\rightarrow\!\!$&  H$_2$S   &$\!\! 1.0\!\times\! 10^{-10}$ & assumed \\
 
 \refstepcounter{reaction}R\arabic{reaction}   & S   + H$_2$    + M & $\!\!\!\rightarrow\!\!$ &  H$_2$S  + M &$\!\! 1.4\!\times\! 10^{-31} \left(T/298 \right)^{-1.9}e^{ -8140/T}$ &  see text \\
    & S  + H$_2$ + M & $\!\!\!\rightarrow\!\!$ &  H$_2$S + M &$\!\! 1.4\!\times\! 10^{-31} \left(T/298 \right)^{-1.9}e^{ -2300/T}$ & alternate rate, see text\\
             & S   + H$_2$  &$\!\!\!\rightarrow\!\!$&  H$_2$S   &$\!\! 1.0\!\times\! 10^{-11}$ & assumed \\

 \refstepcounter{reaction}R\arabic{reaction}   & H + H$_2$S  & $\!\!\!\rightarrow\!\!$ &  HS   + H$_2$  & $\!\! 3.7\!\times\! 10^{-12} \left(T/298\right)^{1.94}e^{-455/T}$ & \citet{Pen1999}\\
 
 \refstepcounter{reaction}R\arabic{reaction}   & H$_2$S       + S & $\!\!\!\rightarrow\!\!$ &  HS + HS  & $\!\! 1.4\!\times\! 10^{-10}e^{ -3720/T}$ & \citet{Shiina1996}\\

 \refstepcounter{reaction}R\arabic{reaction}   & O+ H$_2$S      &$\!\!\!\rightarrow\!\!$ &  HS+ OH& $\!\! 9.2\!\times\! 10^{-12} e^{ -1800/T}$ & \citet{DeMore1997}\\
 \refstepcounter{reaction}R\arabic{reaction}   & OH     + H$_2$S  &$\!\!\!\rightarrow\!\!$ &  H$_2$O + HS & $\!\! 6.1\!\times\! 10^{-12} e^{  -81/T}$ & \citet{Atkinson2004} \\

 \refstepcounter{reaction}R\arabic{reaction}  & HS  + HCO  &$\!\!\!\rightarrow\!\!$ &  H$_2$S  + CO  & $\!\! 5.0\!\times\! 10^{-11}$ & like R20\\
 \refstepcounter{reaction}R\arabic{reaction}   & CH$_2$  + H$_2$S  &$\!\!\!\rightarrow\!\!$ &  CH$_3$ + HS & $\!\! 2.5\!\times\! 10^{-11} e^{  -750/T}$ & \citet{Darwin1995}\\

 \refstepcounter{reaction}R\arabic{reaction}   & H$_2$S   + CH$_3$   &$\!\!\!\rightarrow\!\!$ &  HS + CH$_4$   & $\!\! 2.1\!\times\! 10^{-13} e^{ -1160/T}$ & \citet{Perrin1988}\\

\hline
 \refstepcounter{reaction}R\arabic{reaction}  & O + HS &$\!\!\!\rightarrow\!\!$ &  SO + H  & $\!\! 7.0\!\times\! 10^{-11}$ & \citet{Sander2003}\\
 \refstepcounter{reaction}R\arabic{reaction}  & S   + OH &$\!\!\!\rightarrow\!\!$ &  H  + SO & $\!\! 6.6\!\times\! 10^{-11}$ & \citet{DeMore1997}\\
 \refstepcounter{reaction}R\arabic{reaction}  & O + S$_2$ &$\!\!\!\rightarrow\!\!$ &  SO  + S & $\!\! 1.1\!\times\! 10^{-11}$ & \citet{Hills1987}\\
 \refstepcounter{reaction}R\arabic{reaction}   & S+ O$_2$ & $\!\!\!\rightarrow\!\!$ &  SO  + O & $\!\! 1.5\!\times\! 10^{-13} \left(T/298\right)^{ 2.11}e^{  -730/T}$ & \citet{Lu2004}\\
 \refstepcounter{reaction}R\arabic{reaction}  & S$_3$+ O &$\!\!\!\rightarrow\!\!$ &  S$_2$ + SO  & $\!\!  2.0\!\times\! 10^{-11}e^{  -500/T}$ & \citet{Moses1995}\\
 \refstepcounter{reaction}R\arabic{reaction}   & S$_4$ + O  &$\!\!\!\rightarrow\!\!$ &  S$_3$ + SO   & $\!\! 2.0\!\times\! 10^{-11} e^{  -500/T}$ & \citet{Moses1995}\\

\hline
 \refstepcounter{reaction}R\arabic{reaction}   & S  + CO +M  &$\!\!\!\rightarrow\!\!$&  OCS  + M &$\!\! 3.6\!\times\! 10^{-34}\left(T/298 \right)^{-0.57}$ & see note \\ 
           & S            + CO          &$\!\!\!\rightarrow\!\!$&  OCS        &$\!\! 3.0\!\times\! 10^{-14}$ & see note \\ 
  \refstepcounter{reaction}R\arabic{reaction}   & O   + OCS         &$\!\!\!\rightarrow\!\!$ &  CO           + SO         & $\!\! 7.8\!\times\! 10^{-11} e^{ -2620/T}$ & \citet{Singleton1988}\\

 \refstepcounter{reaction}R\arabic{reaction}   & HS + CO &$\!\!\!\rightarrow\!\!$ &  OCS + H  & $\!\! 4.2\!\times\! 10^{-14} e^{ -7660/T}$ & \citet{Kurbanov1995}\\

 \refstepcounter{reaction}R\arabic{reaction}   & OCS  + S  & $\!\!\!\rightarrow\!\!$ &  CO + S$_2$  & $\!\! 1.5\!\times\! 10^{-13} \left(T/298\right)^{ 2.57}e^{ -1180/T}$ & \citet{Lu2006}\\
 \refstepcounter{reaction}R\arabic{reaction}   & O + OCS   &$\!\!\!\rightarrow\!\!$ &  S + CO$_2$ & $\!\! 8.3\!\times\! 10^{-11} e^{ -5530/T}$ & \citet{Singleton1988}\\
  \refstepcounter{reaction}R\arabic{reaction}   & OCS   + OH &$\!\!\!\rightarrow\!\!$ &  CO$_2$ + HS  & $\!\! 1.1\!\times\! 10^{-13} e^{ -1200/T}$ & \citet{Atkinson2004}\\
 \refstepcounter{reaction}R\arabic{reaction}  & S    + HCO   &$\!\!\!\rightarrow\!\!$ &  OCS + H  & $\!\! 6.0\!\times\! 10^{-11}$ & \citet{Moses1995}\\
 \refstepcounter{reaction}R\arabic{reaction}   & CO + S$_3$  &$\!\!\!\rightarrow\!\!$ &  S$_2$  + OCS   & $\!\! 1.0\!\times\! 10^{-11} e^{-10000/T}$ & see note\\
\hline
 \refstepcounter{reaction}R\arabic{reaction}   & O   + CS  &$\!\!\!\rightarrow\!\!$ &  CO  + S   & $\!\!  2.7\!\times\! 10^{-10} e^{  -760/T}$ & \citet{Atkinson2004}\\
 \refstepcounter{reaction}R\arabic{reaction}  & S   + CH  &$\!\!\!\rightarrow\!\!$ &  CS + H  & $\!\!  2.0\!\times\! 10^{-11}$ & assumed \\
\refstepcounter{reaction}R\arabic{reaction}  & OH  + CS   &$\!\!\!\rightarrow\!\!$ &  OCS  + H  & $\!\!  2.0\!\times\! 10^{-13}$ & assumed \\
\refstepcounter{reaction}R\arabic{reaction} & CH$_2$   +  S &$\!\!\!\rightarrow\!\!$& CS   +  H$_2$    &$\!\! 2.0\!\times\! 10^{-11}$ & assumed\\
\hline
 \refstepcounter{reaction}R\arabic{reaction}    & H  + S$_4$ + M  &$\!\!\!\rightarrow\!\!$ &  HS$_4$  + M & $\!\!  7.0\!\times\! 10^{-30}$  & assumed \\
            & H        + S$_4$       &$\!\!\!\rightarrow\!\!$&  HS$_4$   &$\!\!  7.0\!\times\! 10^{-11}$ & assumed\\
\refstepcounter{reaction}R\arabic{reaction}    & H  + S$_8$   &$\!\!\!\rightarrow\!\!$ &  HS$_4$   + S$_4$  & $\!\! 3.0\!\times\! 10^{-12} e^{  -5000/T}$ & assumed, see text \\
             & H  + S$_8$   &$\!\!\!\rightarrow\!\!$ &  HS$_4$   + S$_4$  & $\!\! 3.0\!\times\! 10^{-11} e^{  -2500/T}$ & alternate rate \\
\refstepcounter{reaction}R\arabic{reaction}   & H  + HS$_4$   &$\!\!\!\rightarrow\!\!$ &  H$_2$  + S$_4$  & $\!\! 1.0\!\times\! 10^{-10} e^{ -500/T}$ & assumed, see text\\
\refstepcounter{reaction}R\arabic{reaction}   & H  + HS$_4$   &$\!\!\!\rightarrow\!\!$ &  H$_2$S  + S$_3$  & $\!\! 1.0\!\times\! 10^{-10} e^{ -2500/T}$ & assumed, see text\\
\refstepcounter{reaction}R\arabic{reaction}   & OH  + HS$_4$   &$\!\!\!\rightarrow\!\!$ & H$_2$O + S$_4$  & $\!\! 3.0\!\times\! 10^{-11}e^{ -500/T} $ & assumed, see text \\
\refstepcounter{reaction}R\arabic{reaction} & NH$_2$ + HS$_4$  &$\!\!\!\rightarrow\!\!$ & NH$_3$ + S$_4$ & $\!\!  3.0\!\times\! 10^{-11}e^{ -500/T} $ & assumed, see text \\
\refstepcounter{reaction}R\arabic{reaction}  & HS + HS$_4$  &$\!\!\!\rightarrow\!\!$ &  H$_2$S  + S$_4$  & $\!\!  3.0\!\times\! 10^{-11} e^{ -2500/T}$ & assumed, see text \\
\refstepcounter{reaction}R\arabic{reaction}  & HS$_4$ + HS$_4$ &$\!\!\!\rightarrow\!\!$ & H$_2$ + S$_8$ & $\!\! 3.0\!\times\! 10^{-12} e^{ -2500/T}$ & assumed, see text\\
\refstepcounter{reaction}R\arabic{reaction}  & S + HS$_4$  &$\!\!\!\rightarrow\!\!$ &  HS  + S$_4$  & $\!\!  3.0\!\times\! 10^{-11} e^{ -1000/T}$ & assumed, see text \\
\hline
\refstepcounter{reaction}R\arabic{reaction} & C$_2$H + C$_2$H$_2$  &$\!\!\!\rightarrow\!\!$ & C$_4$H$_2$ + H & $\!\! 1.25\times 10^{-10} \left(T/298 \right)^{0.24}e^{ 37/T}$    & \citet{Eiteneer2003} \\
\refstepcounter{reaction}R\arabic{reaction} & C$_4$H$_2$ + H +\! M  &$\!\!\!\rightarrow\!\!$ & C$_4$H$_3$ + M & $\!\!  5.9\!\times\! 10^{-25}\! \left(T/298 \right)^{-8.9}\! e^{ -1260/T}$    & \citet{Klippenstein2005}\\
                    & C$_4$H$_2$ + H + \!M  &$\!\!\!\rightarrow\!\!$ & C$_4$H$_3$ + M &   $\!\! 5.2\times 10^{-11} \left(T/298 \right)^{1.2} e^{ -882/T}$  & \citet{Klippenstein2005}\\  
\refstepcounter{reaction}R\arabic{reaction} & C$_4$H$_3$ + H  &$\!\!\!\rightarrow\!\!$ & C$_4$H$_2$ + H$_2$ &  $\!\! 5\times 10^{-11} e^{-500/T}$   & see text\\
\refstepcounter{reaction}R\arabic{reaction} & C$_4$H$_3$ + H  &$\!\!\!\rightarrow\!\!$ & C$_2$H$_2$  +\!  C$_2$H$_2\!$ & $\!\! 5\times 10^{-11} e^{-2000/T} $   & see text \\
\refstepcounter{reaction}R\arabic{reaction} & C$_4$H$_3$ + H  &$\!\!\!\rightarrow\!\!$ & C$_4$H$_4$  & $\!\! 1.8\times 10^{-10} $   & \citet{Harding2007}\\

\hline
\hline
\multicolumn{6}{l}{ $^{\ast}$ M refers to the background atmosphere, principally H$_2$ and He; units of density [cm$^{-3}$].}\\
\multicolumn{6}{l}{R2-R8. These assumed rates are generally consistent with those of \citet{Moses2002,Yung2009}.}\\
\multicolumn{6}{l}{R5. Reported rates are $1\times 10^{-29}$ \citep{Langford1973}, $2.2\times 10^{-29}$ \citep{Nicholas1979}.}\\
\multicolumn{6}{l}{R8. Reported rates are $4\times 10^{-11}$ \citep{Schofield1973}, $<5\times 10^{-12}$ \citep{Nicholas1979}.}\\
\multicolumn{6}{l}{R9r. A blend of \citet{Woiki1995a} and \citet{Shiina1998}.}\\
\multicolumn{6}{l}{R21. This is the reverse of \citet{Shiina1998} upper bound on $\mathrm{H}_2\mathrm{S} + \mathrm{M} \rightarrow \mathrm{H}+\mathrm{HS} + \mathrm{M} $.}\\
\multicolumn{6}{l}{R36. These are reverses of \citet{Oya1994} rate for $\mathrm{OCS} + \mathrm{M} \rightarrow \mathrm{CO}+\mathrm{S} + \mathrm{M} $ and of }\\
\multicolumn{6}{l}{\phantom{R36. }\citet{Schofield1973} high pressure limit $\mathrm{OCS} \rightarrow \mathrm{CO}+\mathrm{S} $.}\\
\multicolumn{6}{l}{R43. Exothermic, but the analogous $\mathrm{CO} + \mathrm{O}_3 \rightarrow \mathrm{CO}_2+\mathrm{O}_2$ has an upper limit at 298 K of $4\times 10^{-25}$.}\\
\multicolumn{6}{l}{R45-R47. There is little information re reactions of CS.}\\
\multicolumn{6}{l}{R48-R61. HS$_4$ and C$_4$H$_3$ are invoked to create sinks on S$_8$ and C$_4$H$_2$, respectively.}\\
\end{longtable}


\begin{thebibliography}{00}

\bibitem[Abelson(1966)]{Abelson1966} 
Abelson, P. (1966).
Chemical events on the primitive Earth.
{\em Proc.\ Natl.\ Acad.\ Sci.\ USA 55,} 1365-1372.

\bibitem[Ag{\'u}ndez et al.(2012)]{Agundez2012}
Ag{\'u}ndez, M., Venot, O., Iro, N., Selsis, F., Hersant, F., H{\'e}brard, E., Dobrijevic, M., (2012).
The impact of atmospheric circulation on the chemistry of the hot Jupiter HD 209458b.
{\em Astron.\ Astrophys.\ 548,} A73.

\bibitem[Ag{\'u}ndez et al.(2014a)]{Agundez2014a}
Ag{\'u}ndez, M., Parmentier, V., Venot, O., Hersant, F., Selsis, F., (2014).
Pseudo 2D chemical model of hot-Jupiter atmospheres: application to HD 209458b and HD 189733b.
{\em Astron.\ Astrophys.\ 564,} A73.

\bibitem[Ag{\'u}ndez et al.(2014b)]{Agundez2014b}
Ag{\'u}ndez, M., Venot, O., Selsis, F., Iro, N., (2014).
 The Puzzling Chemical Composition of GJ 436b's Atmosphere: Influence of Tidal Heating on the Chemistry.
{\em Astrophys.\ J.\  781,} 781.

\bibitem[Anders and Grevasse(1989)]{Anders1989}
Anders, E., Grevasse, N., (1989).
Abundances of the elements: Meteoritic and solar.
{\em Geochim.\ Cosmochim.\ Acta 53,} 197-214.

\bibitem[Atkinson et al.(2004)]{Atkinson2004}
Atkinson, R., Baulch, D.L., Cox, R.A., Crowley, J.N., Hampson, R.F., Hynes, R.G., Jenkin, M.E., Rossi, M.J., Troe, J. (2004),
 Evaluated kinetic and photochemical data for atmospheric chemistry: Volume I - gas phase reactions of Ox, HOx, NOx and SOx species. 
 {\it Atmos. Chem. Phys.  4,} 1461-1738.

\bibitem[Benneke(2015)]{Benneke2015}
  Benneke, B., (2015).
  Strict Upper Limits on the Carbon-to-Oxygen Ratios of Eight Hot Jupiters from Self-Consistent Atmospheric Retrieval.
  ArXiv e-print 1504.07655.

\bibitem[Benson(1978)]{Benson1978}
Benson, S.W. (1978).
Thermochemistry and kinetics of sulfur-containing molecules and radicals.
{\em Chem.\ Rev.\ 78,} 23-35.  DOI: 10.1021/cr60311a003.

\bibitem[Bowman and Dodge(1977)]{Bowman1977}
Bowman, C.T., Dodge, L.G., (1977).
Kinetics of the Thermal Decomposition of Hydrogen Sulfide Behind Shock Waves.
{\em 16$^{th}$ Symp.\ (Int.) Combust.\ 16,} 971.

\bibitem[Darwin \& Moore(1995)]{Darwin1995}
Darwin, D.C., Moore, C.B. (1995).  Reaction rate constants (295 K) for $^3$CH2 with H2S, SO2, and NO2: upper bounds for rate constants with less reactive partners. {\em J. Phys. Chem. 99,} 13467-13470.

\bibitem[DeMore et al.(1997)]{DeMore1997}
DeMore, W.B., Sander, S.P., Golden, D.M., Hampson, R.F., Kurylo, M.J., Howard, C.J., Ravishankara, A.R., Kolb, C.E., Molina, M.J. (1997). {\em Chemical kinetics and photochemical data for use in stratospheric modeling. Evaluation number 12.}  JPL Publication 97-4.

\bibitem[Du et al.(2008)]{Du2008}
 Du, S.Y., Francisco, J.S., Shepler, B.C., Peterson, K.A. (2008). Determination of the rate constant for sulfur recombination by quasiclassical trajectory calculations. {\em J. Chem. Phys. 128,} 204306.
 
\bibitem[Eiteneer and Frenklach(2003)]{Eiteneer2003}
Eiteneer, B., Frenklach, M., (2003).
  Experimental and Modeling Study of Shock-Tube Oxidation of Acetylene.
  {\em Int.\ J.\ Chem.\ Kinet.\ 35,} 391-414.
 
\bibitem[Fair and Thrush(1969)]{Fair1969}
Fair, R.W., Thrush, B.A. (1969).  Mechanism of S2 chemiluminescence in the reaction of hydrogen atoms with hydrogen sulphide. { \em Trans. Faraday Soc. 65,} 1208-1218.


\bibitem[Freytag et al.(2010)]{Freytag2010}
Freytag, B., Allard, F., Ludwig, H.-G., Homeier, D., Steffen, M. (2010). 
The role of convection, overshoot, and gravity waves for the transport of dust in M dwarf and brown dwarf atmospheres.
{\em Astron.\ Astrophys.\ 513,} A19 (14 pp.).

\bibitem[Griffith et al.(1998)]{Griffith1998}
Griffith, C.A., Yelle, R.V., Marley, M.S. (1998).
The Dusty Atmosphere of the Brown Dwarf Gliese 229B.
{\em Science 282,} 2063-2065.

\bibitem[Hansen et al.(2006)]{Hansen2006}
Hansen, N., Klippenstein, S.J., Taatjes, C.A., Miller, J.A., Wang, J.,
Cool, T.A., Yang, B., Yang, R., Wei, L., Huang, C., Wang, J., Qi, F., Law, M.E., Westmoreland, P.R., (2006).
Identification and Chemistry of C4H3 and C4H5 Isomers in Fuel-Rich Flames.
{\em J.\ Phys.\ Chem.\ A, 110,} 3670-3678.

\bibitem[Harding et al.(2007)]{Harding2007}
Harding, L.B., Klippenstein, S.J., Georgievskii, Y., (2007).
On the combination reactions of hydrogen atoms with resonance-stabilized hydrocarbon radicals.
{\em J.\ Phys.\ Chem.\ A 111,} 3789-3801.

\bibitem[Harrington et al.(2004)]{Harrington2004}
Harrington, J, de Pater, I., Brecht, S.H., Deming, D., Meadows, V., Zahnle, K.J., Nicholson, P. (2004).
Lessons from Shoemaker-Levy 9 about Jupiter and Planetary Impacts.
In {\em Jupiter: The Planet, Satellites and Magnetosphere,} F. Bagenol, T. Dowling, W. McKinnon, eds.
Cambridge Univ. Press. pp. 158-184.

\bibitem[Helling and Casewell(2014)]{Helling2014}
Helling, C., Casewell, S. (2014).
Atmospheres of brown dwarfs.
{\em Astron.\ Astrophys.\ Rev.\ 22,} 80 (45 pp.).

\bibitem[Hills et al.(1987)]{Hills1987}
Hills, A.J., Cicerone, R.J., Calvert, J.G., Birks, J.W. (1987).
Kinetics of the reactions of S2 with O, O2, O3, N2O, NO, and NO2. 
{\em J. Phys. Chem. 91,} 1199-1204.

\bibitem[Hu and Seager(2014)]{Hu2014}
Hu, R., Seager, S., (2014).
Photochemistry in Terrestrial Exoplanet Atmospheres. III.
Photochemistry and Thermochemistry in Thick Atmospheres on Super Earths and Mini Neptunes.
{\em Astrophys.\ J.\  784,} 63 (25pp). 

\bibitem[Hu et al.(2014)]{Hu2013}
Hu, R., Seager, S., Bains, W. (2013).
Photochemistry in terrestrial exoplanet atmospheres.
II. H$_2$S and SO$_2$ photochemistry in anoxic atmospheres.
{\em Astrophys.\ J.\  769,} 6 (14pp) 

\bibitem[Hubeny and Burrows(2007)]{Hubeny2007}
Hubeny, I., Burrows, A. (2007). 
A systematic study of departures from chemical equilibrium in the atmospheres of substellar mass objects.
{\em Astrophys.\ J.\  669,} 1248-1261.

\bibitem[Karan et al.(1999)]{Karan1999}
Karan, K., Mehrotra, A.K., Leo A. Behie, L.A., (1999).
On reaction kinetics for the thermal decomposition of hydrogen sulfide.
{\em Amer.\ Inst.\ Chem.\ Eng.\ J.\ 45,} 383-389.

\bibitem[Kasting et al.(1989)]{Kasting1989}
Kasting, J.F., Zahnle, K.J., Pinto, J., Young, A. (1989). 
Sulfur, ultraviolet radiation, and the early evolution of life.
{\em Origins of Life 19,} 95-108.

\bibitem[Klippenstein and Miller(2005)]{Klippenstein2005}
Klippenstein, S.J., Miller, J.A., (2005).
The addition of hydrogen atoms to diacetylene and the heats of formation of i-C4H3 and n-C4H3.
{\em J.\ Phys.\ Chem.\ A 109,} 4285-4295.

\bibitem[Kopparapu et al.(2012)]{Kopparapu2012}
Kopparapu, R.K., Kasting, J.F., Zahnle, K.J., (2012){{Kopparapu}, R.~k. and {Kasting}, J.~F. and {Zahnle}, K.~J.},
A Photochemical Model for the Carbon-rich Planet WASP-12b.
{\em Astrophys.\ J.\  745,} 77. 

\bibitem[Koskinen et al.(2013)]{Koskinen2013}
Koskinen, T.T., Harris, M.J., Yelle, R.V., Lavvas, P., (2013).
The escape of heavy atoms from the ionosphere of HD 209458b. I. A photochemical-dynamical model of the thermosphere.
{\em Icarus 226,} 1678-1694.

\bibitem[Kurbanov et al.(1995)]{Kurbanov1995}
Kurbanov, M.A., Mamedov, Kh.F.  (1995). The role of the reaction CO + SH $\rightarrow$ COS + H in hydrogen formation in the course of interaction between CO and H2S.  {\em Kinet. Catal.  36,} 455-457.

\bibitem[Langford and Oldershaw(1973)]{Langford1973}
Langford, R.B., Oldershaw, G.A., (1973).
Mechanism of Sulfur Formation in the Flash Photolysis of Carbonyl Sulphide.
{\em J.\ Chem.\ Soc.\ Faraday Trans.\ 69,} 1389.

\bibitem[Langford and Oldershaw(1972)]{Langford1972}
Langford, R.B., Oldershaw, G.A., (1972).
Flash Photolysis of H$_2$S.
{\em J.\ Chem.\ Soc.\ Faraday Trans.\ 68,} 1550.

\bibitem[Lavvas et al.(2014)]{Lavvas2014}
Lavvas, P., Koskinen, T., Yelle, R.V., (2014).
Electron Densities and Alkali Atoms in Exoplanet Atmospheres.
{\em Astrophys.\ J.\ 796,} (15 pp).

\bibitem[Lee et al.(1977)]{Lee1977}
Lee, J.H., Stief, L.J., Timmons, R.B. (1977). Absolute Rate Parameters for the Reaction of Atomic Hydrogen with Carbonyl Sulfide and Ethylene Episulfide,  {\em J. Chem. Phys. 67,} 1705-1714.

\bibitem[Liang et al.(2003)]{Liang2003}
Liang, M.C., Parkinson, C.D., Lee, A.Y.-T., Yung, Y.L., Seager, S. (2003).
Source of atomic hydrogen in the atmosphere of HD 209458b.
{\em Astrophys.\ J.\ 596,} L247-L250.

\bibitem[Liang et al.(2004)]{Liang2004}
Liang, M.C., Seager, S., Parkinson, C.D., Lee, A.Y.-T., Yung, Y.L. (2003).
On the Insignificance of Photochemical Hydrocarbon Aerosols in the Atmospheres of Close-in Extrasolar Giant Planets
Astrophys.\ J.\ 605, L61-L64.

\bibitem[Line et al.(2010)]{Line2010}
 Line, M.R., Liang, M.-C., Yung, Y.L. (2010). 
 High-Temperature Photochemistry in the Atmosphere of HD 189733b.
  {\em Astrophys.\ J.\ 717,} 496-502.

\bibitem[Line et al.(2011)]{Line2011}
Line, M.R., Vasisht, G., Chen, P., Angerhausen, D., Yung, Y.L. (2011).
Thermochemical and Photochemical Kinetics in Cooler Hydrogen-dominated Extrasolar Planets: A Methane-poor GJ436b? 
{\em  Astrophys.\ J.\ 738,} article id. 32, 14 pp.


\bibitem[Lodders and Fegley(2006)]{Lodders2006}
Lodders, K., Fegley, B., (2006).  In {\em Chemistry of Low Mass Substellar
Objects.} Ed. J.W.\ Mason (Berlin: Springer), pp.\ 1-31.

\bibitem[Lu et al.(2004)]{Lu2004}
Lu, C.W., Wu, Y.J., Lee, Y.P., Zhu, R.S., Lin, M.C. (2004). Experimental and theoretical investigations of rate coefficients of the reaction S($^3$P) + O2 in the temperature range 298-878 K.  {\em J.\ Chem.\ Phys.\ 121,} 8271-8278

\bibitem[Lu et al.(2006)]{Lu2006}
Lu, C.W.,Wu, Y.J.,Lee, Y.P.,Zhu, R.S.,Lin, M.C. (2006).  Experimental and theoretical investigation of rate coefficients of the reaction S($^3$P) + OCS in the temperature range of 298-985 K.  {\em J.\ Chem.\ Phys.\ 125,} 164329.

\bibitem[Lyons(2008)]{Lyons2008}
Lyons, J.R. (2008). 
An estimate of the equilibrium speciation of sulfur vapor over solid sulfur and implications for planetary atmospheres.
{\em J.\ Sulfur Chem.\ 29,} 269-279.

\bibitem[Macintosh et al.(2015)]{Macintosh2015}
Macintosh, B., Graham, J.R., Barman, T., De Rosa, R.J., Konopacky, Q., Marley, M.S., Marois, C., Nielsen, E.L., Pueyo, L., Rajan, A., and 72 coauthors (2015).  Discovery and spectroscopy of the young jovian planet 51 Eri b with the Gemini Planet Imager. {\em Science 350,} 64-67.

\bibitem[Marley et al.(2007)]{Marley2007}
 Marley, M.S., Fortney, J.J., Seager, S., Barman, T. (2007).
  An Imposing and Comprehensive Title. 
  In {\em Protostars and Planets V.} Eds.\ B.~Reipurth, D.~Jewitt, and K.~Keil.
 Univ.\ Ariz.\ Press, Tucson, pp.\ 733--747.
 
 \bibitem[Meyer(1976)]{Meyer1976}
 Meyer, B., (1976). Elemental sulfur.
 {\em Chemical Reviews 76,} 367-387.

\bibitem[Miguel and Kaltenegger(2014)]{Miguel2014}
Miguel, Y., Kaltenegger, L., (2014).
Exploring Atmospheres of Hot Mini-Neptunes and Extrasolar Giant Planets Orbiting Different Stars with Application to HD 97658b, WASP-12b, CoRoT-2b, XO-1b, and HD 189733b,
{\em Astrophys.\ J.\  780,} 166. 

\bibitem[Miguel et al.(2015)]{Miguel2015}
Miguel, Y., Kaltenegger, L., Linsky, J.L., Rugheimer, S., (2015)
The effect of Lyman $\alpha$ radiation on mini-Neptune atmospheres around M stars: application to GJ 436b.
{\em Mon.\ Not.\ Roy.\ Astron.\ Soc.\  446,} 345-353. 

\bibitem[Millar et al.(1997)]{Millar1997}
Millar, T.J., Farquhar, P.R.A., Willacy, K., (1997).
The UMIST database for astrochemistry 1995.
{\em Astron.\ Astrophys.\ Suppl.\ Ser.\ 121,} 139-185.
 
\bibitem[Miller-Ricci Kempton et al.(2012)]{Miller-Ricci2012}
Miller-Ricci Kempton, E., Zahnle, K., Fortney, J.J. (2011). 
 The Atmospheric Chemistry of GJ 1214b: Photochemistry and Clouds.
 {\em Astrophys.\ J.\ 745}, 3 (13pp).
 
 \bibitem[Morley et al.(2012)]{Morley2012}
 Morley, C.V., Fortney, J.J., Visscher, C., Marley, M.S., Sauman, D., Legett, S.K. (2012).
 Neglected clouds in T and Y dwarf atmospheres.
 {\em Astrophys.\ J.\ 756,} article id.\ 172, 17 pp.

\bibitem[Morley et al.(2013)]{Morley2013}
Morley, C.V., Fortney, J.J., Kempton, E.M.R., Marley, M.S., Visscher, C., Zahnle, K.J. (2013).
Quantitatively Assessing the Role of Clouds in the Transmission Spectrum of GJ 1214b.
{\em Astrophys.\ J.\ 775,} article id.\ 33, 13 pp.

\bibitem[Morley et al.(2015)]{Morley2015}
Morley, C.V., Fortney, J.J., Kempton, E.M.R., Marley, M.S., Visscher, C., Zahnle, K.J. (2013).
Quantitatively Assessing the Role of Clouds in the Transmission Spectrum of GJ 1214b.
{\em Astrophys.\ J.\ 775,} article id.\ 33, 13 pp.

\bibitem[Moses and Nash(1991)]{Moses1991}
Moses, J.I., Nash, D.B., (1991).
Phase transformations and the spectral reflectance of solid
sulfur: can metastable sulfur aIIotropes exist on Io?
{\em Icarus 89,} 277-304.

\bibitem[Moses et al.(1995)]{Moses1995}
Moses, J.I. Allen, M., Gladstone, G.R. (1995).
Post-SL9 sulfur photochemistry on Jupiter.
{\em Geophys.\ Res.\ Lett.\ 22,} 1597-1600.

\bibitem[Moses et al.(2002)]{Moses2002}
Moses, J.I., Zolotov, M.Y.,  Fegley, B. (2002).
 Photochemistry of a volcanically driven atmosphere on Io: Sulfur and oxygen species from a Pele-type eruption.
{\em Icarus 156,} 76-106.

\bibitem[Moses et al.(2005)]{Moses2005}
Moses, J.I., Fouchet, F., B{'e}zard, B., Gladstone, G.R., Lellouch, E., Feuchtgruber, H. (2005).
Photochemistry and diffusion in Jupiter's stratosphere: Constraints from ISO observations and comparisons with other giant planets.
{\em J.\ Geophys.\ Res.\ 110,} E08001 (45 pp.).

\bibitem[Moses et al.(2011)]{Moses2011} 
Moses, J.I., Visscher, C., Fortney, J.J., Showman, A.P., Lewis, N.K., Griffith, C.A., Klippenstein, S.J., Shabram, M., Friedson, A.J., Marley, M.S., Freedman, R.S. (2011).
Disequilibrium Carbon, Oxygen, and Nitrogen Chemistry in the Atmospheres of HD 189733b and HD 209458b. 
{\em Astrophys.\ J.\ 737,} article id. 15, 37 pp.

\bibitem[Moses et al.(2013a)]{Moses2013a} 
Moses, J.I., Madhusudhan, N., Visscher, C., Freedman, R.S. (2013).
Chemical Consequences of the C/O Ratio on Hot Jupiters: Examples from WASP-12b, CoRoT-2b, XO-1b, and HD 189733b.
{\em Astrophys.\ J.\ 763,} article id. 25, 26 pp.

\bibitem[Moses et al.(2013b)]{Moses2013b} 
Moses, J.I., Line, M.R., Visscher, C., Richardson, M.R., Nettelmann, N., Fortney, J.J., Barman, T.S., Stevenson, K.B., Madhusudhan, N. (2013).
Compositional Diversity in the Atmospheres of Hot Neptunes, with Application to GJ 436b.
{\em Astrophys.\ J.\ 777,} article id. 34, 23 pp.

\bibitem[Moses(2014)]{Moses2014} 
 Moses, J.I. (2014).
 Chemical Kinetics on Extrasolar Planets. 
 {\em Phil.\ Trans.\ Roy.\ Soc.\ A 372,} 20130073. DOI: 10.1098/rsta.2013.0073.
 
\bibitem[Nicholas et al.(1979)]{Nicholas1979}
 Nicholas, J.E., Amodio, C.A., Baker, M.J. (1979).
Kinetics and Mechanism of the Decomposition of H$_2$S, CH$_2$SH and (CH$_3$)$_2$S in a Radio-frequency Pulse Discharge
{\em J.\ Chem.\ Soc.\ Faraday Trans.\ 75,} 1868.
 
\bibitem[Olschewski et al.(1994)]{Olschewski1994}
Olschewski, H.A., J. Troe, Wagner, H.Gg., (1994).
UV absorption study of the thermal decomposition reaction $\mathrm{H}_2\mathrm{S} \rightarrow \mathrm{H}_2 + \mathrm{S}(^3\mathrm{P})$.
{\em J.\ Phys.\ Chem.\ 98,} 12964-12967.
 
\bibitem[Oya et al.(1994)]{Oya1994}
Oya, M., Shiina, H., Tsuchiya, K., Matsui, H. (1994).
Thermal decomposition of COS.
{\em Bull.\ Chem.\ Soc.\ Japan 67,} 2311-2313.

\bibitem[Pauling(1970)]{Pauling1970}
Pauling, L. (1970).
{\em General Chemistry.} Dover.

\bibitem[Pavlov and Kasting(2002)]{Pavlov2002}
Pavlov, A.A., Kasting, J.F., (2002).
{\em Astrobiology}

\bibitem[Pen et al.(1999)]{Pen1999}
Pen, J., Hu, X., Marshall, P. (1999). 
Experimental and {\it ab initio} investigations of the kinetics of the reaction of H atoms with H$_2$S.
{\em J.\ Phys.\ Chem.\ A, 103,} 5307-5311.

\bibitem[Perrin et al.(1988)]{Perrin1988}
Perrin, D., Richard, C., Martin, R. (1988). Etude cinetique de la reaction thermique du pentene-2 cis vers 500$^{\circ}$C. III - Influence de H$_2$S. {\em  J. Chim. Phys. 85,} 185-192.

\bibitem[Pinto et al.(1980)]{Pinto1980}
Pinto, J., Gladstone, R., Yung, Y., (1980).
Photochemical production of formaldehyde in the Earth's primitive atmosphere. 
{\em Science 210,} 183-185.

\bibitem[Rimmer and Helling(2016)]{Rimmer2016}
 Rimmer, P.B., Helling, C., (2016).
 A Chemical Kinetics Network for Lightning and Life in Planetary Atmospheres.
 {\em Astrophys.\ J.\ Supp.,} {\it accepted.} arXiv:1510.07052.


\bibitem[Roth et al.(1982)]{Roth1982}
Roth, P., Lohr, R., Braner, U., (1982).
Thermal Decomposition of Hydrogen Sulfide at Low Concentrations.
{\em Combust.\ Flame 45,} 273.

\bibitem[Sander et al.(2003)]{Sander2003}
Sander, S.P., Friedl, R.R.,  Ravishankara, A.R., Golden, D.M., Kolb, C.E., Kurylo, M.J., Huie, R.E., Orkin, V.L., Molina, M.J., Moortgat, G.K., Finlayson-Pitts, B.J.\ (2003).
{\em Chemical Kinetics and Photochemical Data for Use in Atmospheric Studies. Evaluation Number 14.} JPL Publication 02-25.

\bibitem[Schofield(1973)]{Schofield1973}
Schofield, K.  (1973).  Evaluated chemical kinetic rate constants for various gas phase reactions.  {\em J. Phys. Chem. Ref. Data 2,} 25-84.

\bibitem[Shiina et al.(1996)]{Shiina1996}
Shiina, H., Oya, M., Yamashita, K., Miyoshi, A., Matsui, H. (1996). Kinetic studies on the pyrolysis of H$_2$S. {\em  Phys. Chem. 100,} 2136-2140.

\bibitem[Shiina et al.(1998)]{Shiina1998}
Shiina, H., Miyoshi, A., Matsui, H. (1998). Investigation on the insertion channel in the S($^3$P) + H$_2$ reaction. 
{\em J. Phys. Chem. A 102,} 3556 - 3559.

\bibitem[Shum \& Benson(1985)]{Shum1985}
Shum, L.G.S., Benson, S.W., (1985). 
The pyrolysis of dimethyl sulfide, kinetics and mechanism.
{\em Int. J. Chem. Kinet. 17,} 749.

\bibitem[Singleton and Cvetanovic(1988)]{Singleton1988}
Singleton, D.L., Cvetanovic, R.J.  (1988). Evaluated chemical kinetic data for the reactions of atomic oxygen O($^3$P) with sulfur containing compounds.  {\em J. Phys. Chem. Ref. Data 17,} 1377-1399.

\bibitem[Spencer et al.(2000)]{Spencer2000}
Spencer, J.R., Jessup, K.L., McGrath, M.A., Gilda E. Ballester, G.E., Roger Yelle, R.V. (2000).
Discovery of gaseous S$_2$ in Io's Pele plume.
{\em Science 288,} 1208-1210.

\bibitem[Tesner et al.(1990)]{Tesner1990}
Tesner, P.A., Nemirovskii, M.S., Motyl, D.N. (1990).
Kinetics of the thermal decomposition of hydrogen sulfide at 600-1200$^{\circ}$C.
{\em Kinet.\ Catal.\ 31,} 1081-1083.

\bibitem[Tsuchiya et al.(1996)]{Tsuchiya1996}
Tsuchiya, K.,Yamashita, K., Miyoshi, A., Matsui, H. (1996).
Studies on the reactions of atomic sulfur ($^3$P) with H$_2$, D$_2$, CH$_4$, C$_2$H$_6$, C$_3$H$_8$, n-C$_4$H$_{10}$, and
i-C$_4$H$_{10}$.
{\em J.\ Phys.\ Chem.\ 100,} 17202-17206.

\bibitem[Tsuchiya et al.(1997)]{Tsuchiya1997}
Tsuchiya, K., Kamiya, K., Matsui, H., (1997).
 Studies on the Oxidation Mechanism of H$_2$S Based on Direct Examination of the Key Reaction.
 {\em Int.\ J.\ Chem.\ Kinet.\ 29,} 57.
 
\bibitem[Venot et al.(2012)]{Venot2012}
Venot, O., H{\'e}brard, E., Ag{\'u}ndez, M., Dobrijevic, M., Selsis, F., Hersant, F., Iro, N., Bounaceur, R., (2012).
A chemical model for the atmosphere of hot Jupiters.
{\em Astron.\ Astrophys.\ 546,} A43.

\bibitem[Venot et al.(2013)]{Venot2013}
Venot, O., Fray, N., B{\'e}nilan, Y., Gazeau, M.-C., H{\'e}brard, E., Larcher, G., Schwell, M., Dobrijevic, M., Selsis, F., (2013).
High-temperature measurements of VUV-absorption cross sections of CO$_{2}$ and their application to exoplanets.
{\em Astron.\ Astrophys.\ 551,} A131.

\bibitem[Venot et al.(2014)]{Venot2014}
Venot, O., Ag{\'u}ndez, M., Selsis, F., Tessenyi, M., Iro, N., (2014).
The atmospheric chemistry of the warm {Neptune} {GJ} 3470b: {Influence} of metallicity and temperature on the {CH$_{4}$/CO} ratio.
{\em Astron.\ Astrophys.\ 562,} A51.

\bibitem[Venot et al.(2015)]{Venot2015}
Venot, O., H{\'e}brard, E., Ag{\'u}ndez, M., Decin, L., Bounaceur, R., (2015).
New chemical scheme for studying carbon-rich exoplanet atmospheres.
{\em Astron.\ Astrophys.\ 577,} A33.

\bibitem[Visscher et al.(2006)]{Visscher2006}
Visscher, C.W., Lodders, K., Fegley, B. (2006).
Atmospheric chemistry in giant planets, brown dwarfs, and low-mass dwarf stars. ii. Sulfur and phosphorus.
{\em Astrophys.\ J.\ 648,} 1181-1195.

\bibitem[Visscher and Moses(2011)]{Visscher2011}
Visscher, C.W., Moses, J.I. (2011).
Quenching of carbon monoxide and methane in the atmospheres of cool brown dwarfs and hot Jupiters. 
{\em Astrophys.\ J.\ 738,} article id. 72, 12 pp.

\bibitem[Visscher(2012)]{Visscher2012}
Visscher, C.W. (2012).
Chemical timescales in the atmospheres of highly eccentric exoplanets.
{\em Astrophys.\ J.\ 757,} article id. 5, 8 pp.

\bibitem[Vuitton et al.(2012)]{Vuitton2012}
 Vuitton, V., Yelle, R.V., Lavvas, P., Klippenstein, S.J. (2012).
Rapid association reactions at low pressure: impact on the formation of hydrocarbons on Titan.
{\em Astrophys.\ J.\ 744,} article id. 11, 7 pp.

\bibitem[Woiki and Roth(1994)]{Woiki1994}
Woiki, D.; Roth, P. (1994).
Kinetics of the high-temperature H$_2$S decomposition.
{\em J.\ Phys.\ Chem.\ 98,} 12958-12963.

\bibitem[Woiki and Roth(1995a)]{Woiki1995a}
Woiki, D., Roth, P., (1995a).
A shock tube study of the reaction $\mathrm{S} + \mathrm{H}_2 = \mathrm{SH} + \mathrm{H}$ in pyrolysis and photolysis systems.
{\em Int.\ J.\ Chem.\ Kinet.\ 27,}  547-553.

\bibitem[Woiki and Roth(1995b)]{Woiki1995b}
Woiki, D., Roth, P. (1995b).
Oxidation of S and SO by O$_2$ in high-temperature pyrolysis and photolysis reaction systems.
{\em Int.\ J.\ Chem.\ Kinet.\ 27,} 5-71.

\bibitem[Young et al.(1983)]{Young1983}
Young A.T. (1983).
Venus cloud microphysics.
 {\em Icarus 56,} 568.

\bibitem[Yung et al.(2009)]{Yung2009}
Yung, Y.L., Liang, M.C., Jiang, X., Shia, R.L., Lee, C., B{\'e}zard B.,  Marcq, E., (2009).
Evidence for carbonyl sulfide (OCS) conversion to CO in the lower atmosphere of Venus.
{\em J.\ Geophys.\ Res.\ 114,} E00B34. 

\bibitem[Zahnle et al.(1995)]{Zahnle1995} 
Zahnle, K.J, Mac Low, M.-M., Lodders, K., B. Fegley, B. (1995). 
Sulfur chemistry in the wake of Comet Shoemaker-Levy 9.
{\em Geophys.\ Res.\ Lett.\ 22,} 1593-1596.

\bibitem[Zahnle et al.(2009)]{Zahnle2009}
Zahnle, K.J., Marley, M.S., Freedman, R.S., Lodders, K., Fortney, J.J. (2009).
Atmospheric sulfur chemistry on hot Jupiters.  {\em Astrophys.\ J.\ Lett.\ 701,} L20-L24.

\bibitem[Zahnle and Marley(2014)]{Zahnle2014}
Zahnle, K.J., Marley, M.S. (2014).
Methane, Carbon Monoxide, and Ammonia in Brown Dwarfs and Self-Luminous Giant Planets.
{\em Astrophys.\ J.\ 797,} article id.\ 41, 19 pp.

 \end{thebibliography}
 \end{document}